\documentclass{emulateapj} 
\usepackage{hyperref}

\newcommand{\kms}{{\hbox {km\thinspace s$^{-1}$}}}
\newcommand{\Lsun}{{\hbox {L$_\odot$}}}
\newcommand{\Msun}{{\hbox {M$_\odot$}}}

\newcommand{\cmd}{{\hbox {cm$^{-2}$}}}

\newcommand{\hdo}{{\hbox {H$_{2}$O}}}

\newcommand{\tdust}{{\hbox {$T_{\mathrm{dust}}$}}}
\newcommand{\taud}{{\hbox {$\tau_{\mathrm{100}}$}}}

\slugcomment{To appear in ApJ}

\shorttitle{OH outflows in local ULIRGs}
\shortauthors{Gonz\'alez-Alfonso et al.}

\begin{document}


\title{Molecular outflows in local ULIRGs: 
energetics from multi-transition OH analysis}


\author{E. Gonz\'alez-Alfonso\altaffilmark{1,2}, 
        J. Fischer\altaffilmark{3}, 
        H. W. W. Spoon\altaffilmark{4},
        K. P. Stewart\altaffilmark{3},
        M. L. N. Ashby\altaffilmark{2},
        S. Veilleux\altaffilmark{5}, 
        H. A. Smith\altaffilmark{2},
        E. Sturm\altaffilmark{6},
        D. Farrah\altaffilmark{7},
        N. Falstad\altaffilmark{8}, 
        M. Mel\'endez\altaffilmark{5,9,10}, 
        J. Graci\'a-Carpio\altaffilmark{6},
        A. W. Janssen\altaffilmark{6},
        V. Lebouteiller\altaffilmark{11}
}
\affil{$^1$Universidad de Alcal\'a, Departamento de F\'{\i}sica
     y Matem\'aticas, Campus Universitario, E-28871 Alcal\'a de Henares,
     Madrid, Spain}
\affil{$^2$Harvard-Smithsonian Center for Astrophysics, 60 Garden Street,
  Cambridge, MA 02138, USA} 
\affil{$^3$Naval Research Laboratory, Remote Sensing Division, 4555
     Overlook Ave SW, Washington, DC 20375, USA}
\affil{$^4$Cornell University, Cornell Center for Astrophysics and Planetary
Science, Ithaca, NY 14853, USA} 
\affil{$^5$Department of Astronomy and Joint
Space-Science Institute, University of Maryland, College Park, MD
  20742, USA} 
\affil{$^6$Max-Planck-Institute for Extraterrestrial Physics (MPE),
  Giessenbachstra{\ss}e 1, 85748 Garching, Germany}
\affil{$^{7}$Department of Physics, Virginia Tech, Blacksburg, VA 24061, USA} 
\affil{$^8$Department of Earth and Space Sciences, Chalmers University of
  Technology, Onsala Space Observatory, Onsala, Sweden } 
\affil{$^9$NASA Goddard Space Flight Center,  Greenbelt, MD  20771, USA}
\affil{$^{10}$KBRwyle Science, Technology and Engineering Group
1290 Hercules Avenue  Houston, TX  77058, USA}
\affil{$^{11}$Laboratoire AIM - CEA/Saclay, Orme des Merisiers, 91191
  Gif-sur-Yvette, France}







\begin{abstract}
We report on the energetics of molecular outflows in 14 local Ultraluminous
Infrared Galaxies (ULIRGs) that show unambiguous outflow signatures (P-Cygni
profiles or high-velocity absorption wings) in the far-infrared lines of OH
measured with the {\it Herschel}/PACS spectrometer. All sample galaxies
  are gas-rich mergers at various stages of the merging process. Detection of
both ground-state (at 119 and 79 $\mu$m) and one or more radiatively-excited
(at 65 and 84 $\mu$m) lines allows us to model the nuclear gas ($\lesssim300$
pc) as well as the more extended components using spherically symmetric
radiative transfer models. Reliable models and the
  corresponding energetics are found in 12 of the 14 sources.
The highest molecular outflow velocities are found in buried sources, 
in which slower but massive expansion of the nuclear gas is also observed.
With the exception of a few outliers, the outflows have momentum fluxes of 
$(2-5)\times L_{\mathrm{IR}}/c$ and mechanical luminosities of $(0.1-0.3)$\% of
$L_{\mathrm{IR}}$.  
The moderate momentum boosts in these sources ($\lesssim3$) suggest that
the outflows are mostly momentum-driven by the combined effects of AGN and 
nuclear starbursts, as a result of radiation pressure, winds, and supernovae
remnants. 
In some sources ($\sim20$\%), 
however, powerful ($10^{10.5-11}$ \Lsun) AGN feedback and (partially)
energy-conserving phases are required, 
with momentum boosts in the range $3-20$.
These outflows appear to be stochastic, strong-AGN feedback events that occur
throughout the merging process. 
In a few sources, the outflow activity in the innermost regions has subsided
in the last $\sim1$ Myr. While OH traces the molecular outflows at sub-kpc
scales, comparison of the masses traced by OH with those previously
inferred from tracers of more extended outflowing gas suggests that most mass
is loaded (with loading factors of $\dot{M}/\mathrm{SFR}=1-10$) from the
central galactic cores ($\mathrm{a \, few} \times 100$ pc), qualitatively
  consistent with an ongoing inside-out quenching of star formation. Outflow
depletion timescales are $<10^8$ yr, shorter than the gas
consumption timescales by factors of $1.1-15$, and are anti-correlated
  with the AGN luminosity.   
\end{abstract}



\section{Introduction} \label{intro}


The correlations found between the masses of supermassive
black holes (SMBHs) and the velocity dispersions, masses, luminosities, light
concentrations, and S\'ersic indices 
of the spheroidal components of their host galaxies 
\citep[e.g.,][]{mag98,geb00,fer00,gra01,gra07,tre02,mar03,fer05,bei12,sha16} 
suggest a fundamental link between SMBH growth and stellar mass assembly. In
addition, the color distribution of local galaxies
\citep[e.g.,][]{str01,bal04,sch14}, 
with the blue galaxies actively forming stars and red-and-dead early-type
galaxies evolving passively, suggests that the color of red early-type
galaxies must have evolved rapidly,  
with star formation terminated on short timescales
\citep[e.g.,][]{hop06a,sch14}. Spatially resolved observations of
  $z\sim2.2$ massive galaxies show an inside-out quenching of star formation,
  on timescales of $<1$ Gyr in the inner regions \citep{tac15}.
An appealing way to explain these observations is via a self-regulated 
feedback model involving the morphological transformation of 
late-type to early-type galaxies through mergers, which first funnel large
amounts of gas into the circumnuclear regions\footnote{We use the term
  ``nuclear region'' and ``nuclear starburst'' for spatial scales
  $\lesssim300$ pc, to  differentiate from the term ``circumnuclear'' that is
  more generally used for scales of $\lesssim1-2$ kpc.} 
of the system, leading to both a circumnuclear starburst and the growth of a
SMBH. Above a SMBH critical mass, the energy or momentum released by the
SMBH limits efficiently the accretion onto the SMBH and quenches the
starbursts through the expulsion of the interstellar gas from which stars are
formed (negative feedback), ultimately yielding the SMBH-$\sigma$ 
relationship  
\citep{sil98,fab99,fab12,kin03,kin05,mat05,spr05,mur05,hop06b,kin15}. 
Violent relaxation of the stellar component deeply changes the morphology from
pre-merger disk galaxies to a coalescenced system with a spheroidal component,
which dominates over the re-formed disk in case of major mergers with limited
gas fraction \citep[e.g.,][]{hop09}.  
Observational evidence for the quenching of star formation by AGN feedback
have been reported recently \citep[e.g.,][]{far12,ala15}.
There are alternative explanations of the $M_{\mathrm{BH}}-M_{\mathrm{bulge}}$
correlation, e.g., gas accretion onto the BH from a
viscous inner disk with limited replenishment by star formation in an outer
disk \citep{bur01}, gravitational collapse of the inner regions of an
isothermal bulge \citep{ada01}, and hierarchical assembly of BH and
stellar mass through cycles of galaxy merging \citep{pen07,jah11}. Other ways
to account for the bimodality of galaxy colors and the quenching of star
formation include suppression of cold inflows of gas 
\citep{dek06} and gravitational heating of the intracluster medium
in mergers \citep{kho08}. While these environment mechanisms of quenching, as
well as ``strangulation'' on long time scales \citep{pen15}, may
dominate the secular evolution of late-type galaxies, a much more rapid
morphological and quenching ($t_{\mathrm{quench}}\lesssim250$ Myr) evolution 
is favored for producing early-type galaxies \citep{sch14}.

While feedback processes such as superwinds in ULIRGs and starbursts have been
observed for decades in lines of ionized and neutral atomic gas
\citep[e.g.,][]{hec90,vei05,rup02,rup05a,rup05b,rup05c,lip05,lip09,spo09}, 
outflows observed in lines that trace the molecular medium are also key to
understanding and quantifying this process, because the molecular phase
may carry a significant or even dominant fraction of the momentum and mass
outflow rates in buried stages.
Far-IR spectroscopy with {\it Herschel Space Observatory} Photoconductor 
Array Camera and Spectrometer (PACS) \citep{pil10,pog10}  
has indeed revealed powerful molecular outflows in ULIRGs traced by OH, with
velocities exceeding 1000\,km\,s$^{-1}$ in some sources and mass outflow rates
of several hundreds M$_{\odot}$\,yr$^{-1}$ \citep[][hereafter
  GA14]{fis10,stu11,spo13,vei13,gon14a}. The 
high-velocity outflows discovered with {\it Herschel} were found to be 
ubiquitous and thus inferred to be wide-angle in local ULIRGs
\citep{vei13,sto16}. These 
investigations also revealed a correlation between the 
outflow velocity and the AGN luminosity. In the far-IR, the
outflows are also traced by the line wing emission of the [C {\sc ii}] 158
$\mu$m transition \citep{jan16}. High-velocity molecular outflows are
detected at (sub)millimeter wavelengths in lines of CO, HCN, and
HCO$^{+}$ 
\citep[e.g.,][]{fer10,fer15,cic12,cic14,aal12,aal15,gar15,lin16}. 
Lower-velocity molecular outflows are also detected in millimeter lines
of the above species and in CS \citep{sak09,bol13,tun15,ala15,mar16}. 
The (sub)millimeter lines are now routinely observed with high angular
resolution and in some sources trace the outflowing gas out to kpc scales.  

Far-IR molecular observations can provide key and unique insight into the
outflow phenomenon: $(i)$ the strength and optical depth of the far-IR
continuum generates P-Cygni line profiles in some lines, unambiguously
indicating the presence of outflowing gas, discarding other alternatives such
as high turbulence or non-circular rotation motions 
\citep[e.g.,][]{gui15,dia16}; $(ii)$ blueshifted absorption can be 
traced to low velocities, probing low-velocity outflows that may be
missed from pure emission lines due to confusion with the line core; 
$(iii)$ despite the relatively poor spatial resolution of far-IR telescopes,
multi-transition observations including high-lying transitions, provide a 
robust means to quantify the main outflow parameters (mass outflow rate, 
momentum flux, etc).

OH in particular is an excellent tracer of these molecular outflows 
\citep[][GA14]{fis10,stu11,spo13,vei13}, with high abundances in 
active regions including photodissociated regions (PDRs), cosmic-ray 
dominated regions (CRDRs), and X-ray dominated regions (XDRs)
\citep[e.g.,][]{goi02,goi11,mei11,gon12,gon13}. 
Owing to the spin-orbit interaction, its rotational level structure is
characterized by two ladders ($^2\Pi_{3/2}$ and $^2\Pi_{1/2}$), with the
cross-ladder transitions much more optically thin than intra-ladder and thus
enabling the estimation of column densities. With high $A-$Einstein
coefficients, the high-lying transitions along the $^2\Pi_{3/2}$ ladder,
observed in absorption in galaxies, require strong far-IR radiation densities
to be excited. Specifically, the high-lying OH doublet at 65\,$\mu$m is an
excellent tracer of the shortlived but most active and buried phase of (U)LIRG
evolution \citep[][hereafter GA15]{gon15}, 
tracing the gas that lost its angular momentum during the
merger, falling onto the $\sim100$ pc nuclear region where it generates a
nuclear starburst \citep{hop09}. It is just in such an obscured phase
when most BH accretion is expected to occur, as indicated by
the cosmic X-ray background \citep{fab99b}, and when outflows are expected to
be most efficiently accelerated \citep{ste16}. 

\section{Guide to the astrophysical issues addressed in this paper} 
\label{over}

Here we present the results of our study of molecular outflows 
in a sample of local ULIRGs observed in OH with {\it Herschel}/PACS.
Our multi-transition analysis aims  
$(i)$ to infer and describe the presence of outflowing gas in high-lying
OH transitions, which probe the nuclear regions of ULIRGs that are first
exposed to the feedback action of an AGN and a nuclear starburst, and compare
and relate the line profiles of these high-lying lines which those of the
ground-state lines; 
$(ii)$ to analyse quantitatively the OH observations to
infer the main physical parameters and energetics of the molecular outflows 
(outflowing mass, mass outflow rate, momentum flux, and energy flux),
comparing our results with those obtained from other tracers and establishing
depletion time scales; 
$(iii)$ to give insight into the accelerating mechanism of these outflows
by assesing the relative roles of an AGN and a nuclear starburst, the 
relative contribution to the acceleration by radiation pressure on dust 
grains 
and winds, and whether the observed outflows are found in momentum- or
energy-conserving phases. 


The observations, data treatment, and the sample selection criteria are
described in \S\ref{obs}, where we define a modeling sample of 14 ULIRGs
observed in at least three far-infrared OH doublets.  
While all them are gas-rich mergers or interacting systems, they show a high
morphological diversity in merging stage. The sample spans AGN luminosities of
$(0.3-2)\times10^{12}$ \Lsun\ and star forming rates of $50-350$ \Msun/yr.

In \S\ref{results} we present the observational results of our
multi-transition profile analysis. We identify the presence of gas
outflowing from the nuclear regions ($<300$ pc) traced by radiatively-excited,
high-lying OH transitions. We show that the highest molecular
outflow velocities, traced by the OH doublet at 119 $\mu$m
\citep{spo13,vei13}, are found in buried sources that show strong absorption
in the OH 65 $\mu$m transition at systemic velocities. The highly
excited line profiles also show  evidence for slower expansion of the nuclear
regions of the ULIRGs that show P-Cygni profiles in the ground-state OH
doublet at 119 $\mu$m.

In \S\ref{rt} we describe our library of spherically symmetric radiative
transfer models, our methods of achieving best-fit solutions to the line
profiles, and the way of estimating the energetics of the outflows (outflowing
mass, mass outflow rate, momentum flux, and energy flux). 

The results of our 
quantitative analysis are presented in \S\ref{modres}. We compare our results
with the maximum estimated
momentum and energy rates that the AGN and the starburst can provide. We find
that while nuclear starbursts can provide a non-negligible contribution to the
observed outflows, they are most likely unable to drive them alone. An AGN is
required in most cases and clearly dominates in some ULIRGs.  
Outflow depletion timescales are $<10^8$ yr, significantly shorter
than the gas consumption timescales, and show an anti-correlation with
  $L_{\mathrm{AGN}}$. We also find that the nuclear outflowing 
activity has recently subsided in a few sources of the sample. 
The outflowing masses inferred from OH, which probes the outflows at sub-kpc
scales, are similar to those obtained from other tracers of more
extended outflowing gas.

The inferred outflow energetics are further interpreted in terms of simple
dynamical models in \S\ref{discussion}.
An important question discussed in the literature, and related to the growth
of the SMBH and to the normalization of the $M_{\mathrm{BH}}-\sigma$
correlation, is whether the outflows are 
momentum-conserving (e.g., driven by the ram pressure of winds and radiation
pressure on dust grains) or energy-conserving (e.g., driven by the thermal
pressure of a hot bubble that cannot cool on the timescale of its expansion).
We find that the total momentum deposited into the interstellar medium
(ISM) by the combined effects of the AGN and starburst is enough to
explain the outflows in most sources, but significant momentum boosts
are apparently required in some ($\sim20$\%) ULIRGs. The inferred momentum
boosts of the most powerful outflows depend on whether the gravitational
potential well is assumed to be balanced by rotation and on the assumed
geometrical dilution of the outflowing gas, and are estimated to be in the
range $3-20$. These are, in most cases, lower than the maximum momentum boosts
predicted by theoretical studies on energy-conserving outflows.
Partially energy-conserving phases, which we
find associated with compact outflow components with high columns, appear to
represent short stages with high AGN luminosities, and are uncorrelated 
with the merger stage. Our analysis indicates that although
radiation pressure on dust grains may (nearly) support the gas in the 
direction of the rotation axis, it cannot drive the outflows and thus
winds are required.  

Our main conclusions are summarized in \S\ref{conclusions}; 
we describe in more detail the individual sources and model fits in
Appendix~\ref{indsour}; 
a comparison of the present results with those obtained from ionized lines in
starburst galaxies \citep{hec15} is given in Appendix~\ref{heck15}; 
radiation pressure support is evaluated in Appendix~\ref{radp};
and some discussion of the velocity fields in the outflows is given in 
Appendix~\ref{vfield}.

\section{Observations and data analysis} \label{obs}

\subsection{Selection of the ULIRG Outflow Modeling Sample}
\label{sample}

We based the selection of our outflow modeling sample on all  
{\it Herschel}/PACS OH
observations  of local ULIRGs for which at least these three doublets were
available: the OH $^2\Pi_{3/2}$ $J=5/2-3/2$ doublet at 119 $\mu$m (hereafter
OH119), the cross-ladder $^2\Pi_{1/2}-^2\Pi_{3/2}$ $J=1/2-3/2$ doublet at 79
$\mu$m (OH79), and the OH $^2\Pi_{3/2}$ $J=9/2-7/2$ doublet at 65 $\mu$m
(OH65). In most sources, the $^2\Pi_{3/2}$ $J=7/2-5/2$ doublet at 84 $\mu$m
(OH84) has also been observed. An energy level diagram of OH showing these
transitions can be found in GA14. The OH119 and OH79 lines are ground-state
transitions, with OH119 40$\times$ more optically thick than OH79
\citep{fis10}. The OH84 and OH65 transitions have excited lower levels with
energies of $\approx120$ and $\approx300$\,K above the ground state,
respectively.  

Five Herschel observing programs included spectroscopic velocity-resolved
observations of ULIRGs in the above rotational transitions of OH: 
the Herschel guaranteed time key program SHINING (PI:
E. Sturm); the open time program HerMoLIRG (PI: E. Gonz\'alez-Alfonso); a
program that provided the full far-infrared spectra of two (U)LIRGs (PI:
J. Fischer); the HERUS program \citep[PI: D. Farrah;][]{far13}; and
a director’s discretionary  program focused on the two outflow
sources of the HERUS sample with the most prominent OH119 absorption at
  high velocities 
\citep[PI: H. Spoon;][]{spo13}. The first three programs 
observed three or more OH lines in all 20 ULIRGs in the Revised Bright Galaxy
Sample \citep[RBGS,][]{san03}. The latter two programs observed 
the ground-state OH119 and OH79 transitions in an additional complete set of
24 more distant ULIRGs out to $z<0.2$ and followed up with profiles of the
excited OH84 and OH65 transitions in IRAS~03158+4227 and IRAS~20100$-$4156.
In all, 22 ULIRGs at $z<0.2$ were observed in three or more OH transitions.   

To define the ULIRG outflow modeling sample discussed in
\S\ref{rt}-\S\ref{discussion}, we further constrained
the sample described above to the ULIRGs in which $(i)$ a P-Cygni profile or a
high-velocity blue wing in OH119 is detected \citep{vei13,spo13};
$(ii)$ the excited OH84 and/or OH65 doublets are detected. 
Of all ULIRGs reported in the
OH65 doublet (GA15), the former condition rules out the following sources:
Arp~220, IRAS~15250+3609, IRAS~F17207$-$0014, and IRAS~F22491$-$1808, where
the OH119 doublet is dominated by foreground absorption at redshifted
velocities; and IRAS~07251$-$0248, where no OH119 redshifted emission feature
is seen and the very strong and flat absorption feature does not clearly
indicate outflowing gas. Modeling these sources requires an specific
approach with more than the 3-components scheme used in \S\ref{rt}.
We also do not model IRAS~12112+0305 and IRAS~19542+1110, both showing
a P-Cygni profile in OH119 \citep{vei13}, because of a
mispointing in the three observed doublets 
(OH119, OH79, and OH65) resulting in the placement of the source near the edge
of a spaxel, which causes  a skewing of the instrumental profile. 
This results in a modeling sample of 14 local ULIRGs, which are listed in
Table~\ref{tbl-1}, together with some basic properties of the sources and the
observation identification numbers (OBSIDs) of the OH observations. 
The excluded sources have similar luminosities to the modeled ones, 
but some have lower AGN contributions that may be lower limits due do
high far-IR extinction 
\citep[Arp~220, IRAS~07251$-$0248, and IRAS~F17207$-$0014;][]{vei13}. 

The modeling sample, individually described in Appendix~\ref{indsour}, is
  somewhat biased to the most prominent outflowing sources, but may be still
  considered a good representation of the diversity of the molecular outflow
  phenomenon in local ULIRGs. All sample sources are morphologically
  classified in the optical as mergers or interacting systems, albeit with a
  high diversity in evolving stages (App.~\ref{indsour}): from widely
  separated galaxies ($\sim50$ kpc, IRAS~03158+4227), to double nuclei systems
  with projected separation $<10$ kpc (IRAS~08572+3915, Mrk~273,
  IRAS~14348$-$1447, IRAS~20100$-$4156), multiple colliding systems
  (IRAS~10565+2448, IRAS~19297$-$0406), and advanced mergers showing a
  single nucleus with tidal tails (IRAS~05189$-$2524, IRAS~09022$-$3615,
  Mrk~231, IRAS~13120$-$5453, IRAS~14378$-$3651, IRAS~20551$-$4250,
  IRAS~23365+3604). They also have a diversity in the AGN contribution to the
  luminosity (Table~\ref{tbl-1}) resulting in estimated AGN luminosities in
  the range $(0.3-2)\times10^{12}$ \Lsun\ (due to high extinction at
  far-IR wavelengths, GA15, some values of $L_{\mathrm{AGN}}$ could be
  underestimated). Assuming Eddington luminosities, the implied SMBH
  masses are $10^7-10^8$ \Msun.

\subsection{Data analysis}

For the basic analysis developed in \S\ref{results} (subsections
\S\ref{cov}-\S\ref{inflows}), we have used the OH observations of these
ULIRGs together with those of Luminous Infrared Galaxies (LIRGs,
$L_{\mathrm{IR}}>10^{11}$ \Lsun) as well as the nearby Bright Infrared Galaxy
(BIRG) NGC 4945, also detected in excited OH transitions (GA15).
In all these sources, we also used additional observations 
of the [C {\sc ii}] 158\,$\mu$m, 
[O {\sc i}]   63\,$\mu$m, and 
[O {\sc i}]  145\,$\mu$m lines from the programs listed above
\citep{gra11,far13}. 

All lines were observed in the highly sampled, range-mode of the PACS
spectrometer, 
with the exception of NGC 4945 which was observed in line spectroscopy
  mode with Nyquist sampling.    
Most of our analysis was based on pipeline-processed spectra created
by the Herschel Science Centre (HSC) using the Herschel Data Processing 
system; the Standard Product Generation version was HIPE
14.0.1, with calibration tree version 72.
We downloaded the level-2 data 
products\footnote{\url{http://herschel.esac.esa.int/Docs/PACS/html/pacs_om.html}} 
as needed from the Herschel Science Archive (HSA). 
Because both the molecular absorption lines and the continua are basically
point-like in the sources studied in this work, we
have used the point source calibrated spectra ``c129'', which scale the
emission from the central $\approx9''\times9''$ spatial pixel to the total
emission from the central $3\times3$ spaxels (``c9''), which is itself scaled
according to the point-source correction. The absolute flux scale is 
robust to potential pointing jitter, with continuum flux reproducibility
  of $\pm15$\%. The PACS spectral resolution is 290,
160, 145, and 190 \kms\ at 119, 79, 84, and 65 $\mu$m.

In some specific cases, the HSA spectra showed significant fluctuations in the
continuum level close to or within the wavelength range where the line wings
could be expected. We reprocessed those data by applying polynomial fits to
the flat-fielding with masking windows around the observed absorption and
emission features, which usually significantly improved the quality of the
baselines. 
Nevertheless, the uncertainty in the equivalent width and flux in the
line wings is usually dominated by the fitted baseline, and in some cases
may be up to $\sim25$\%.  

The observation of the OH119 doublet in IRAS~03158+4227 was mispointed by
about half a spaxel \citep{spo13}. We have examined spaxels 12 (central) and
13 where the OH119 absorption/emission is detected, and found a similar
wing with velocities exceeding $\sim1500$ \kms\ in both spaxels,
though the actual value is relatively uncertain due to insufficient baseline
on the blue side of the spectra. The redshifted emission feature is, however,
slightly stronger in the central spaxel, so the central spaxel profile is used
here for subsequent analysis.

Baseline fitting was subsequently performed with polynomials 
of order 1-2, and the fits were used to generate continuum-normalized
spectra and to extract the far-IR continuum flux densities.
Gaussians were fitted to absorption and emission features to obtain the peak
velocities and line strengths, as discussed in \S\ref{results}.

\begin{deluxetable*}{llccccccc}
\tabletypesize{\scriptsize}
\tablecaption{Modeling sample galaxies and {\it Herschel} OBSIDs of the OH
  doublets} 
\tablewidth{0pt}
\tablehead{
\colhead{Galaxy} & 
\colhead{$z_{\mathrm{CII}}$} &
\colhead{$D$} &
\colhead{$L_{\mathrm{IR}}$} &
\colhead{$\alpha_{\mathrm{AGN}}$} &
\colhead{OH119} &
\colhead{OH79} &
\colhead{OH84} &
\colhead{OH65}  \\
\colhead{name} & 
\colhead{} &
\colhead{(Mpc)} & 
\colhead{($10^{12}$ \Lsun)} & 
\colhead{} &
\colhead{OBSID}  &
\colhead{OBSID}  &
\colhead{OBSID}  &
\colhead{OBSID}  \\
\colhead{(1)} & 
\colhead{(2)} & 
\colhead{(3)} & 
\colhead{(4)} &
\colhead{(5)} &
\colhead{(6)} &
\colhead{(7)} &
\colhead{(8)} &
\colhead{(9)}
}
\startdata
 IRAS F03158+4227 & 0.13459&  $658 $ & $4.27 $ & $0.47$ & 1342238963 & 1342263478& 1342262940& 1342263479\\
 IRAS F05189$-$2524 & 0.04272&  $186 $ & $1.38 $ & $0.72$ & 1342219441 & 1342219442& 1342248556& 1342219445\\ 
 IRAS F08572+3915 & 0.05824&  $261 $ & $1.32 $ & $0.70$ & 1342208956 & 1342184687\tablenotemark{a}& 1342253600& 1342208954\\
 IRAS 09022$-$3615  & 0.05963&  $268 $ & $1.92 $ & $0.55$ & 1342209402 & 1342209403& & 1342209406 \\
 IRAS F10565+2448 & 0.04309&  $193 $ & $1.14 $ & $0.47$ & 1342207787 & 1342207788& 1342254243& 1342207790\\
 Mrk 231          & 0.04218&  $186 $ & $3.37 $ & $0.67$ & 1342186811& 1342186811& 1342253537& 1342207782\\ 
 IRAS 13120$-$5453  & 0.03107&  $136 $ & $1.86 $ & $0.33$ & 1342214628 & 1342214629& 1342248348& 1342214630\\ 
 Mrk 273          & 0.03780&  $166 $ & $1.45 $ & $0.34$ & 1342207801& 1342207802& 1342257293& 1342207803\\  
 IRAS F14348$-$1447 & 0.08257&  $376 $ & $2.09 $ & $0.17$ & 1342224243& 1342224242& & 1342224244\\   
 IRAS F14378$-$3651 & 0.06812&  $304 $ & $1.46 $ & $0.21$ & 1342204337& 1342204338& 1342250130& 1342204339\\  
 IRAS F19297$-$0406 & 0.08558&  $383 $ & $2.46 $ & $0.23$ & 1342208890& 1342208891& & 1342208893\\  
 IRAS F20100$-$4156 & 0.12971&  $632 $ & $4.68 $ & $0.27$ & 1342216371& 1342216371& 1342267942& 1342268108\\
 IRAS F20551$-$4250 & 0.04295&  $185 $ & $1.02 $ & $0.57$ & 1342208933& 1342208934& 1342253748& 1342208936\\ 
 IRAS F23365+3604 & 0.06449&  $281 $ & $1.43 $ & $0.45$ & 1342212514& 1342212515& 1342257685& 1342212517    
\enddata
\label{tbl-1}
\tablecomments{(1) Galaxy name; (2) Redshifts inferred from Gaussian fits to
  the [C {\sc ii}] 158 $\mu$m line; 
  (3) Distance to the galaxy; adopting a flat Universe with $H_0=71$ km
   s$^{-1}$ Mpc$^{-1}$ and $\Omega_{\mathrm{M}}=0.27$. 
  (4) IR luminosity ($8-1000$ $\mu$m), estimated using the fluxes in the four
  {\it IRAS} bands \citep{san03,sur04}; (5) Estimated AGN contribution to the
  bolometric luminosity, as derived from the $f15/f30$ ratio
  \citep{vei09,spo13}. This method yields $\alpha_{\mathrm{AGN}}=0.8$ in
    Mrk~231, but we have reduced the value in this source to $2/3$ because
    the nuclear starburst has an estimated contribution to the bolometric
    luminosity of $25-40$\% \citep{dav04}. Our adopted value, while basically
    consistent with the $f15/f30$ diagnostic, enables easy comparison with the
    works by \cite{cic14} and \cite{fer15}, who adopted just half that value
    \citep[$\alpha_{\mathrm{AGN}}\approx1/3$, based on][]{nar10}.
   We have adopted $L_{\mathrm{bol}}=1.15L_{\mathrm{IR}}$
  \citep{vei09,vei13}. The AGN and starburst luminosities are  
   $L_{\mathrm{AGN}}=\alpha_{\mathrm{AGN}}L_{\mathrm{bol}}$ and 
   $L_{\mathrm{*}}=(1-\alpha_{\mathrm{AGN}})L_{\mathrm{bol}}$. 
  (5)-(8) OBSIDs of the OH119, OH79, OH84, and OH65.
}
\tablenotetext{a}{The line was also targeted in instrument verification
  observations with OBSIDs 1342184688, 1342184689, 1342184690, 1342184691,
  1342184692, and the averaged spectrum is used. 
}
\end{deluxetable*}

\begin{figure*}
\begin{center}
\includegraphics[angle=0,scale=.9]{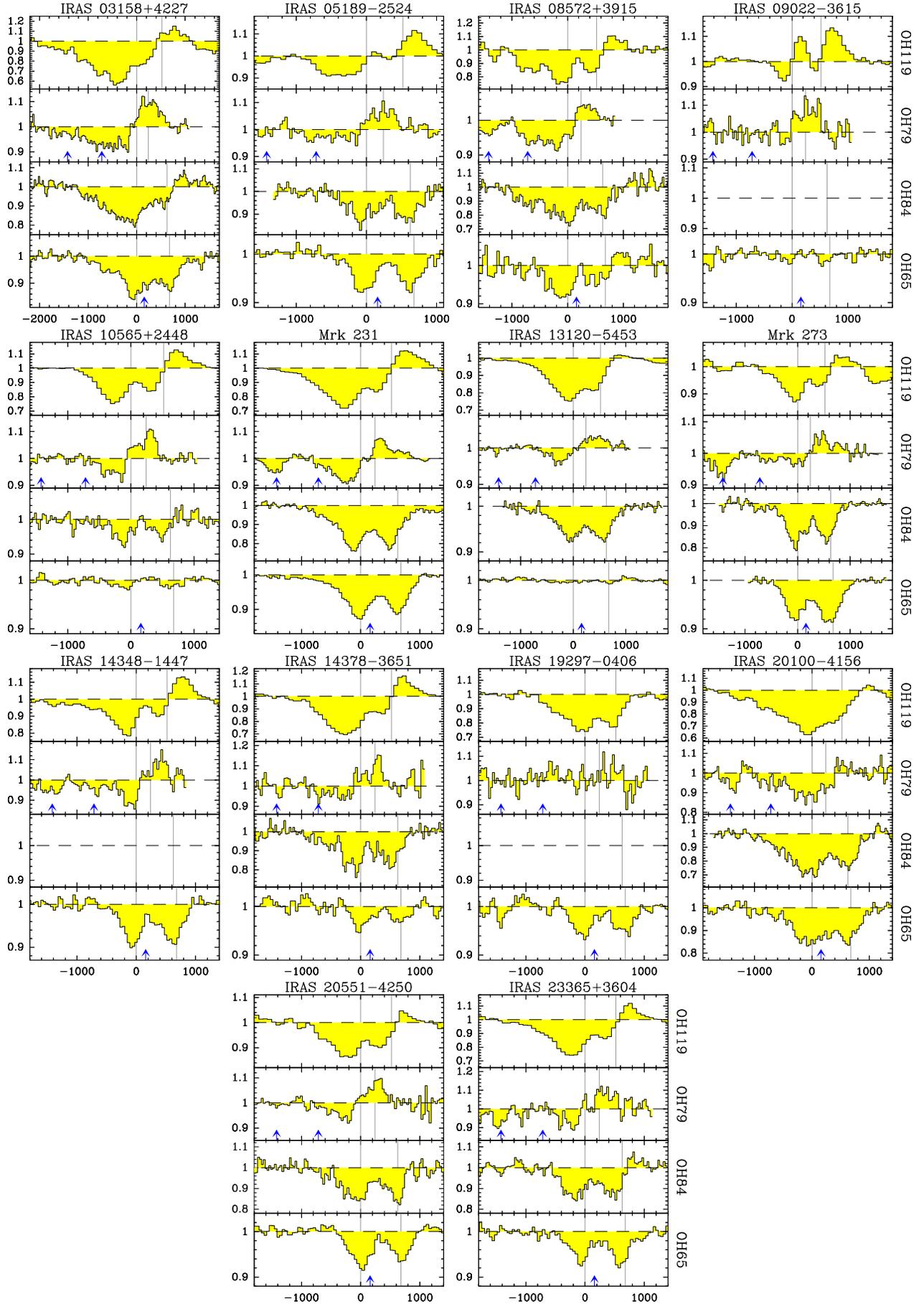}
\end{center}
\caption{The continuum-normalized spectra of OH in the 14 sources that are
  modeled in \S\ref{rt}. Abscissa shows velocities referred to the blue
  component of each doublet, and the two 
vertical grey lines indicate the rest position of the two doublet
components. Redshifts are obtained from Gaussian fits to the [C {\sc ii}] 158
$\mu$m line (Table~\ref{tbl-1}). The blue arrows indicate the position of
\hdo\ lines that could potentially contaminate the OH absorption (see text for
details). Note that the ordinate scale varies from line to line and from
source to source, and that the plotted velocity range also varies with source.
}   
\label{ohspec}
\end{figure*}

\section{Results}
\label{results}

\subsection{Description of the OH spectra} \label{overall}

Figure~\ref{ohspec} shows the continuum-normalized OH spectra in all 14
sources of our modeling sample. The OH119 and some OH79 and higher excitation
OH profiles were previously presented by \cite{fis10}, \cite{vei13},
\cite{spo13}, \cite{stu11}, and GA14 based on earlier HIPE pipeline
reductions.   
The velocity scale in Fig.~\ref{ohspec} uses the redshifts obtained from
Gaussian fits to the high SNR [C {\sc ii}] 158 $\mu$m line, which is
strongly dominated by the core of the line \citep{jan16}. In all sources,
the excited [O {\sc i}] 145 $\mu$m line was also observed with
{\it Herschel}/PACS, giving velocity shifts relative to the [C {\sc ii}] line
of $\leq40$ \kms, and in most (nine) sources $\leq20$ \kms.

 The blue arrows in Fig.~\ref{ohspec} indicate the location of possible
 contamination by lines 
 of \hdo, dozens of which appear in the far-IR spectra of some (U)LIRGs
 \citep[e.g.,][hereafter GA12]{fal15,gon12}. The strongest of these lines is
 the $4_{23}-3_{12}$ ($E_{\mathrm{lower}}=250$ K) centered at $\approx-1400$
 \kms\ in the OH79 spectra. The line is most likely contaminating the OH79
 blueshifted wing in IRAS~03158+4227 at 
 velocities $\lesssim-1400$ \kms. It is
clearly detected in other sources in which the OH79 wing 
does not blend with it. Potentially more problematic are instances of
possible contamination of OH79 spectra by the \hdo\ $6_{15}-5_{24}$ at
$\approx-720$ \kms. However, this line is much higher in energy
($E_{\mathrm{lower}}\approx600$ K), and it is a factor $2-7$ weaker than the
lower energy line in the extremely buried sources NGC~4418 and Arp~220,
respectively (GA12). It is likely that this line significantly contaminates
the OH79 spectra of IRAS~14348$-$1447 and IRAS~20100$-$4156. In some cases,
the OH65 doublet may be slightly contaminated by the \hdo\ $6_{25}-5_{14}$
line ($E_{\mathrm{lower}}\approx580$ K) at $+160$ \kms, which could have the
effect of shifting the velocity of peak 
absorption of the OH65 blue component to less negative velocities. There are
other possible contaminating lines as well. In the
OH119 profiles of IRAS~03158+4227, Mrk~231, Mrk~273, IRAS~14348$-$1447,
IRAS~20100$-$4156, and IRAS~23365+3604, there is substantial absorption 
by CH$^+$ $J=3-2$ and $^{18}$OH at velocities $>1000$ km/s ($>500$ km/s
relative to the red component of the OH119 doublet), which may
significantly weaken the emission feature of the OH119 doublet.

The high-velocity absorption wings or P-Cygni profiles observed in OH119
unambiguously indicate outflowing gas in all sample galaxies. 
The maximum blueshifted velocities observed in OH119 range from the very
moderate $\sim250$ \kms\ in IRAS~09022$-$3615, to $\sim1800$ \kms\ in 
IRAS~03158+4227, while most sources show maximum velocities in the range
$600-1300$ \kms\ \citep[see][]{stu11,spo13,vei13}. With the exception of
IRAS~19297$-$0406, all sources in Fig.~\ref{ohspec} also show P-Cygni profiles
or high-velocity blueshifted wings in the cross-ladder OH79
transition. Relative to the absorption strength, the 
redshifted emission feature is usually stronger in OH79 than in OH119. 
This is most likely because OH79 emits both 
through direct absorption and reemission of 79 $\mu$m continuum
photons and via absorption of 53 and 35 $\mu$m photons and cascade down to the
ground-state through the $\Pi_{1/2}$ $J=1/2$ level (GA12), while the
OH119 doublet only efficiently scatters through direct
absorption and reemission of 119 $\mu$m continuum photons.  

In contrast, P-Cygni profiles in the excited OH84 and OH65 doublets are only
observed in IRAS~03158+4227 (OH84), IRAS~08572+3915 (OH84 and OH65), and
  IRAS~23365+3604 (OH84). This can be attributed to the fact that the excited
  lines are formed closer to the optically thick far-IR source responsible for
  their excitation (GA14), and thus their emission from the far side is
  obscured. Nevertheless, OH84 blueshifted line wings are observed up to
  velocities of $\sim1000$ \kms\ in IRAS~03158+4227,
  IRAS~08572+3915, Mrk~231, IRAS~14378$-$3651, and IRAS~20100$-$4156, while
  significant wings with velocities $<400$ \kms\ are also observed in 
  IRAS~05189$-$2524, IRAS~13120$-$5453, Mrk~273, IRAS~20551$-$4250, and
  IRAS~23365+3604. The OH65 transition requires high columns to be
  excited (GA15), thus high-velocity absorption wings in this line are only
  found in the most extreme sources: IRAS~03158+4227, IRAS~08572+3915,
  Mrk~231, and IRAS~20100$-$4156. We argue in \S\ref{rt} that a high velocity
  wing in OH65 is the most reliable indicator of powerful AGN feedback.

  It is also worth noting that in most sources in our sample, the main
  absorption features of the OH84 and OH65 doublet components, while peaking
  at around central velocities, are slightly blueshifted relative to the
  systemic velocities as measured in the [C {\sc ii}] 158 $\mu$m line. This
  systematic effect is further explored in \S\ref{oh65vshift}, where we
  conclude that high columns of gas (if not most of the gas) are
  slowly expanding from the nuclear regions of ULIRGs, thus potentially
  changing the morphology of these regions and shifting the star
  formation to increasing radii.

\subsection{Comparing outflow properties as traced by both OH and CO} \label{secohco}

\begin{figure*}
\begin{center}
\includegraphics[angle=0,scale=.6]{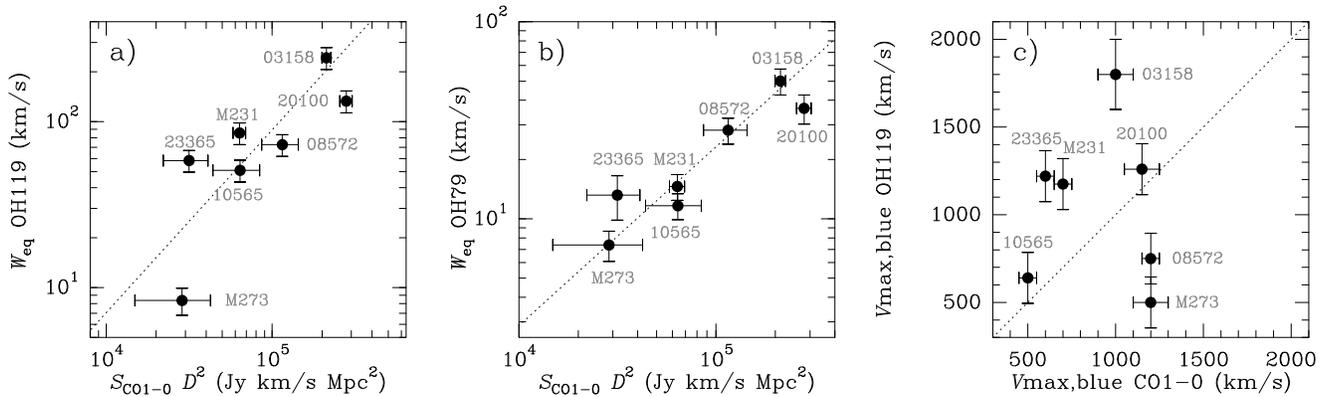}
\end{center}
\caption{Comparison of the observed OH and CO outflow properties.
  The equivalent width of the (a) OH119, and (b) OH79 doublets, as a
  function of the CO flux in the blueshifted line wing multiplied by the
  square of distance. The OH $W_{\mathrm{eq}}$'s have been measured over the
  same velocity ranges as the CO fluxes. The dotted lines in a) and b)
    indicate chi-square fitting in the log-log plane, giving
  $\log_{10}(W _{\mathrm{eq}}\,\mathrm{OH119}) = (-3.56\pm0.53)+(1.10\pm0.10)
  \log_{10}(S_{\mathrm{CO1-0}}\,D^2)$ (correlation coefficient of 0.79), and
  $\log_{10}(W _{\mathrm{eq}}\,\mathrm{OH79}) = (-3.17\pm0.47)+(0.91\pm0.09)
  \log_{10}(S_{\mathrm{CO1-0}}\,D^2)$ (correlation coefficient of 0.93).
  c) Comparison of
  the maximum blueshifted velocity of OH119 (corrected for the PACS
    spectral resolution at 119 $\mu$m) and CO (1-0). The dotted line
  indicates equal velocities. Abbreviated source names are indicated. The CO
  data are taken from \cite{cic12,cic14} and A. Gowardhan et al. (in
    prep).  
}   
\label{ohco}
\end{figure*}

It is instructive to compare some observational parameters of molecular
outflows as derived independently from OH and CO observations. Studies of
outflows in local ULIRGs based on CO observation have been carried out for 
5 of our sources \citep{fer10,cic12,cic14,fer15}:
Mrk~231, Mrk~273, IRAS~08572+3915, IRAS~23365+3604, and IRAS~10565+2448. 
More recently, CO observations have been also obtained in two additional OH
outflow sources, IRAS~03158+4227 and IRAS~20100$-$4156 (A. Gowardhan et al. in
prep). Given that the CO luminosity in the line wings is expected to be
proportional to the gas mass of the molecular outflow, comparison between OH
and CO sheds light on the physical properties associated with the OH
doublets. In Fig.~\ref{ohco}ab we plot the equivalent widths of OH119 and OH79
as a function of the CO (1-0) flux in the blueshifted linewing multiplied by
the square of distance (a quantity proportional to the outflow mass). The OH
equivalent widths have been calculated for the same velocity ranges as used
for the CO blueshifted linewings given in \cite{cic12,cic14} and A. Gowardhan
et al. (in prep). There is a hint of correlation between  $W_{\mathrm{eq}}$
(OH119) and $S_{\mathrm{CO1-0}}\,D^2$ (correlation coefficient of 0.79) though
with notable dispersion. Since the OH119 absorption strength is basically
measuring the covering factor of the 119 $\mu$m continuum by the outflow
($f_{119}$, see \S\ref{cov}), the trend in Fig.~\ref{ohco}a generally
indicates higher outflow mass with increasing covering factor, but the
correlation is limited most likely due to the high optical
depth of OH119 at most velocities.
A better linear correlation is indeed found
between $W_{\mathrm{eq}}$ of the more optically thin OH79 absorption strength 
and $S_{\mathrm{CO1-0}}\,D^2$ (slope of $0.91\pm0.09$
in the log-log plane, and correlation coefficient of 0.93).  
Therefore, the absorption strength of the OH79 doublet
is sensitive to the mass of the outflow, from which we may expect to be able
to infer reliable values for the outflow energetics.
A trend was also found between the
outflowing gas mass inferred from the [C {\sc ii}] 158 $\mu$m line wings and
from OH \citep[discussed in \S\ref{rt}, see][]{jan16}. 



Although OH119 might be expected to have moderate optical depths at the
highest outflow velocities, 
the maximum outflow velocities observed on the blue sides of the OH119
and CO lines show intriguing differences (Fig.~\ref{ohco}c),
with OH showing higher blueshifted velocities than CO in IRAS~03158+4227,
Mrk~231 and IRAS~23365+3604, and lower velocities in Mrk~273. This is
consistent with the fact that the two species 
may probe, at least at the highest outflowing 
velocities, somewhat different components of the galaxies. We may expect that 
OH generally traces more inner regions than CO:
as the expanding gas breaks into clumps and moves away from the
central region, the collisionally excited CO emission will still be observed
if the density remains high enough within the clumps (i.e., if they are
efficiently confined by the interclump-ionized medium), but these clumps 
will cover a decreasing fraction of the far-IR 
continuum and will thus produce lower absorption and emission in
the OH molecules that are radiatively excited. 
On the other hand, CO is less sensitive to a compact outflowing shell with
small radius and mass, even with enough column density to ensure detection
in OH119.


\subsection{The covering factor of the 119 $\mu$m continuum} \label{cov}

\begin{figure}
\begin{center}
\includegraphics[angle=0,scale=.5]{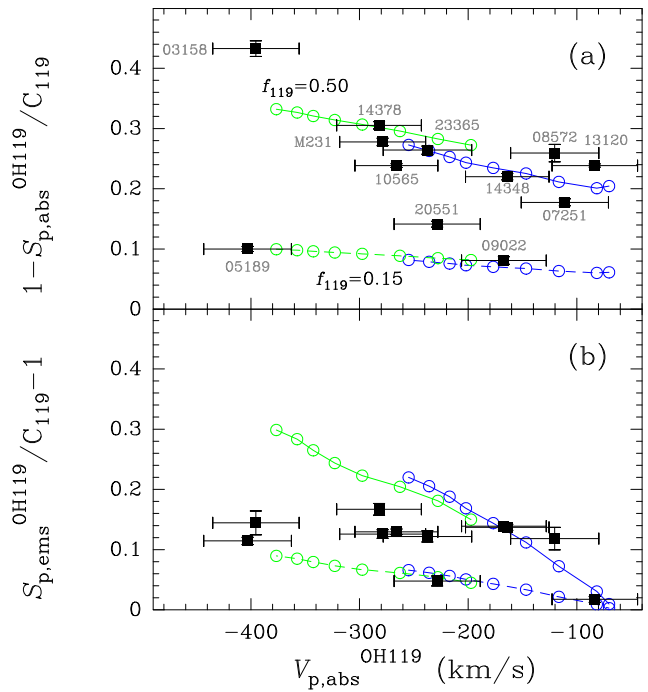}
\end{center}
\caption{The strength, relative to the continuum at 119\,$\mu$m
    (C$_{119}$), of the peak absorption feature 
  ($1-S_{\mathrm{p,abs}}^{\mathrm{OH119}}/C_{119}$, panel a), 
  and of the peak emission feature ($S_{\mathrm{p,ems}}^{\mathrm{OH119}}/C_{119}-1$,
  panel b) of the OH119 doublet as a function of the
  velocity of the peak absorption feature
  ($V_{\mathrm{p,abs}}^{\mathrm{OH119}}$). Values are obtained from
    Gaussian fits to the emission and absorption features. Only sources with
    $V_{\mathrm{p,abs}}^{\mathrm{OH119}}<-80$ \kms\ are included; abbreviated
    names are indicated. The green ($v_{\mathrm{out}}=200$ \kms) and blue
    ($v_{\mathrm{out}}=30$ \kms) curves 
  show the results of single-component radiative transfer models for two
  covering factors of the 119\,$\mu$m continuum by the outflowing OH,
  $f_{119}=0.5$ (solid curves) and $f_{119}=0.15$ (dashed, see text for
    details).  
}   
\label{oh119absemis}
\end{figure}

Given that the OH119 doublet is optically thick, its absorption strength 
constrains the covering factor of the 119\,$\mu$m continuum 
by the outflowing OH. Fig.~\ref{oh119absemis} shows the peak
absorption (panel a) and emission (b) strengths as a function of the OH119
peak absorption velocity ($V_{\mathrm{p,abs}}^{\mathrm{OH119}}$) for all
the observed ULIRGs that show the peak absorption more blueshifted 
than $-80$ \kms. We thus exclude here the sources that peak in OH119 at
systemic velocities, because the strength of the peak absorption in these
galaxies is determined by non-outflowing gas components.
Figure~\ref{oh119absemis}a indicates that the strength of the peak absorption
in OH119 attains a maximum of 44\% in IRAS~03158+4227, and is as low as
$\approx10$\% in IRAS~05189$-$2524. Most sources have, however, peak
  absorption troughs of $20-30$\% of the continuum (see also 
Fig.~\ref{ohspec}), with hints of increasing absorption strength with
increased blueshifted velocity.

The absorption strengths are significantly lower
than those predicted from simple models of an outflowing spherical shell
surrounding and fully covering a source of far-IR emission (GA14, see also
\S\ref{rt}). For instance, a shell with a velocity field
varying linearly with radius, with gas velocities of $v_{\mathrm{int}}=400$
and $v_{\mathrm{out}}=200$ \kms\ at the
inner and outer radius, and velocity dispersion of $\Delta V=100$ \kms,
generates in OH119 a P-Cygni profile with peak
absorption of $\approx60$\% of the 119 $\mu$m continuum at 
$V_{\mathrm{p,abs}}^{\mathrm{OH119}}=-260$ \kms. This is about twice the
observed absorption observed in Mrk~231, IRAS~14378$-$3651, and
IRAS~23365+3604, where the OH119 absorption peaks at similar velocities
(Fig.~\ref{oh119absemis}a). 
Since the OH119 doublet is optically thick, and thus insensitive to
the OH column density at least at velocities close to the maximum absorption
trough, the discrepancy indicates that OH only covers a fraction (half in the
example above) of the observed 119 $\mu$m continuum. Hence, the covering
factor of the galaxy 119\,$\mu$m continuum by the outflowing OH can be
estimated. 

\begin{figure*}
\begin{center}
\includegraphics[angle=0,scale=.60]{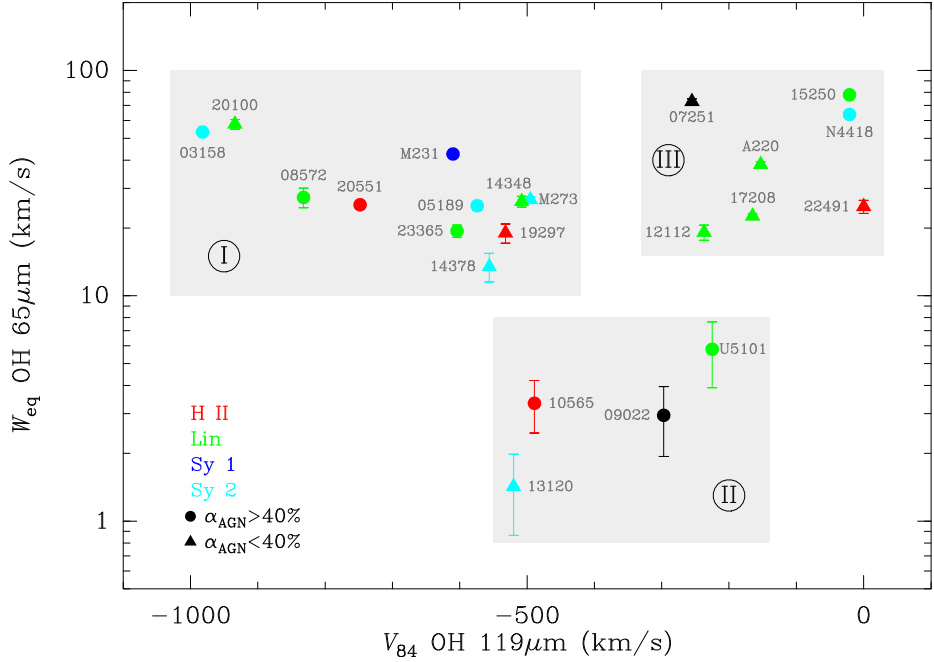}
\end{center}
\caption{The equivalent width of the high-lying OH65 doublet at central
  velocities (between $-200$ and $+200$ \kms,
  $W_{\mathrm{eq}}(\mathrm{OH65})$; GA15) as a 
  function of $V_{84}$, the velocity below which 84\% of the absorption in the
  OH119 doublet takes place \citep[from][]{vei13}. We do not include NGC~6240
  in this plot because of the presence of a blueshifted emission  
  feature beyond the OH119 P-Cygni profile (J. Fischer et al., in prep.).
  Red, green, blue,
  light-blue, and black colors indicate H{\sc ii}, LINER, Seyfert-1,
  Seyfert-2, and unclassified optical spectral types, respectively
  \citep[from][or NED/SIMBAD]{vei95,vei99,vei09,ver06,rup05b,gar06,kim98}. 
  Circles and triangles indicate sources with fractional AGN
  contribution to the 
  bolometric luminosity of $\alpha_{\mathrm{AGN}}\ge40$\% and $<40$\%,
  respectively, as derived from $f15/f30$ \citep{vei09}. Three regions (I, II,
  and III) marked with shaded rectangles are identified in this plane (see
  text). The highest outflow velocities, as measured from the sensitive and
  ground-state OH119 doublet, are found in sources with strong OH65 absorption
  (i.e., in buried and warm sources), but the reciprocal does not apply:
  sources with high OH65 absorption are also found with low $|V_{84}|$. In
  addition, two sources with weak OH65 absorption have relatively high
  $|V_{84}|\sim500$ \kms.}   
\label{oh65v84}
\end{figure*}


The observed values in Fig.~\ref{oh119absemis} are then compared with
simple modeling results (colored lines and symbols) described in
\S\ref{rt}. Rather than the multi-component models generated in
  \S\ref{rt} to fit simultaneously all observed OH line profiles, we use here
  single models to explain the regularities observed in the data set of
  Fig.~\ref{oh119absemis}, with no attempt to fit the OH119 line profile. 
We adopt a central continuum source with $T_{\mathrm{dust}}=55$ K and 
$\tau_{100}=1$, 
surrounded by an extended shell of outflowing
gas, with $v_{\mathrm{out}}=30$ and 200 \kms\ (blue and green curves,
respectively), variable $v_{\mathrm{int}}\leq900$ \kms, and velocity
dispersion $\Delta V=100$ \kms. From the emergent
theoretical spectra, convolved with the PACS resolution,
$V_{\mathrm{p,abs}}^{\mathrm{OH119}}$ and the strengths of the peak absorption
and emission were calculated and multiplied by the appropriate factor
$f_{119}$, the covering factor of the 119 $\mu$m continuum by the outflowing
OH, to match the observations. Figure~\ref{oh119absemis} indicates that most
of the observed OH119 absorption and emission strengths can be explained with
$f_{119}$ ranging from $0.15$ (dashed curves) to $0.50$ (solid curves).

P-Cygni profiles occur when the absorbing and approaching gas in front of
  a continuum source, and the emitting and receding gas behind it, 
radiatively decouple from one another owing to the expansion. This explains
the increase of the peak strengths with increasing outflow velocities for
fixed $f_{119}$. In spite of the varying covering factors that may be expected
to be present in our sample, the effect of this decoupling is apparent in the
data. For high outflow 
velocities, and an outflow significantly more extended than the continuum
source, the absorption and emission features are expected to be similar, as
the OH merely re-distributes the continuum in velocity space. The strength of
the emission feature is decreased with decreasing velocities due to
extinction at 119 $\mu$m by the continuum source, which accounts for the fact
that the absorption feature is stronger than the emission one. For low
velocities, only the absorption feature is expected to be detectable. The
modeling values in Fig.~\ref{oh119absemis} do not sensitively depend on fixed
parameters such as \tdust, $\tau_{100}$, or $N_{\mathrm{OH}}$, but depend 
on the gas velocity dispersion and of course on geometry, as our values apply
to spherical symmetry.

The inferred range $f_{119}=0.15-0.50$ is similar to the range obtained from
the multi-component analysis in \S\ref{modres} ($0.17-0.67$; the 
highest value is found in IRAS~03158+4227). These values are 
significantly lower than 1, indicating
that the outflowing OH covers in most sources only a fraction of the 
observed $119$ $\mu$m continuum. On the other hand, 
$f_{119}\sim0.50$ appears to be too high to be produced by a collimated jet,
thus favoring wide-angle coverage as recently found in the outflow of
  Mrk~231 from high angular resolution CO $2-1$ observations \citep{fer15}. In
  IRAS~17208-0014, the outflow 
  observed in CO $2-1$ at the highest blueshifted velocities ($v<-450$ \kms)
  appears to be relatively collimated \citep{gar15}, while the OH119
  absorption at the same velocities is shallow ($\leq2$\%). This independently
  suggests that OH119 is primarily sensitive to wide-angle outflows.
The sources in Fig.~\ref{oh119absemis}a with relatively low absorption
troughs, IRAS~05189$-$2524 and IRAS~20551$-$4250 (also Mrk~273; 
IRAS~09022$-$3615 is optically thin in the far-IR), 
however, may have a molecular outflow significantly collimated.
In any case, the fractional absorption  
may be attributed to clumpiness of the outflowing gas, biconical
(two-lobed)  structure, and to the fact that
some far-IR emitting regions of the host galaxy are not affected by the
outflow. 

\subsection{Outflowing gas and buried sources} \label{oh65of}

GA15 showed that OH65 is a unique tracer of warm and optically thick cores
that account for a significant fraction (if not most) of the (U)LIRG's
luminosity, confirming early indications that ULIRGs are optically thick
  in the far-IR \citep{dow93,sol97}.  
The extreme properties inferred for these cores may 
imply that they represent the most deeply buried stages of AGN-starburst
co-evolution. The OH outflows are expected to emanate from these active
central cores.  For this reason we examine here the relationship between the
OH65 absorption (at central velocities)
and the velocity of the outflows as
measured with the most sensitive ground-state OH doublet, OH119. 
$W_{\mathrm{eq}}(\mathrm{OH65})\gtrsim20$ \kms\ indicates warm
($T_{\mathrm{dust}}\gtrsim60$ K) and optically thick 
(continuum optical depth at 100 $\mu$m $\gtrsim1$, or
$N_{\mathrm{H}}\gtrsim10^{24}$ \cmd) cores.
Results are shown in Fig.~\ref{oh65v84} where the outflow velocity is
characterized by $V_{84}$(OH119), the velocity
below which 84\% of the absorption in the OH119 doublet takes place
\citep{vei13,sto16}. 


In the  $V_{84}$(OH119)-$W_{\mathrm{eq}}$(OH65) plane, galaxies occupy three
distinct regions, indicated in Fig.~\ref{oh65v84} with shaded rectangles: 
galaxies with the highest $|V_{84}|\gtrsim500$ \kms\ are also strong in OH65
(region I), establishing the connection between the buried and warm
cores and the high-velocity outflows. Since the maximum outflowing
  velocity in OH is correlated with the AGN luminosity
  \citep{stu11,spo13,vei13,sto16}, this relationship suggests 
  that the warm material probed by the OH65 doublet ultimately provides the
  conditions for copious black hole accretion and strong AGN feedback.

Nevertheless, the converse does not occur, 
because a number of sources with very strong OH65, including three very
extreme sources (IRAS~$15250+3609$, IRAS~$07251-0248$, and the LIRG NGC~4418),
have low $|V_{84}|$ (region III). Although most ULIRGs in region III are
associated with a low attributed contribution of the AGN to the luminosity,
the lack of high OH velocities in these sources does not fully preclude 
  an energetically dominant AGN, because the velocities are
  expected to decrease with increasing column of gas that is accelerated, and
  the optically thick cocoons in these sources may cover a large solid 
angle.   

In addition, at least two sources 
have relatively high $|V_{84}|\sim500$ \kms\ but are weak in OH65 (region II; 
IRAS~$10565+3609$ and IRAS~$13120$-$5453$). 
The optical spectral types, denoted by symbol color in Fig.~\ref{oh65v84}, and
the AGN fractions inferred from the mid-IR continua, denoted by symbol shape,
do not account for the differences between sources in the three
regions, although AGN dominated sources are more clearly identified in
regions I and II, than in region III. 
Most region III sources have a redshifted OH119 and OH79 peak
  absorption, and/or redshift absorption in the [O {\sc i}] 63 $\mu$m line
  (IRAS~15250+3609, NGC~4418, IRAS~F17207$-$0014, Arp~220,
  IRAS~22491$-$1808), indicating a 
  complex velocity field with ground-state far-IR molecular line shapes
  dominated by non-outflowing foreground gas that, at least in some cases,
  appears to be inflowing (see \S\ref{inflows}). 
We thus focus our modeling in \S\ref{rt} on  
sources in regions I and II (see also \S\ref{sample}). 


\subsection{Expansion of the nuclear regions} \label{oh65vshift}

\begin{figure*}
\begin{center}
\includegraphics[angle=0,scale=.75]{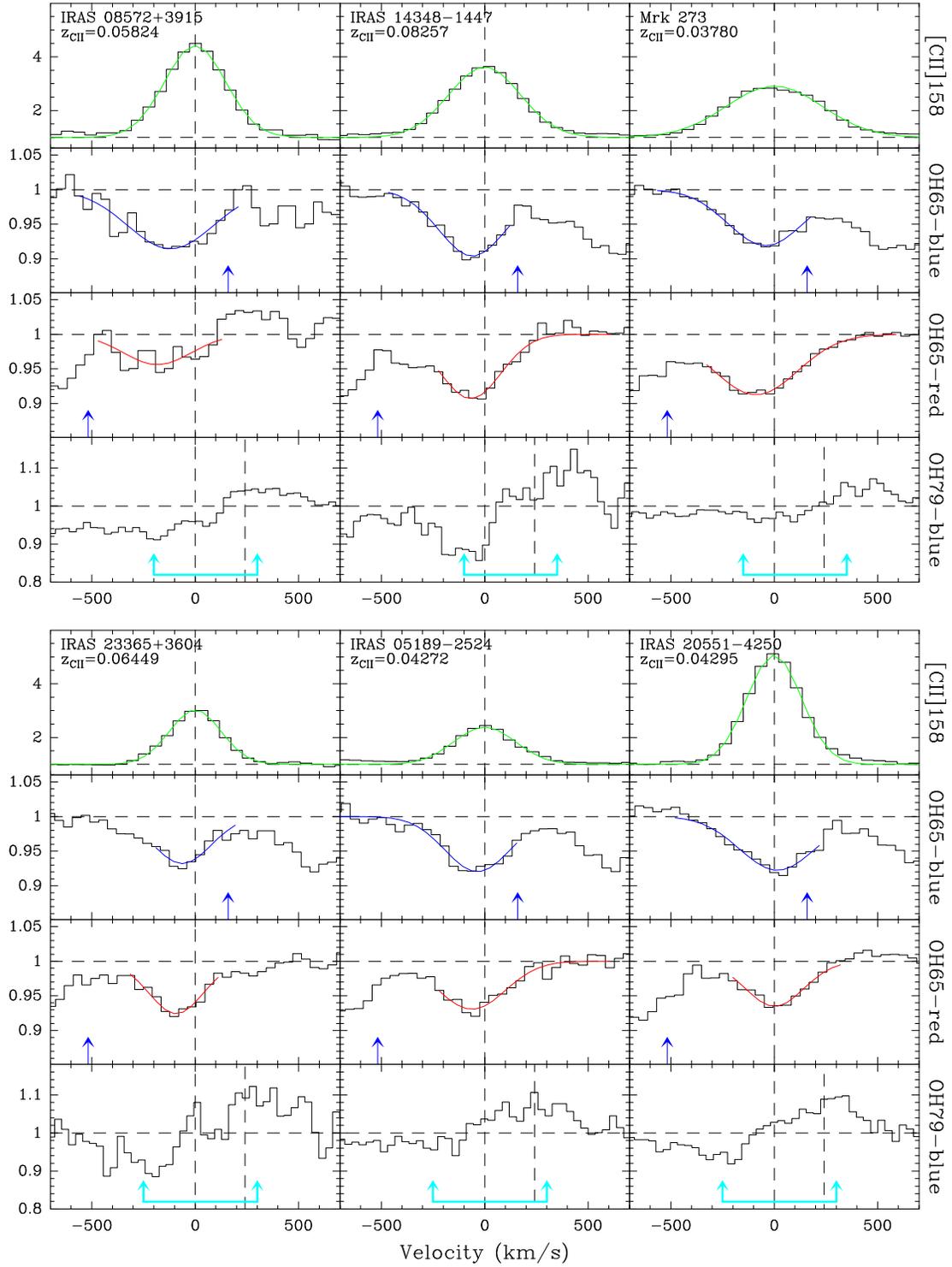}
\end{center}
\caption{The [C {\sc ii}]158, the blue and red
  components of OH65, and the OH79 line shapes in six sources where evidence
  for nuclear outflowing gas is found. Redshifts used for all profiles 
  have taken in all cases from Gaussian fits to [C {\sc ii}]158 $\mu$m (in
  green). The OH65 blue and red components have also been fitted with
  Gaussians (blue and red curves), showing significant velocity blue shifts
  or asymmetries relative to [C {\sc ii}]158 in all these cases. 
  The blue arrows indicate the  
  position of the \hdo\ $6_{25}-5_{14}$ line, which could be contaminating the
  blue component of some OH65 profiles. The light-blue arrows in the lower
    panels indicate the approximate peaks of absorption and emission in the
    OH79 P-Cygni line profiles, while the OH79 absorption features have their
    counterpart in the blueshifted OH65 doublet absorption features.
}   
\label{ciioh65prof}
\end{figure*}

In this section we examine the evidence for low-velocity outflows
  detected in the excited OH65 and OH84 doublets.
Because OH65 is a key tracer of warm and optically thick nuclear regions
of galaxies (GA15), outflowing signatures in OH65 may be expected to 
indicate expansion motions 
of the regions preferentially exposed to feedback from a SMBH or an extreme
starburst. However, one crucial point when studying velocity shifts in OH65 is
the redshift, or velocity center, of the nuclear region. Because OH65 is
a very high-lying transition, it is only excited in regions with a strong
far-IR field and thus with high extinction in the far-IR, and hence the
redshifted reemission feature from behind the continuum source 
is mostly obscured and not
detected. Only one source, IRAS~08572+3915, shows direct evidence for a true
P-Cygni in OH65 (Fig.~\ref{ohspec}).  For all the other sources we must infer
the nuclear molecular outflows by measuring velocity shifts and identifying 
absorption wings in OH65.

\begin{figure*}[t]
\begin{center}
\includegraphics[angle=0,scale=.6]{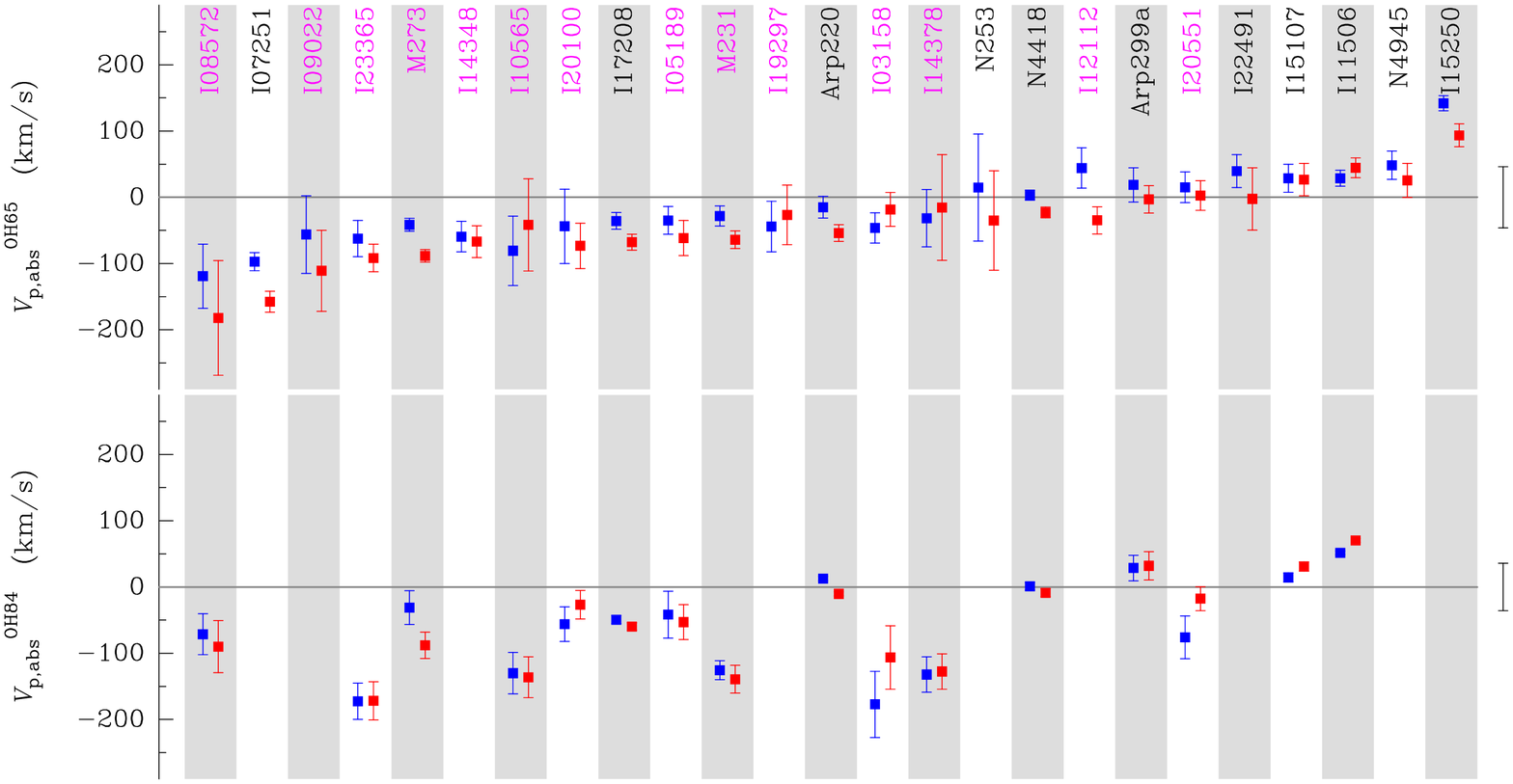}
\end{center}
\caption{The velocity of peak absorption of the OH65
  ($V_{p,abs}^{\mathrm{OH65}}$, upper) and OH84 ($V_{p,abs}^{\mathrm{OH84}}$, lower)
  doublets relative to the velocity of the [C {\sc ii}]158 $\mu$m line, in all
  sources where the OH65 doublet is detected (GA15), including the 14
  ULIRGs of the modeling sample. Blue and red symbols
  indicate peak absorption velocities of the blue and red components of the
  doublets, obtained from Gaussian fits to each doublet
    component. 
$V_{\mathrm{p,abs}}^{\mathrm{OH65}}$ of the blue component is systematically
    lower than that of the red component most likely due 
to contamination of the blue component by the high-lying 
\hdo\ $6_{25}-5_{14}$ line.
Errorbars are 2$\sigma$, and the errorbars on the right-hand
    side indicate potential ``skewness'' error from pointing 
 drifts, which we adopt as $0.01$ $\mu$m 
(i.e., $\approx45$ \kms\ for OH65
See Section 4.7.2 of the PACS Observer's Manual,
\url{http://herschel.esac.esa.int/Docs/PACS/html/pacs_om.html}). 
Abbreviated source names are indicated; those in magenta indicate
  that clear signatures of outflowing gas (P-Cygni or wings) in the
  ground-state OH119 doublet are observed. Sources are sorted 
  according to the component-average value of $V_{p,abs}^{\mathrm{OH65}}$.
}
\label{oh65vpeak}
\end{figure*}

In Fig.~\ref{ciioh65prof}, we compare the profiles of [C {\sc ii}] 158 $\mu$m
line, the blue and red components of the OH65 doublet separately, and the OH79
doublet in six sources that show indications of nuclear outflowing gas.  
Velocity profiles for OH65 are then presented for both
the blue and red components in separate panels, while the profiles
for OH79 are presented for the blue component of the doublet (the zero
velocity of the red component is also indicated by a dashed vertical line at
$\approx230$ \kms). To infer the velocity of peak 
absorption in both the blue and red components of OH65
($V_{\mathrm{p,abs}}^{\mathrm{OH65}}$), ``truncated'' Gaussian fits have been
applied to each of the OH65 components in such a way that contamination 
by the other $l-$doubling component and by high-velocity wings are minimized,
and are also shown in Fig.~\ref{ciioh65prof} in blue and red, respectively. We
note that $V_{\mathrm{p,abs}}^{\mathrm{OH65}}$ of the blue component may
actually be lower (more negative) than the fitted value in some sources due
to contamination by the high-lying \hdo\ $6_{25}-5_{14}$ line, the position of
which is indicated with blue arrows. Therefore, we mostly rely on the
velocities inferred from the OH65 red component to infer the detection of
outflowing gas in OH65. The bottom panels in Fig.~\ref{ciioh65prof} show the
OH79 profiles in these sources, which all show P-Cygni line shapes.

The OH65 profiles are broad, extending all the way to $-500$
\kms\ on the blue wing, and in some cases beyond the limits of the 
[C {\sc ii}] emission at zero intensity. 
(Note, however, that the [C {\sc ii}] line wings observed in \cite{jan16}
  are hardly perceptible in Fig.~\ref{ciioh65prof} due to the linear scale
  used for the ordinates.) In addition, clear velocity shifts 
of $50-200$ \kms\ between the OH65 peak absorption and the [C {\sc ii}] peak
emission are seen in Fig.~\ref{ciioh65prof}. In IRAS~08572+3915,
IRAS~14348$-$1447, Mrk~273, and IRAS~23365+3604, the OH65 feature is
blueshifted relative to [C {\sc ii}]. In IRAS~20551$-$4250, OH65 peaks at 
the systemic velocity but shows an asymmetric blue wing in both OH65
components up to $-400$ \kms.

Blueshifts in OH65 relative to [C {\sc ii}] can be attributed to 
$(i)$ differences between the nuclear redshift(s) and that of the
bulk of the host galaxy as traced by [C {\sc ii}], possibly due to nuclear
motions associated with the merger; $(ii)$ non-circular motions in the
nuclear region due to bar-like or oval distortions and warps
\citep[leading to elliptical orbits; e.g.][]{san89}; or $(iii)$ OH65
absorption that is tracing expanding (radial) 
gas flows in the nuclear region. While the first 
two possibilities should be considered, particularly in ULIRGs
where the rotating, merging nuclei might indeed generate such velocity shifts,
comparison of the OH65 and OH79 profiles suggests that 
the third possibility is clearly favored in some sources. The OH79 line
profiles show P-Cygni profiles 
in all sources presented in Fig.~\ref{ciioh65prof} and, in
IRAS~08572+3915, IRAS~14348$-$1447, and more tentatively in Mrk~273, the
blueshifted absorption features in OH79 have counterparts both in emission at
positive velocities in OH79 (indicated by the light-blue arrows) and in
absorption at the same velocities in the OH65 doublet. 
It is this OH79-OH65 correspondence in the blue part of the
profiles, together with the P-Cygni in OH79 
(as well as the preponderance of OH65 blueshifted absorption as discussed
below), that enables us to distinguish between rotating and outflowing motions
in the nuclear regions from the OH65 profiles. The inferred outflows do
not preclude, however, the simultaneous incidence of a non-outflowing
component in OH65.

Fig.~\ref{oh65vpeak} shows $V_{\mathrm{p,abs}}^{\mathrm{OH65}}$ in all sources
that are detected in OH65 (GA15, thus including more than the 14 sources
of our modeling sample), as well as the corresponding quantity for
OH84 in those sources where the doublet is available. 
With the exceptions of IRAS~07251$-$0248 and IRAS~17208$-$0014 
\citep[the latter showing evidence for outflowing gas in
  CO,][]{gar15}, all other sources in Fig.~\ref{oh65vpeak} with significant
blueshifted OH65 absorption show a P-Cygni profile in OH119, which is
indicated in Fig.~\ref{oh65vpeak} by magenta colored source names, while the
incidence of a P-Cygni OH119 line shape in sources with no OH65 blueshift is
significantly lower. 
Three sources, IRAS~14378$-$3651, IRAS~20551$-$4250, and
IRAS~12112+0305, show P-Cygni profiles in OH119 but their OH65 doublets
peak at central velocities.  IRAS~20551$-$4250, however, has a 
blueshifted wing in OH65 that is also attributable to outflowing gas
(Fig.~\ref{ciioh65prof}). 
The intermediate OH84 doublet confirms
the velocity shifts seen in OH65 in most cases. 
Specifically, IRAS~10565+2448 is weak in OH65 (but detected, GA15) and
its $V_{\mathrm{p,abs}}^{\mathrm{OH65}}$ value has large
uncertainties. However, shifts of $\sim130$ \kms\ are clearly seen 
in OH84. In IRAS~23365+3604, the velocity shifts in OH84 are larger than in
OH65, as is also the case in Mrk~231 (GA14). Identifying outflowing gas in
excited OH transitions may miss some molecular outflow sources, such as
e.g., Arp~220 where clear indication of low-velocity outflowing molecular gas
is seen in several species including OH \citep{sak09,gon12,tun15,mar16}.

\begin{figure*}[t]
\begin{center}
\includegraphics[angle=0,scale=.65]{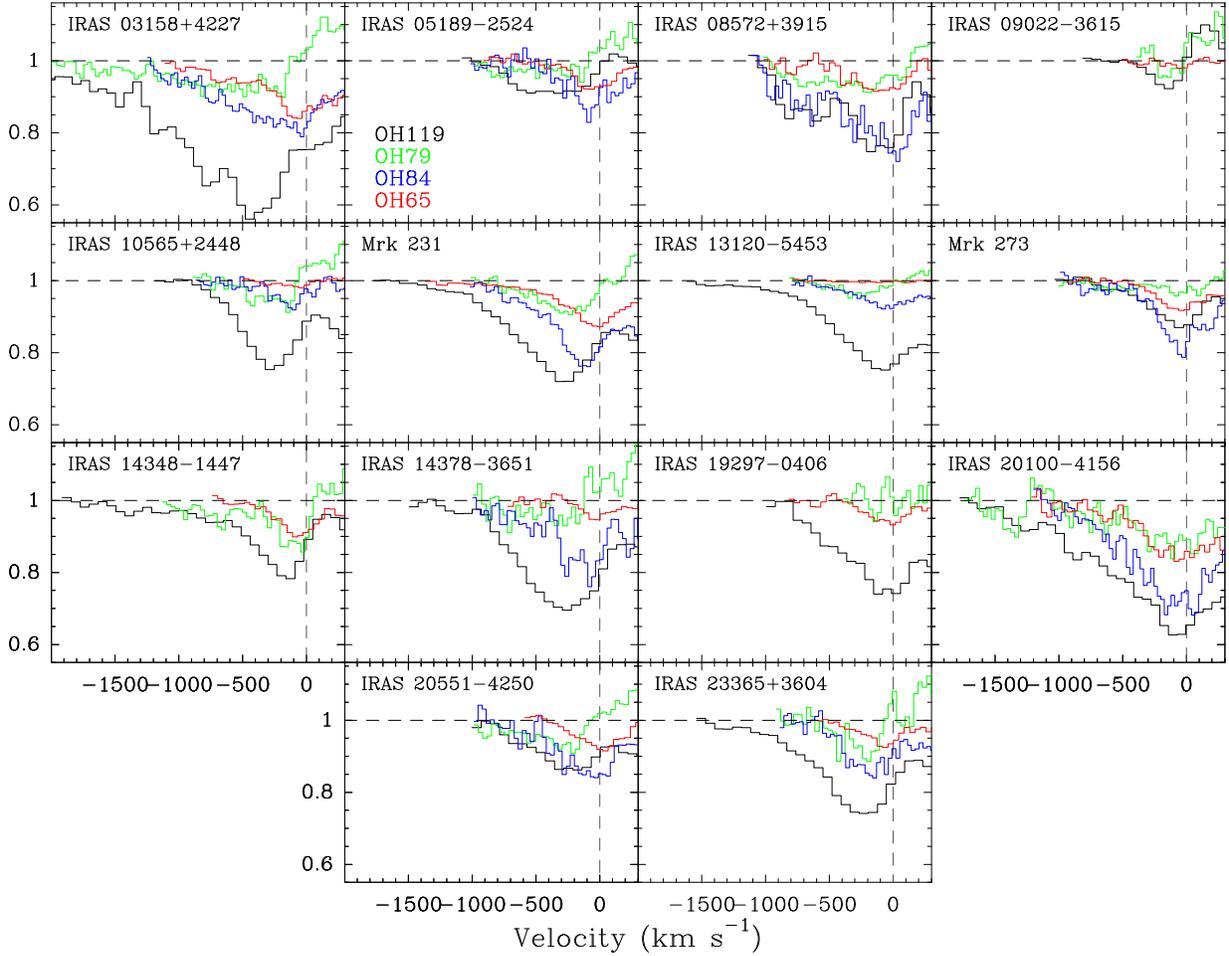}
\end{center}
\caption{Detailed comparison of the blueshifted side of the OH119 (black
  histograms), OH79 (green), OH84 (blue), and OH65 (red) line profiles
  observed in the 14 sources that are modeled in \S\ref{rt}. We only
  show the relevant velocity ranges of each doublet for clarity. The
  \hdo\ $4_{23}-3_{12}$ line is most likely producing absorption at velocities
  $\lesssim-1400$ \kms\ in the OH79 spectra of IRAS~03158+4227 and 
  IRAS~20100$-$4156.
}   
\label{ohblue}
\end{figure*}

A detailed comparison of the blueshifted absorption wings in OH119, OH79,
OH84, and OH65 is shown in Fig.~\ref{ohblue} for the 14 sources 
modeled in \S\ref{rt}. Besides Mrk~231, IRAS~03158+4227,
  IRAS~20100$-$4156, and IRAS~08572+3915, where the high velocity
outflowing gas in OH65 is evident, Mrk~273 and IRAS~14348$-$1447 show similar
line shapes in OH65 and in the P-Cygni OH119-OH79 doublets, indicating that
OH65 is also tracing outflowing gas. Similarly, the OH65 line wing in
IRAS~23365+3604 has its counterpart in OH119 and OH84, both showing P-Cygni
profiles (Fig.~\ref{ohspec}). We conclude that, in addition to Mrk~231
(GA14), IRAS~03158+4227, and IRAS~20100$-$4156, the six sources in
Fig.~\ref{ciioh65prof} show evidence for outflowing gas in the high-lying
OH65 transition. In addition, IRAS~09022$-$3615 and IRAS~10565+2448,
although weak in OH65 (GA15), show hints of blueshifted OH65 absorption
as well (Fig.~\ref{ohspec}).  

\begin{figure*}
\begin{center}
\includegraphics[angle=0,scale=.85]{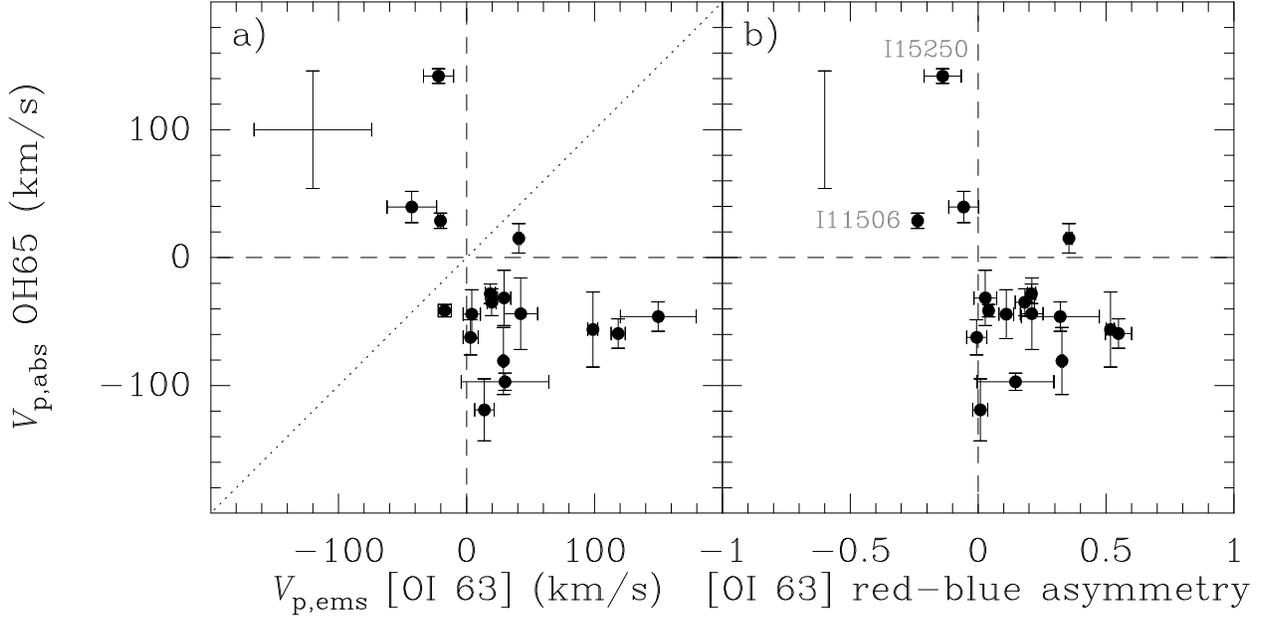}
\end{center}
\caption{The velocity of peak absorption of the OH65 doublet (blue
    component of the doublet) as a function of a) the velocity centroid of 
  the [O {\sc i}] 63 $\mu$m emission line, and b) the [O {\sc i}] 63 $\mu$m
  red-blue asymmetry, which is defined as 
  $(F_+-F_-)/(F_++F_-)$ where $F_+$ ($F_-$) is the flux measured at positive
  (negative) velocities. The dotted line in a) indicates equal velocities.
  The errorbars on the upper-left sides of each panel indicate potential
  ``skewness'' uncertainties due to pointing drifts. The two sources in our
  sample with evidence for inflowing gas, IRAS~11506$-$3851 and
  IRAS~15250+3609 (see \S\ref{inflows}), are labeled in panel b.
}   
\label{oh65oi63}
\end{figure*}

Could these velocity shifts observed in OH65 be associated with
  merging motions, rather than to outflowing or inflowing gas? We compare
  the peak velocities of the OH65 doublet (in absorption) to those of the 
  [O {\sc i}] 63 $\mu$m (hereafter OI63) emission line in
  Fig.~\ref{oh65oi63}a.  These velocities tend to be anticorrelated: when the
  OH65 peak is blueshifted, the OI63 peak tends to be redshifted and vice
  versa. Fig.~\ref{oh65oi63}b shows $V_{\mathrm{p,abs}}^{\mathrm{OH65}}$ as a
  function of the OI63 red-blue asymmetry. Consistent with panel a, panel b
  shows that, when OH65 is blueshifted, the OI63 line usually has a
  ``red-type'' profile (and vice versa), i.e., the profile is distorted by a
  reduction of the intensity on the blue side. This is fully consistent with
  the scenario of atomic oxygen also outflowing with the OH, and thus
  absorbing the continuum and line emission arising from
  behind\footnote{Although some continuum is absorbed, 
    the OI63 line is not seen in absorption because it has an {\it A}-Einstein
    coefficient of $\sim10^{-4}$ s$^{-1}$ (much lower than OH65, $\approx1.2$
    s$^{-1}$) and is thus easily excited through collisions in warm and
    dense nuclear regions. The OI63 line is only seen in absorption
  in sources where high columns of oxygen are found in extended, low-density
  regions in front of the 63 $\mu$m continuum source 
  \citep[GA12,][]{fal15}.}.   
  It also rules out that the OH65 shifts 
  are the result of global nuclear motion relative to the more tenous and
  extended region probed by the [C {\sc ii}] line, because in this latter case
  one would expect the OH65 and OI63 to peak at the same velocities, contrary
  to the observed trends. 


To put these results on a first quantitative framework, the moderate
velocities found here are similar to those predicted by \cite{kin03} in his
analytical modeling of outflows driven by a wind with momentum flux comparable
to the Eddington-limited radiation field ($L_{\mathrm{edd}}/c$), i.e., 
$v_m=(G\,L_{\mathrm{edd}}/2f_g\sigma^2c)^{1/2}$, giving $\approx80$
\kms\ for $L_{\mathrm{edd}}=10^{12}$ \Lsun, velocity dispersion $\sigma=200$
\kms, and gas fraction $f_g=0.16$. The derivation of $v_m$ in \cite{kin03} 
assumes an isothermal sphere of dark matter, involving an outflowing gas
mass of
$3\times10^8\times(\sigma/200\,\mathrm{km\,s^{-1}})^2\times(r/100\,\mathrm{pc})$
\Msun\ and a column density of 
$2\times10^{23}\times(\sigma/200\,\mathrm{km\,s^{-1}})^2\times(r/100\,\mathrm{pc})^{-1}$. It ignores the potential well, implicitely assuming that the gas is
supported by radiation pressure or rotation.

{\it These results show that the nuclear
    regions of galaxies with high velocity outflows (as seen in OH119)
    also show slower nuclear outflows, i.e., the
    far-IR ``photospheres'' where OH65 is formed are (partially or completely)
    expanding at low velocities. Since OH65 requires high columns to be 
    detected, star formation may proceed at significant rates in this
    expanding material, with the consequence that star formation propagates
    outwards.}  

\subsection{Velocity components} \label{velcomp}

\begin{figure*}
\begin{center}
\includegraphics[angle=0,scale=.6]{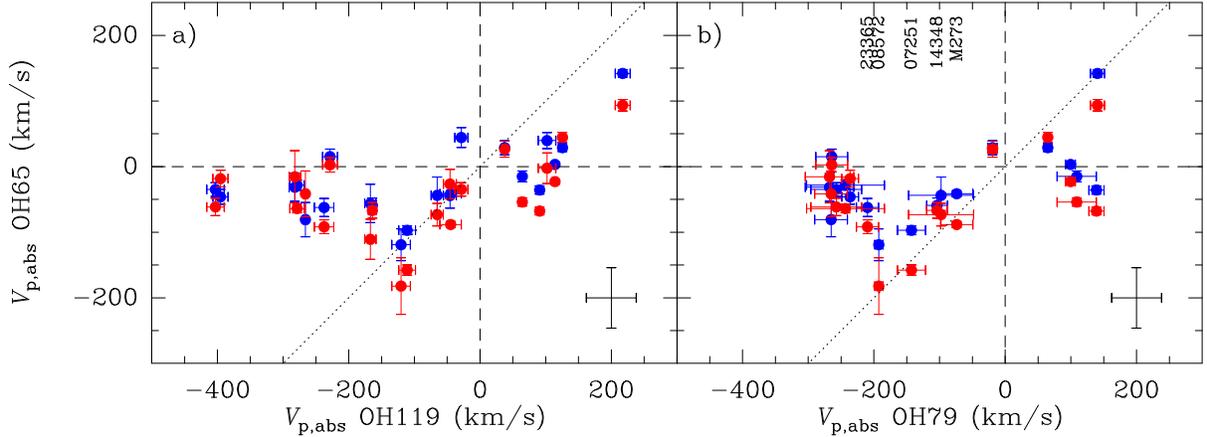}
\end{center}
\caption{a) 
  The velocity of peak absorption of the OH65 doublet,
  $V_{p,abs}^{\mathrm{OH65}}$, as a function of
  a) the velocity of peak absorption of the OH119 doublet (blue transition)
  and b) of the OH79 doublet (blue transition). Blue and red 
  symbols indicate $V_{p,abs}^{\mathrm{OH65}}$ of the blue and red transitions
  of the OH65 doublet. Abbreviated names are indicated for some sources. The
  dotted lines indicate equal velocities,
  $V_{p,abs}^{\mathrm{OH65}}=V_{p,abs}^{\mathrm{OH119}}$ and
  $V_{p,abs}^{\mathrm{OH65}}=V_{p,abs}^{\mathrm{OH79}}$. The errorbars
    on the lower-right side of each panel indicate potential ``skewness''
    uncertainties due to pointing drifts.
}   
\label{ofvelsoh}
\end{figure*}

\begin{figure}
\begin{center}
\includegraphics[angle=0,scale=.65]{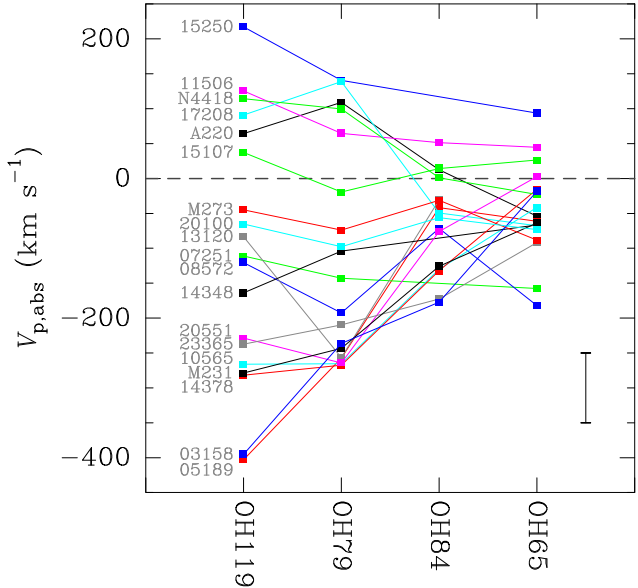}
\end{center}
\caption{a) The velocity of peak absorption ($V_{p,abs}$) of the OH119, OH79,
  OH84, and OH65 doublets in all sources where OH65 and or OH84 are
  detected. $|V_{\mathrm{p,abs}}|$ decreases with increasing excitation
  energy. The errorbar indicates potential ``skewness'' uncertainties
    from pointing drifts. Abbreviated source names are indicated.
}   
\label{vpaoh}
\end{figure}

Figure \ref{ofvelsoh} summarizes the relationship between
$V_{\mathrm{p,abs}}^{\mathrm{OH65}}$ and the velocities inferred from OH79 and
OH119. 
Some sources have 
$V_{\mathrm{p,abs}}^{\mathrm{OH65}}\approx
V_{\mathrm{p,abs}}^{\mathrm{OH119}}\approx V_{\mathrm{p,abs}}^{\mathrm{OH79}}<0$
(Fig.~\ref{ofvelsoh}), indicating the presence of nuclear outflowing gas with
the three lines tracing basically the same gas. In 
other sources, OH65 peaks at different velocities than OH119 and OH79,
indicating that different components are traced by the OH doublets.
It is also apparent from Figs.~\ref{ohblue} and \ref{ofvelsoh} that the
ground-state OH119 doublet usually traces gas at more blueshifted
velocities than the other doublets in most sources.
The peak absorption velocities of the four doublets are
compared in Fig.~\ref{vpaoh} for all sources where OH65 is detected, clearly
showing a tendency for the high-lying OH65 and OH84 lines to trace gas at
lower velocities than the ground-state OH119 and OH79 doublets. 
In the ground-state lines, absorption and reemission at systemic
  velocities take place over large volumes and tend to cancel each other 
\citep[see Fig.~4 in][and Fig.~8 in GA14]{spo13},
  leading to clear P-Cygni profiles due to absorption and reemission at
  higher velocities. 
This cancellation cannot happen in the high-lying lines, in which
  lower velocity shifts indicate higher column densities of gas
  with lower outflowing velocities.


\subsection{Compact and extended outflows} \label{comext}

Despite the relatively low spatial resolution of {\it Herschel}/PACS, the
OH excitation can nevertheless constrain the spatial extent of the outflows. 
In IRAS~08572+3915, Mrk~273, and IRAS~20551$-$4250, the absorption in the 
excited OH84 transition is almost as strong as (or even stronger than) the
OH119 absorption (Fig.~\ref{ohblue}).  In these objects, essentially all
outflowing gas traced by the ground-state and optically thick OH119 is
significantly excited by the far-IR field, indicating relatively small
distances to the nuclear region and thus a compact outflow.  Mrk~273 in
particular was found by \cite{cic14} from high resolution CO
(1-0) observations to have the smallest size among all their detected ULIRGs. 
In the other sources, OH119 is much stronger than OH84 at least at some
negative velocities, indicating the presence of a spatially extended outflow
component. 

\subsection{Inflows} \label{inflows}

\begin{figure*}
\begin{center}
\includegraphics[angle=0,scale=.6]{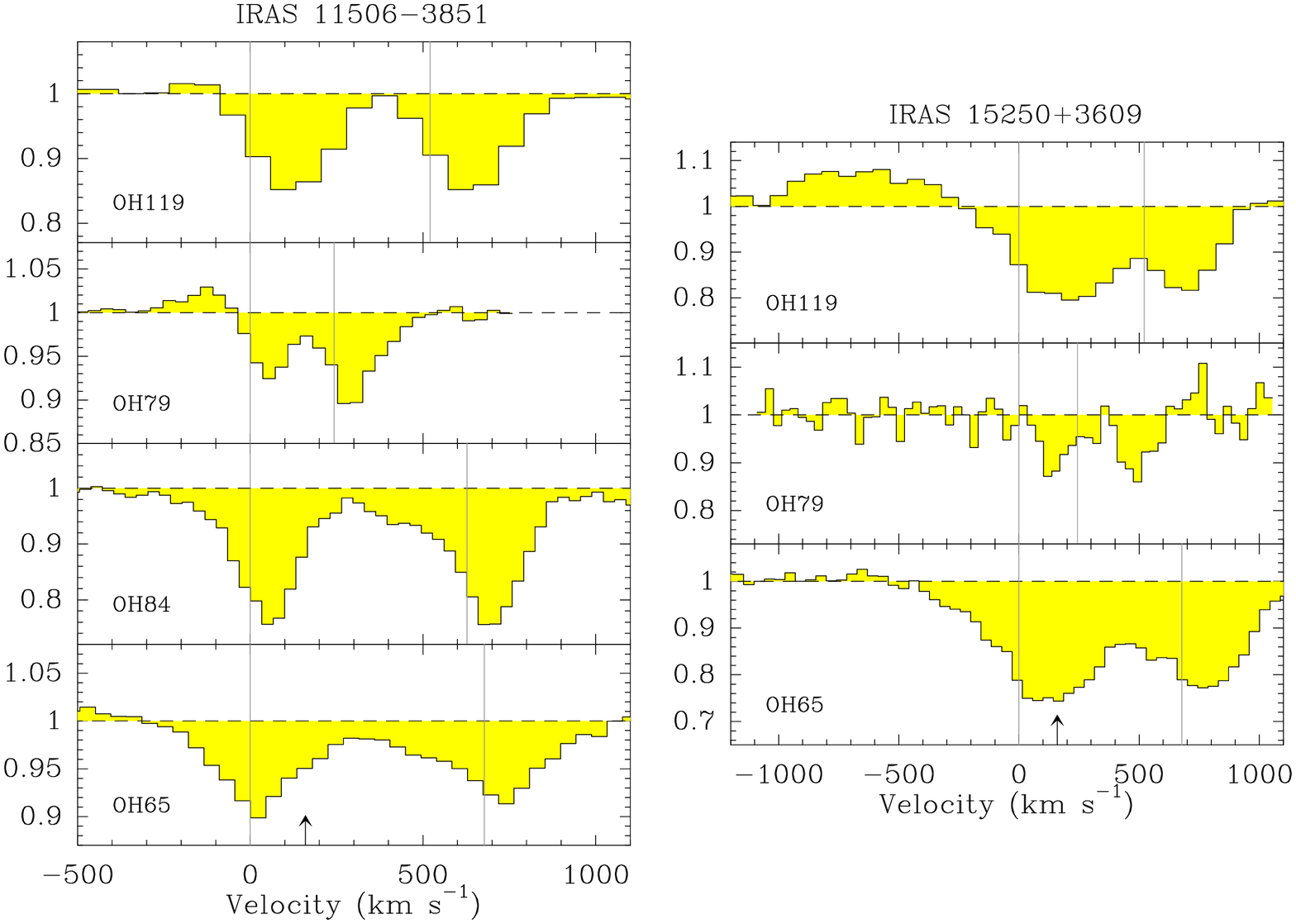}
\end{center}
\caption{The observed OH line shapes in IRAS~11506$-$3851 and IRAS~15250+3609,
  where inflow signatures are detected. The vertical gray lines in
  all panels indicate the rest position of the two $l-$doubling
  components, with the velocity scale taken relative to the blue component of
  each doublet. The arrow in the OH65 panel indicates possible contamination
  by a high-lying \hdo\ line.
}   
\label{inf}
\end{figure*}

Besides the prevalent outflowing motions traced by OH, inflow
signatures are also seen in a few (U)LIRGs as observed with
  Herschel/PACS. In addition to the previously 
studied case of NGC~4418 \citep{gon12}, where the inflow has also been 
inferred from high-angular resolution observations \citep{sak13,cos13}, 
the case of Zw~049 where an inverse P-Cygni profile is also observed in 
the [O {\sc i}]63 $\mu$m line \citep{fal15}, the inverse P-Cygni
profile of OH119 in Circinus \citep{sto16}, and the inflow observed
in OH119 around Arp~299a \citep{fal17}, the clearest examples are
the LIRG IRAS~11506$-$3851 and the ULIRG IRAS~15250+3609. Their OH spectra are
shown in Fig.~\ref{inf}, where the redshifts are obtained from Gaussian fits
to the [C  {\sc ii}]158 $\mu$m profiles. In IRAS~11506$-$3851, the 
OH119, OH79, and OH84 peak absorption velocities are redshifted 
relative to [C {\sc ii}], while the highest-lying OH65 velocity is more
centered than the others but still slightly redshifted 
(see also Fig.~\ref{vpaoh}). The 
strongest evidence for inflow comes from OH79, which shows an inverse P-Cygni
profile with an emission feature at $\sim-100$ \kms\ and a concomitant
asymmetry between the 
red and blue components of the doublet. The emission feature appears to have a
weak counterpart in OH119 as well. In addition, the high-lying OH84 
line shows hints of a blueshifted line wing in absorption, 
possibly indicating the coexistence of an outflow in the nuclear
region, as observed in CO \citep{per16}. In IRAS~15250+3609, the observed 
OH119, OH79, and OH65 doublets are also redshifted relative to [C {\sc ii}],
and OH119 shows a broad emission feature between $\sim-300$ and $-1000$
\kms. Rather than a simple infall around a central warm source, the extreme
velocities in this source rather suggest an approaching double-nucleus system,
with the strongest source of far-IR radiation associated with the nucleus
closer to the observer and illuminating the secondary nucleus
responsible for the OH119 emission feature, which may be disrupted by tidal
forces. Interestingly, this is the source with the highest equivalent
  width in OH65 (GA15, see also Fig.~\ref{oh65v84}), and a  
blueshifted line wing in this doublet also suggests the simultaneous 
occurrence of a nuclear outflow, as observed in HCO$^+$ $3-2$ \citep{ima16}. 
Both IRAS~11506$-$3851 and IRAS~15250+3609 
show a blueshift profile in the OI63 line, i.e., the line shows a reduction 
of the intensity on the red side (Fig.~\ref{oh65oi63}b).
These sources will be studied in more detail in future work.

\section{Radiative transfer models} \label{rt}

\subsection{A library of model components} \label{libr}

\begin{figure}
\begin{center}
\includegraphics[angle=0,scale=.5]{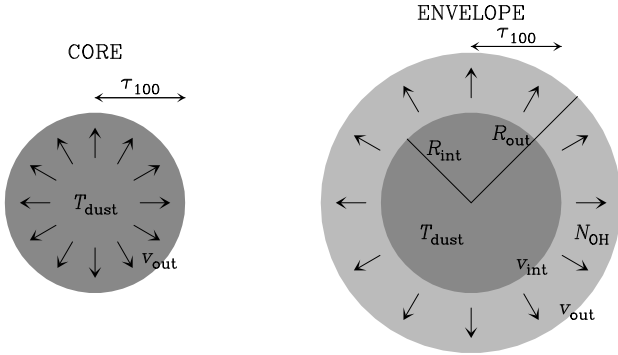}
\end{center}
\caption{Schematic representation of the two types of model components, CORE
  and ENVELOPE and independent parameters (\S\ref{libr}) used to fit the
  observed OH line profiles.  
}   
\label{modeltypes}
\end{figure}

In order to quantitatively estimate the energetics associated with the
observed outflows, we have developed a library of model components with the
radiative transfer code described in \cite{gon97,gon99}. In short, the method
calculates the statistical equilibrium populations in all shells of a
spherically symmetric source, as well as the emergent continuum and velocity
profiles of all lines after convolution with the PACS spectral resolution. 
The approach is
thus non-LTE and non-local, takes into account the radiative pumping by both
local and non-local dust (the dust is mixed with the gas), and accurately 
includes the effects of both line and continuum opacity effects and line
overlap between the $l-$doubling components of OH. The gas velocity field 
can be composed of pure turbulence and radially outflowing motions with
or without velocity gradients. The radial velocity is allowed to vary
continuously within any shell. We developed two types
of models to fit the continuum and line emission of our sample sources: "CORE"
and "ENVELOPE" models (Fig.~\ref{modeltypes}).  They are described below.

{\sl CORE models:} These models are intended to describe the continuum
emission and line absorption at systemic velocities produced in the
nuclear regions of galaxies (GA15). They consist of a source of dust
with temperature \tdust\ and radial optical depth at 100 $\mu$m \taud. The OH,
mixed with the dust, has turbulence  $\Delta V$, and its column is determined
by \taud\ according to eq.~(1) in \cite{gon14b} with an OH abundance
relative to H nuclei of $X_{\mathrm{OH}}=2.5\times10^{-6}$. The adopted
abundance is, within a factor of $\sim3$, consistent with the value inferred
from multi-transition observations of OH in the Galactic Sgr~B2 and Orion KL
outflow \citep{goi02,goi06} and in buried galaxy nuclei
\citep[GA12,][]{fal15}, and with chemical models of dense photodissociation
regions \citep[the peak value,][]{ste95}, and of cosmic-rays and X-rays
dominated regions \citep{mei11}. 
The source structure includes expansion with constant velocity $v_{\mathrm{out}}$
to describe the low-velocity blueshifts of the peak absorption of OH65 that 
indicate outflowing gas in the nuclear region (\S\ref{oh65vshift}), 
and is divided into shells to account for the varying excitation with
  depth into the source. 
The free parameters of the CORE components which are varied from model to
model are then \tdust, \taud, $\Delta V$, and $v_{\mathrm{out}}$.

{\sl ENVELOPE models:} these models also consist of a central source of only 
dust which is {\em surrounded} by an expanding shell of gas and dust (that is
also divided into shells to account for the varying physical conditions with
distance to the central source). The gas velocity is allowed to vary linearly
with radius across the expanding shell (GA14), and thus the free parameters
are \tdust\ and \taud\ of the central source, and 
$R_{\mathrm{out}}/R_{\mathrm{int}}$, $v_{\mathrm{int}}$,
$v_{\mathrm{out}}$, $\Delta V$, and $N_{\mathrm{OH}}$ of the expanding shell.
The density of dust across the shell is determined by the OH density, assuming
also $X_{\mathrm{OH}}=2.5\times10^{-6}$ and a gas-to-dust ratio by mass of 100
(GA14). In all models we assume a constant mass outflow rate, so that
$n_{\mathrm{H}}\,v\,r^2$ is uniform between $R_{\mathrm{int}}$ and
$R_{\mathrm{out}}$ (GA14). 


\subsection{$\chi^2$ fitting} \label{chi2}

Our approach relies on the assumption that the OH profiles can be considered a
superposition of the single-component models described above. This 
is questionable mostly at central velocities, where there may be radiative
coupling between different gas components, but has the overall advantage of
simplicity. 

Up to $N_C=3$ model components were allowed to be combined in order to
simultaneously fit the continuum-normalized OH119, OH79, OH84 (when
available), and OH65 spectra in a given source. For each combination of
components, the modeled continuum-normalized value of line $i$, velocity
channel $j$, is 
\begin{equation}
S_{ij}=1+\sum_{l=1}^{N_C} f_{il} \, s_{ijl},
\end{equation}
where the sum extends to the $N_C$ components of the combination. $f_{il}$ is
the ratio of the continuum flux density of model $l$ to the total (observed)
continuum, at the wavelength of line $i$:
\begin{equation}
f_{il}=C_{il}/C_{i,\mathrm{OBS}},
\label{fil}
\end{equation}
where $C$ stands for the continuum flux density and OBS refers to the observed
values. $s_{ijl}$ is the continuum-normalized model prediction minus $1$: 
\begin{equation}
s_{ijl}=F_{ijl}/C_{il}-1,
\end{equation}
where $F_{ijl}$ and $C_{il}$ are the predicted line flux density and continuum
flux density, respectively, for model component $l$.
Our minimization process calculates, for each combination of model components,
the values of $f_{il}$ that best fit the observed spectra. For a given model
component $l$, however, the values of $f_{il}$ for different lines are related
to each other through 
\begin{equation}
f_{il} = f_{\mathrm{OH119},l} \, 
\frac{C_{il}/C_{\mathrm{OH119},l}}{C_{i,\mathrm{OBS}}/C_{\mathrm{OH119,OBS}}},
\label{fil2}
\end{equation}
and thus only the values of $f_{\mathrm{OH119},l}$ are fitted.
The reduced $\chi^2$ value for a given combination that we aim to minimize is
\begin{equation}
\chi^2=\frac{1}{\sum_{l=1}^{N_C}N_v(i)-N_C} 
\sum_{i=1}^{N_l} \sum_{j=1}^{N_v(i)} \frac{1}{\sigma_i^2} [S_{ij}-E_{ij}]^2,
\label{chi2red}
\end{equation}
where $N_l$ is the number of lines fitted in a given source, $N_v(i)$ and 
$\sigma_i$ are the number of velocity channels of and the rms noise around
line $i$, and $E_{ij}$ is the observed continuum-normalized value of line $i$
at velocity channel $j$. The $N_C$ parameters that are varied to minimize 
$\chi^2$ for each combination are $f_{\mathrm{OH119},l}$, and this is done using
standard methods.

In practice, we used the same ``line-averaged'' value for all $\sigma_i$ in a
given source, and increased the relative weight of channels in the line 
wings, in which we are mostly interested. 
Depending on the observed line shapes in a
given source, we selected ($N_{\mathrm{CORE}}$, $N_{\mathrm{ENVELOPE}}$)
as $(1,2)$, or $(1,1)$. 
The minimization procedure was done for all possible 
combinations allowed by the following restrictions: $(i)$ 
the predicted SED for the components should be as warm or warmer than
the observed SED. As the OH expands, the molecules are subject to a
progressively colder far-IR SED, but not colder than the average galaxy SED as
seen by the observer. $(ii)$ In some sources, a maximum \tdust\ was 
imposed for the CORE models, to avoid overestimating the continuum at short
wavelengths. More than 100 CORE models and more than 400 ENVELOPE models were
generated, giving from $\sim10^5$ to $\mathrm{several}\times10^6$
allowed combinations. 
 
In addition to $\chi^2$, we calculated the corresponding value
$\chi_{\mathrm{BW}}^2$ that uses only the velocity channels at blueshifted
velocities. A first selection was done including all combinations with 
a value of $\chi^2$ within 20\% of the minimum $\chi^2$. Among all these
combinations, the best fit was selected as the combination that gives the
minimum value of $\chi^2_{\mathrm{BW}}$. Figures~\ref{fitsoh119}, \ref{fitsoh79},
\ref{fitsoh84}, and \ref{fitsoh65} show our best fit composite models for
OH119, OH79, OH84, and OH65 in the 14 modeled sources (\S\ref{obs}). 

\subsection{Approaching the minimum $\chi^2$} \label{minimum}

Since the number of input parameters involved in the ENVELOPE models is
  high, the following 3-step strategy to approach the minimum was generally
  adopted: 
  $(i)$ we first developed a coarse grid of models, common to all
    sources, by varying the velocity 
  field, $N_{\mathrm{OH}}$, and \tdust, for two values of
  $R_{\mathrm{out}}/R_{\mathrm{int}}=1.3$ and $1.6$, and keeping $\tau_{100}=1$ 
  and $\Delta V=100$ \kms\ fixed. A first search of the minimum was 
  then performed, and the grid was additionally refined in some regions of 
  parameter space if no satisfactory fit was
  found. $(ii)$ The grid was additionally
  refined by varying the velocity field, $N_{\mathrm{OH}}$, and \tdust\ around
  the minimum and also varying to a greater extent
  $R_{\mathrm{out}}/R_{\mathrm{int}}$ and $\tau_{100}$. A second search of the
  minimum $\chi^2$ was carried out. $(iii)$ The final step consisted in
  further refining the model grid around the minimum, varying also $\Delta V$
  and with special consideration of $N_{\mathrm{OH}}$. The region around the
  minimum was then sufficiently populated to determine the errorbars
  associated with the energetics (see \S\ref{energetics}).


\subsection{Continuum and minimum size} \label{continuum}

Our fitting procedure uses the observed continuum-normalized OH spectra
 and the corresponding theoretical continuum-normalized spectra, as well
as the observed far-IR colors in the OH bands as inputs 
 (eq.~\ref{fil2}). Once the 
 value of $f_{\mathrm{OH119},l}$ is fitted, we use the observed 
{\it absolute} value of the continuum flux density at 119 $\mu$m,
$C_{\mathrm{OH119,OBS}}$, to infer the absolute value of the 119 $\mu$m
continuum flux density associated with model component $l$, 
$C_{\mathrm{OH119},l}=f_{\mathrm{OH119},l} \times C_{\mathrm{OH119,OBS}}$. 
Since the normalized SED for any component is fully determined for given
\tdust\ and $\tau_{100}$, the inferred $C_{\mathrm{OH119},l}$ value gives a
specific prediction for the absolute SED of the underlying far-IR source
associated with component $l$. The absolute calibration of the models is thus
based on the far-IR continuum, which in turn is the source of the OH
excitation. 

With the absolute SED known, the required solid angle $\Delta\Omega$ of
component $l$ is obtained, and hence the {\it minimum} size of the component 
through $R_1=D\sqrt{\Delta\Omega/\pi}$, where $D$ is the
distance to the source.  
This minimum radius implicitly assumes that the OH in each outflow
component is not clumpy (we relax this assumption below when calculating the
energetics), so that $f_{\mathrm{OH119},l}<1$ is due to additional continuum
from the galaxy not associated with the  $l-$component (i.e., associated with
other OH components or to far-IR emission unrelated to OH).   


We compare the predictions for the galaxy continua with the
observed SEDs for all modeled sources in Fig.~\ref{fitscont}, where the red
curves indicate the total continua (i.e., the sum of all components including
the non-outflowing one). Our superposition approach is
consistent with observations as long as the modeled 
continuum does not exceed the observed one at any wavelength (it can
underestimate the observed continuum if some of it is unassociated with
OH). In two sources, IRAS~03158+4227 and IRAS~14348$-$1447, the modeled
continuum slightly exceeds the observed SED at $\lambda<50$ $\mu$m, indicating
that the continua from the different components are not independent in 
these sources. 

\begin{figure*}
\begin{center}
\includegraphics[angle=0,scale=.65]{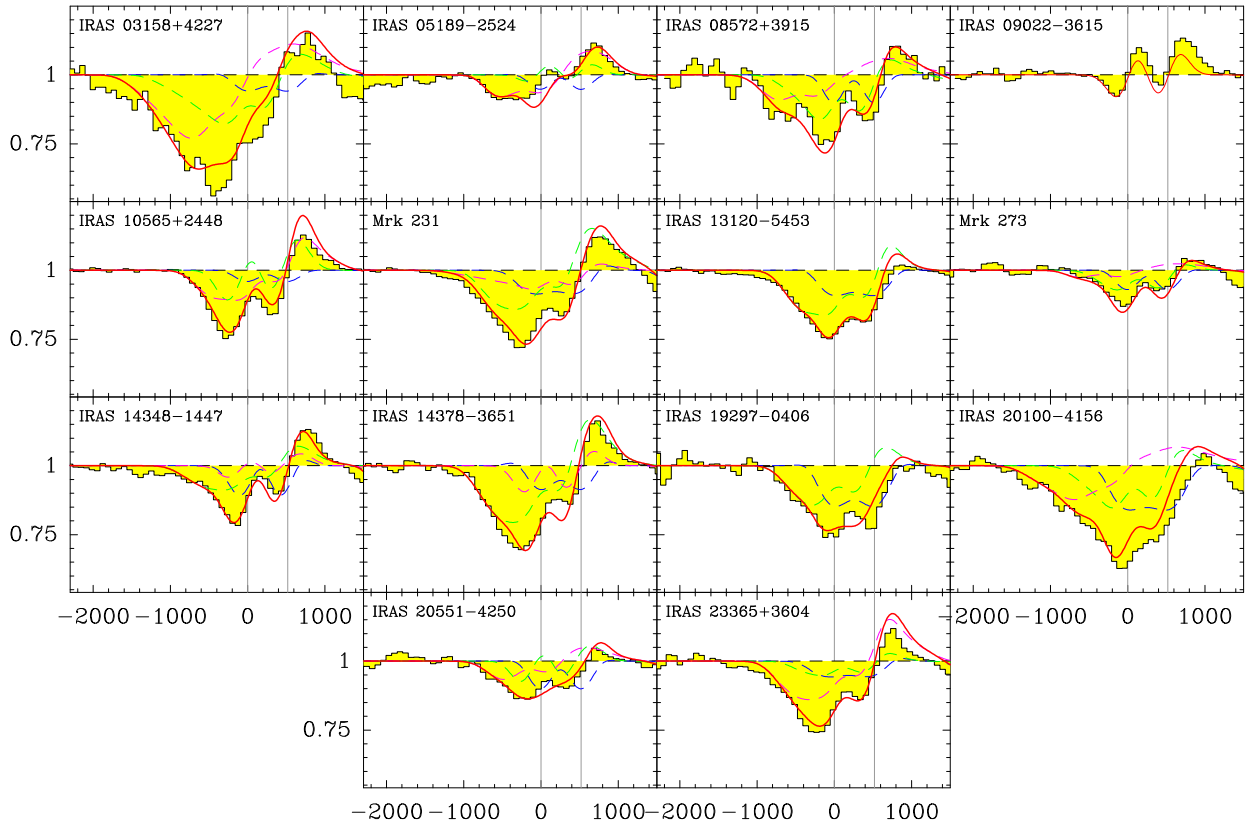}
\end{center}
\caption{Model fits to the observed OH119 doublets (shown as yellow-shaded
  histograms) in 14 sources. The models simultaneously fit the line profiles
    of all (OH119, OH79, OH84 when available, and OH65) doublets observed in
    each source through $\chi^2$ minimization using combinations of model
    components taken from a library. Dashed lines indicate
  different components (up to 3 in each source, in blue, green, magenta), and
  red is total. The two vertical lines in each panel
  indicate the rest position of the doublet transitions. The velocity scale is
  relative to the blue transition of the doublet. The fits to the other OH
  doublets are shown in Figs.~\ref{fitsoh79}-\ref{fitsoh65}, and the predicted
  far-IR continua are compared with the observed SEDs in Fig.~\ref{fitscont}.
}   
\label{fitsoh119}
\end{figure*}
\begin{figure*}
\begin{center}
\includegraphics[angle=0,scale=.65]{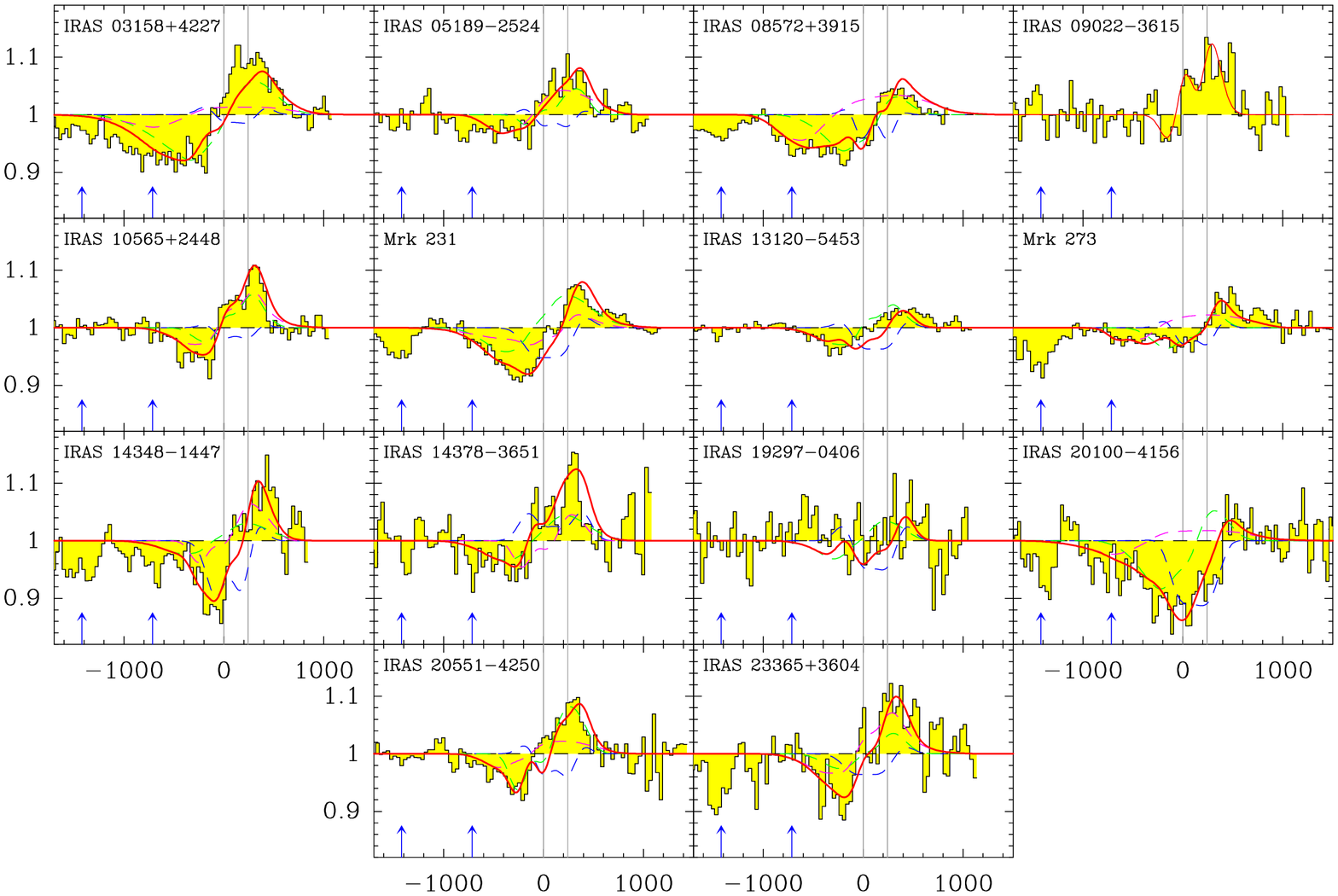}
\end{center}
\caption{Model fits to the OH79 doublets in 14 sources. Blue arrows indicate
  possible contamination by H$_2$O. See also caption of Fig.~\ref{fitsoh119}. 
}   
\label{fitsoh79}
\end{figure*}

\begin{figure*}
\begin{center}
\includegraphics[angle=0,scale=.65]{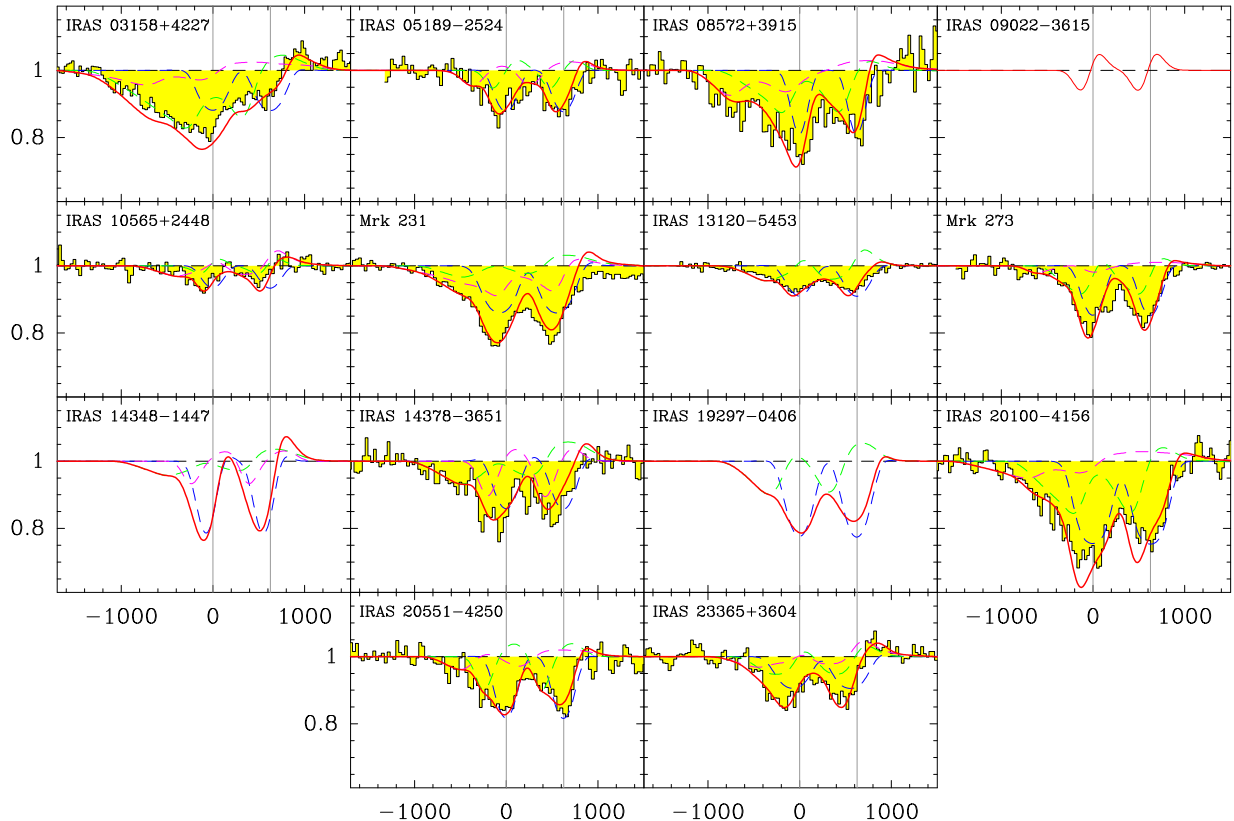}
\end{center}
\caption{Model fits to the OH84 doublets in 14 sources. See also caption of
  Fig.~\ref{fitsoh119}.
}   
\label{fitsoh84}
\end{figure*}
\begin{figure*}
\begin{center}
\includegraphics[angle=0,scale=.65]{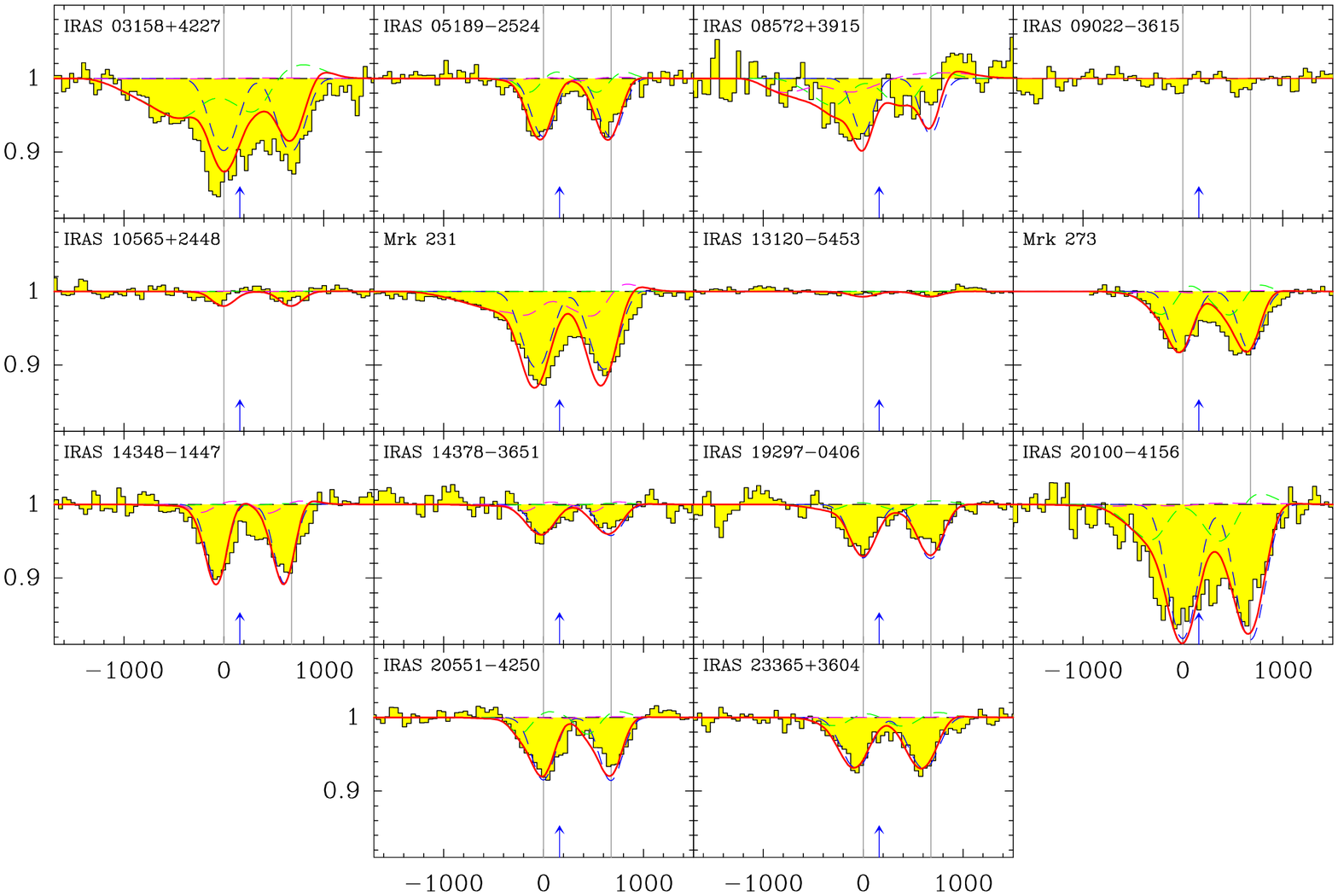}
\end{center}
\caption{Model fits to the OH65 doublets in 14 sources. Blue arrows indicate
  possible contamination by H$_2$O. See also caption of Fig.~\ref{fitsoh119}. 
}   
\label{fitsoh65}
\end{figure*}

\begin{figure*}
\begin{center}
\includegraphics[angle=0,scale=.65]{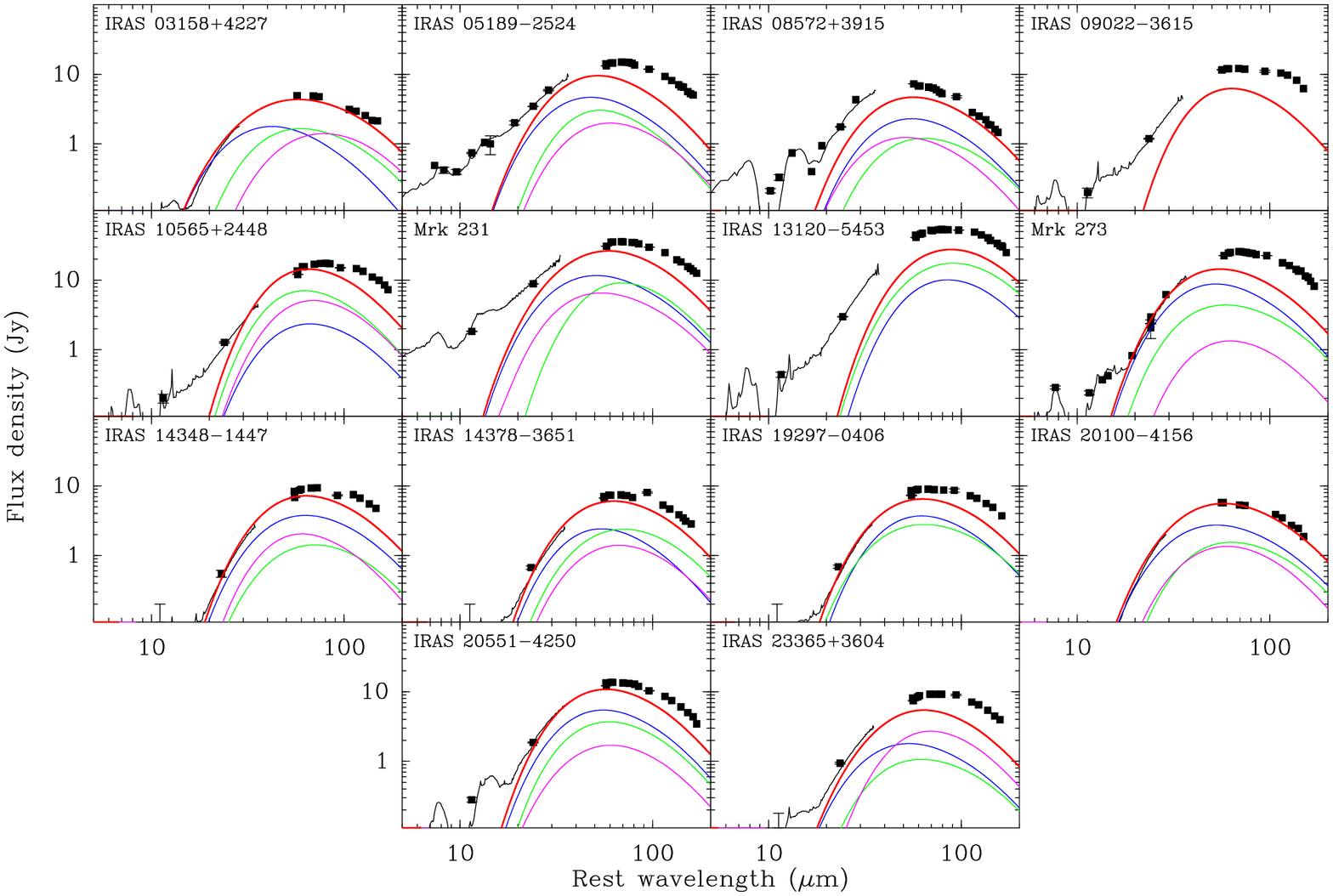}
\end{center}
\caption{Predictions for the continuum of the models for OH shown in 
 Figs.~\ref{fitsoh119}-\ref{fitsoh65}, compared with the observed spectral
 energy distributions. Blue, green, and magenta lines indicate the
 continua associated with the OH components shown with the same colors in
 Figs.~\ref{fitsoh119}-\ref{fitsoh65}, and red is total. In some sources, the
 continuum associated with the OH 
 components closely matches the observed SED between 20 and 50 $\mu$m, but the
 continuum is usually underpredicted at longer wavelengths.
}   
\label{fitscont}
\end{figure*}

\subsection{Estimating the energetics} \label{energetics}

Using the thin-shell approximation, the mass of the outflowing gas associated
with a given component $l$ is
\begin{equation}
M_{\mathrm{out}}=4\pi R_1^2 \mu \, m_{\mathrm{H}} \, N_{\mathrm{H}},   
\label{mshell}
\end{equation}
where $\mu=1.4$ is the mass per H nucleus, and 
$N_{\mathrm{H}}=N_{\mathrm{OH}}/X_{\mathrm{OH}}$. Both $R_1$ 
($=R_{\mathrm{out}}$ for the CORE models, and
$=(R_{\mathrm{int}}+R_{\mathrm{out}})/2$ for the ENVELOPE models)
and $N_{\mathrm{OH}}$ are directly inferred from the fitting procedure.

Since the continuum of each component underpredicts the observed SED, the
observations can also be interpreted in terms of a clumpy distribution, 
where the source radius $R$ is actually larger than $R_1$,
$R=f_c^{-1/2}\,R_1$, and the outflowing OH only covers a fraction
$f_c$ of the underlying continuum source. We determine the minimum value of
$f_c$ such that the modeled flux density does not overpredict the observed SED
{\it at any wavelength}\footnote{Note that $f_c>f_{119}$ in
  warm components, because the predicted flux density at $\lambda<100$ $\mu$m
  is closer to the observed one than at 119 $\mu$m.}. The outflowing gas mass
is, nevertheless, the same as 
in eq.~(\ref{mshell}), which we now express as
\begin{equation}
M_{\mathrm{out}}=4\pi f_c R^2 \mu \, m_{\mathrm{H}} \, N_{\mathrm{H}},
\label{mof}
\end{equation}
but the mass outflow rate and momentum flux (see below) decrease by a factor
of $f_c^{1/2}$. For the CORE models with $v_{\mathrm{out}}>0$, where the whole
source is expanding, we limit $N_{\mathrm{H}}$ to $5\times10^{23}$ \cmd\ if
the column through the source exceeds that value because this is the column
that yields a continuum optical depth of $\sim1$ at 60 $\mu$m.

While we consider $R$ to be the most likely value of the outflow radius, and
thus a clumpy distribution is adopted, the value of $R_1$ is relevant as it
indicates the radius below which the OH is predicted to cover the full $4\pi$
sr. This value is used in our simple dynamical model in \S\ref{discussion}.

There are two limiting approaches to the estimation of the energetics
associated with a given component: the local (maximum) or instantaneous values
and the average (minimum) values \citep[see also][]{rup05c}. The local
estimates use the inferred density $n_{\mathrm{H}}$ of the expanding gas: 
\begin{eqnarray}
\dot{M}_{\mathrm{loc}} & = & f_c \, 4\pi R^2 \mu \, m_{\mathrm{H}} \,
n_{\mathrm{H}} \, v = 
\frac{M_{\mathrm{out}} \, v}{\Delta R}
\\
\dot{P}_{\mathrm{loc}} & = & \dot{M}_{\mathrm{loc}} \, v,
\\
\dot{E}_{\mathrm{loc}} & = & \frac{1}{2} \dot{M}_{\mathrm{loc}} \, v^2,
\end{eqnarray}
where $\Delta R=R_{\mathrm{out}}-R_{\mathrm{int}}$ is the thickness of the
outflowing shell. These local estimates give the mass, momentum, and energy
crossing instantaneously per unit time a spherical surface of radius $R$, and
can only be considered ``time-averaged'' if we assume that the outflowing gas
extends to $r=0$ although we only observe a fraction of it owing to far-IR
opacity (extinction) and line opacity effects. This approach was used in
\cite{stu11} and GA14. The average values assume, by contrast, that there is
no more outflowing gas inside the curtain of dust, so that
\begin{eqnarray}
\dot{M}_{\mathrm{out}} & = & f_c \, 4\pi R^2 \mu \, m_{\mathrm{H}} \, 
\frac{N_{\mathrm{H}} \, v}{R} =
\frac{M_{\mathrm{out}} \, v}{R}
\label{mdot_out}
\\
\dot{P}_{\mathrm{out}} & = & \dot{M}_{\mathrm{out}} \, v
\label{pdot_out}
\\
\dot{E}_{\mathrm{out}} & = & \frac{1}{2}\dot{M}_{\mathrm{out}} \, v^2
\label{edot_out}
\end{eqnarray}

Evidently, the average values are a factor of $\Delta R/R$ lower than the
local values. {\it In this paper, we use the most conservative
$\dot{M}_{\mathrm{out}}$, $\dot{P}_{\mathrm{out}}$, and
  $\dot{E}_{\mathrm{out}}$ values to characterize the outflows}. 
These are the ``time-averaged thin shell''
values in \cite{rup05c}, which have also been used by a number of
authors describing the energetics of the ionized and neutral phases of
outflows \citep[e.g.,][]{rup13a,ara13,bor13,hec15}, and are most appropriate
for comparison with outflow models \citep[e.g.,][]{fau12,ste16,tho15}. In 
some studies, a factor of 3 higher values have been used because the emitting
spherical (or multiconical) volume is assumed to be filled with uniform
density \citep[e.g.,][]{fer10,fer15,mai12,rod13,cic14,har14,gar15}. For a
steady flow with constant velocity, however, we would expect a density at
the outer radius only 1/3 that of the average, thus also yielding the
expression in eq.~(\ref{mdot_out}). 

On the other hand, the assumed flow time scale ($R/v$)
is overestimated in this expression if a fraction of the gas
is loaded (and suddenly accelerated) at some distance from the
center, which is relevant here given the compact sizes we infer below. 
However, if the gas is smoothly accelerated from the center, the flow
  time scale will be longer than $R/v$ by a factor $\sim2$.
Our assumption of spherical symmetry (the $4\pi$ factor) may
overestimate the energetics by a factor $\sim2$ in the case of
bipolar emission with no 
gas flowing along the plane of sky. In systems where gas flows primarily
  along the plane of sky, the energetics will obviously be underestimated.
In the above expressions involving $v$, we use the average of
$v_{\mathrm{int}}$ and $v_{\mathrm{out}}$. 

The {\it total} values that characterize a given source are the sum of 
$M_{\mathrm{out}}$, $\dot{M}_{\mathrm{out}}$, $\dot{P}_{\mathrm{out}}$, and
$\dot{E}_{\mathrm{out}}$ over all components \citep{rup05c}, 
\begin{equation}
(M_{\mathrm{tot}},\dot{M}_{\mathrm{tot}},\dot{P}_{\mathrm{tot}},\dot{E}_{\mathrm{tot}})
=\sum_{\mathrm{comp}} 
(M_{\mathrm{out}},\dot{M}_{\mathrm{out}},\dot{P}_{\mathrm{out}},\dot{E}_{\mathrm{out}})
\label{par_glob}
\end{equation}
$\dot{M}_{\mathrm{tot}}$, $\dot{P}_{\mathrm{tot}}$, and
$\dot{E}_{\mathrm{tot}}$ are plotted in
Fig.~\ref{ener} as a function of $L_{\mathrm{bol}}$, $L_{\mathrm{AGN}}$, and
$L_{\mathrm{*}}$ (see also Fig.~\ref{pdotglb}).

To estimate the uncertainties of the total values in
  eq.~(\ref{par_glob}), 
we calculate the values of $M_{\mathrm{tot}}$, $\dot{M}_{\mathrm{tot}}$,
$\dot{P}_{\mathrm{tot}}$, and $\dot{E}_{\mathrm{tot}}$, for all model combinations 
that yield $\chi^2$ and $\chi^2_{\mathrm{BW}}$ to within 20\% of their minimum
values, and conservatively use the maximum and minimum values of the
energetics of all these combinations to define the error bars. 
Uncertainties usually range from $50$\% of the best
fit value, to more than a factor 2 in some cases, and do not include
systematic uncertainties due to our adopted OH abundance.

We also calculate the mass loading factor of each outflow component,
$\eta=\dot{M}_{\mathrm{out}}/\mathrm{SFR}$, where SFR is the star formation rate
($\mathrm{SFR}=10^{-10}\,L_{\mathrm{*}}$ \Msun\,yr$^{-1}$, where $L_{\mathrm{*}}$ has
units of \Lsun). The starburst and AGN contributions to
$L_{\mathrm{IR}}$ (see Table~\ref{tbl-1}) are estimated from the flux
densities at 15 and 30 $\mu$m according to \cite{vei09,vei13}.
The total value of $\eta$ is also obtained by summing
the contributions from all outflowing components.

%

\section{Modeling results} \label{modres}

\subsection{Overall model fits} \label{modfit}

Our best fits to the OH119, OH79, OH84, and OH65 doublets are shown in
Figs.~\ref{fitsoh119}, \ref{fitsoh79}, \ref{fitsoh84}, and \ref{fitsoh65},
respectively. Table~\ref{tbl-2} lists the parameters of the outflowing
  components for the best fit combination, together with the derived
  energetics. The total energetics of the outflows are
  listed in Table~\ref{tbl-3}.

In IRAS~09022$-$3615, where OH65 is not detected and OH84 is not
available, we use only one model component to fit OH119 and OH79. In
IRAS~13120$-$5453, where OH65 is also undetected, we use two model
components. Similarly, we use only two components in IRAS~19297$-$0406, where
OH79 is undetected. In the rest of the sources, three components are required
to obtain a reasonable model fit. One of them is used to match the absorption
in the excited OH84 and OH65 doublets at central velocities (CORE models),
though the fit sometimes requires expansion velocities of up to $100$
\kms. The other two components are needed to fit line absorption and emission
at higher velocities ($\gtrsim200$ \kms). With our limited number of
components, some very high-velocity wings seen in OH119 (IRAS~03158+4227,
IRAS~13120$-$5453, IRAS~14348$-$1447, IRAS~14378$-$3651, and IRAS~23365+3604)
are not fitted, because the associated energetics from only OH119 are
relatively uncertain.

The overall goodness of the fits to the line shapes indicates that, 
in spite of the two-lobed structure generally observed for the 
  outflows traced by CO \citep[e.g.,][]{ala11,cic14,gar15,fer15},
the spherical symmetry implicit in the model is, as a first
approximation, a reliable approach to the outflow systems, in
line with the wide-angle outflow geometry inferred from OH119 detection
statistics \citep[][see also \S\ref{cov}]{vei13}.
However, some sources show indications of significant departures
from sphericity, which is best seen in the fits to the 
OH119 doublet in Fig.~\ref{fitsoh119}. In strict spherical symmetry, an OH119
redshifted emission feature that is weak relative to the absorption feature
can only be attributed to far-IR extinction of the emission behind the
continuum source, and detection of OH65 (a tracer of optically thick continuum
at 119 $\mu$m) would be consistent with this explanation. In
IRAS~13120$-$5453, 
however, there is essentially no reemission at redshifted velocities, but OH65
is undetected. The modeled OH119 profile indeed strongly overpredicts the
redshifted emission, indicating that the gas responsible for the absorption is
not accompanied by comparable amounts of OH on the other side of the continuum
source. Significant overpredictions of the OH119 redshifted emission are also
seen in IRAS~03158+4227 (absorption by CH$^+$ may reduce the OH119
  emission feature in this source, \S\ref{overall}), 
IRAS~10565+2448, IRAS~19297$-$0406, and
IRAS~23365+3604. On the other hand, the opposite effect is seen in 
IRAS~09022$-$3615, where the OH119 emission feature is stronger than the
absorption feature. This can be explained by either collisional excitation of
the OH119 line, or a relative excess of outflowing gas behind the continuum
source. We consider the energetics inferred from the models of
IRAS~09022$-$3615 and IRAS~19297$-$0406 relatively uncertain,
and we do not include them in the analysis below. As argued in \S\ref{cov},
the molecular outflows in IRAS~05189$-$2524, Mrk~273, and IRAS~20551$-$4250,
in which the OH119 absorption trough is relatively weak, are most likely
relatively collimated. 

\begin{figure*}
\begin{center}
\includegraphics[angle=0,scale=.65]{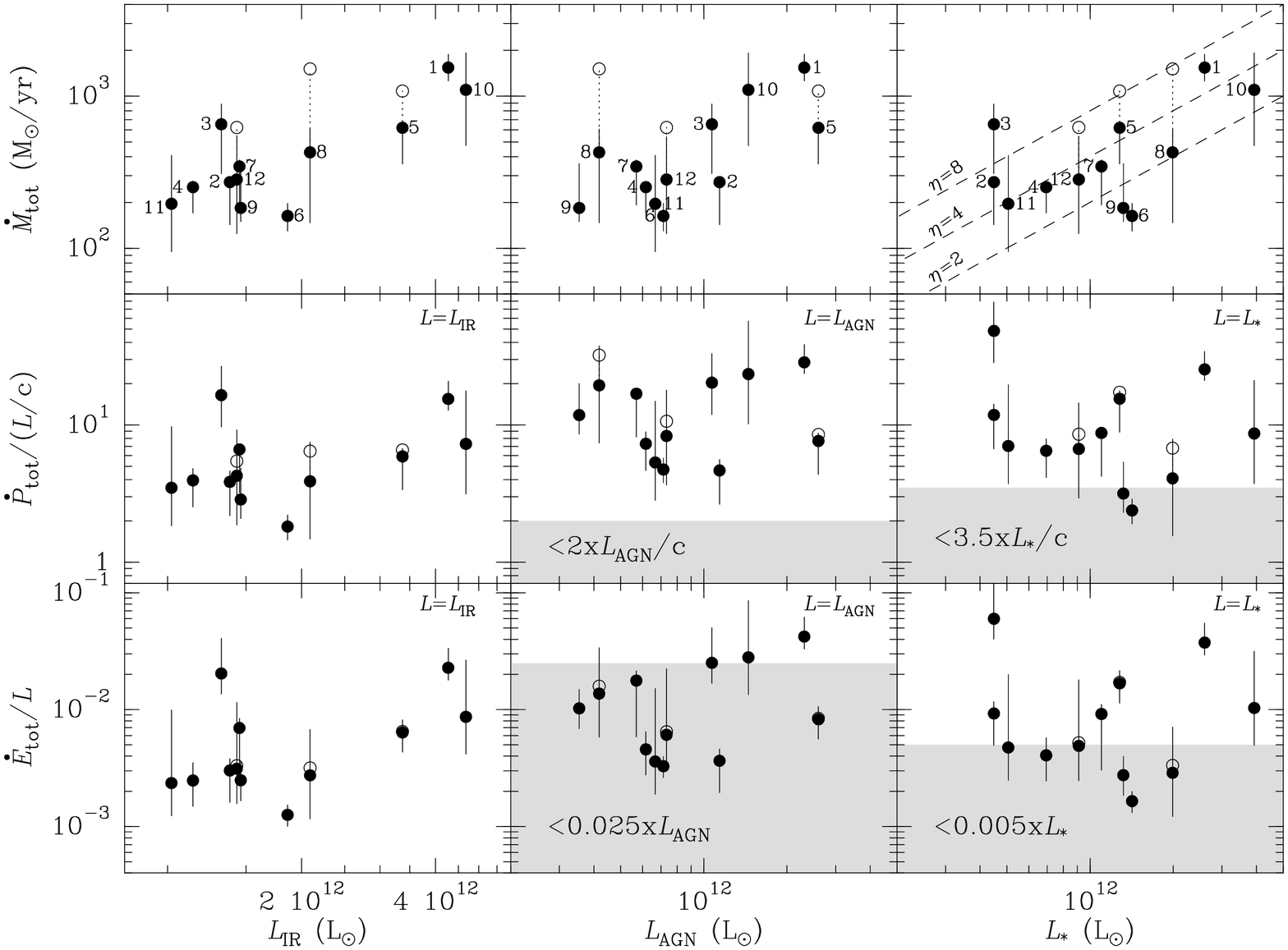}
\end{center}
\caption{Estimated energetics of 12 out of 14 sources modeled in this
  paper; modeling results for IRAS~09022-3615 and IRAS~19297-0406 are
    relatively uncertain and not considered. 
  $\dot{M}_{\mathrm{tot}}$, $\dot{P}_{\mathrm{tot}}$, and 
  $\dot{E}_{\mathrm{tot}}$ are the total values defined in
  eqs.~(\ref{par_glob}) (see \S\ref{energetics}). Filled 
  circles indicate the values obtained by ignoring the (uncertain) low
  velocity ($<200$ \kms) components, while open circles include them.
  Sources are labeled as: 1: IRAS~03158+4227; 2: IRAS~05189$-$2524; 3:
    IRAS~08572+3915; 4: IRAS~10565+2448; 5: Mrk~231; 6: IRAS~13120$-$5453; 7:
    Mrk~273; 8: IRAS~14348$-$1447; 9: IRAS~14378$-$3651; 
    10: IRAS~20100$-$4156 11: IRAS~20551$-$4250; 12: IRAS~23365+3604. 
  The dashed lines in the upper-right panel indicate mass loading factors of
  $\eta=\dot{M}_{\mathrm{tot}}/\mathrm{SFR}=2$, 4, and 8. Shaded rectangles
    mark the momentum and energy rates that can be supplied by an AGN and a
    starburst according to several models discussed in \S\ref{modener}. 
}   
\label{ener}
\end{figure*}

\begin{figure*}
\begin{center}
\includegraphics[angle=0,scale=.68]{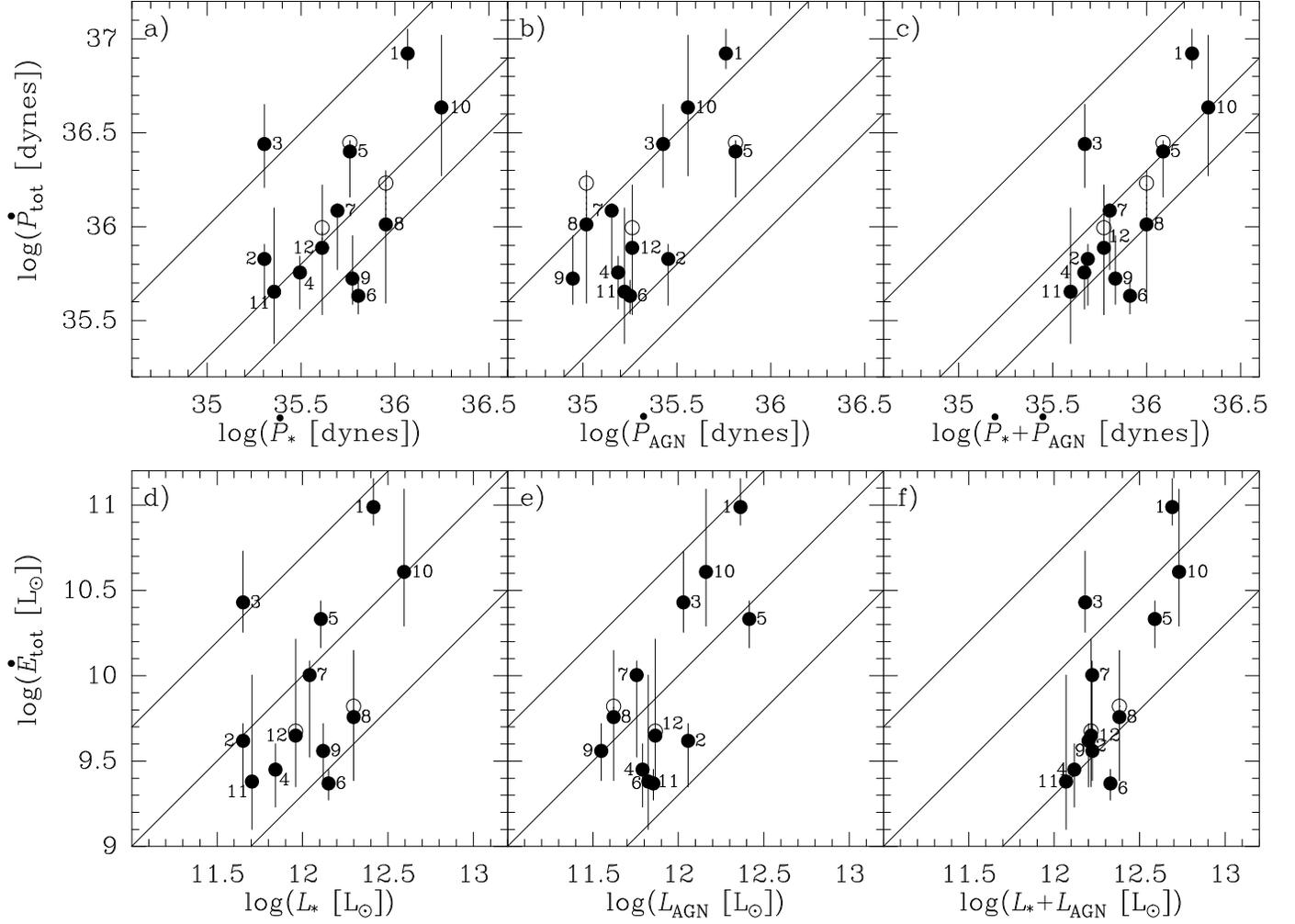}
\end{center}
\caption{{\it Upper panels:} the inferred total momentum fluxes
  ($\dot{P}_{\mathrm{tot}}$, eq.~\ref{par_glob}) as a function of a) the
  estimated momentum rate supplied by the starburst
  ($\dot{P}_{\mathrm{*}}\sim3.5L_{\mathrm{*}}/c$), b) the 
estimated momentum rate supplied by the AGN
($\dot{P}_{\mathrm{AGN}}\sim2L_{\mathrm{AGN}}/c$), and c) the sum of
  both. The gray diagonal lines indicate ratios of
$\dot{P}_{\mathrm{tot}}/\dot{P}_{\mathrm{*,AGN,TOTAL}}=1$, 2, and 10. 
{\it Lower panels:} the estimated total energy fluxes
  ($\dot{E}_{\mathrm{tot}}$) as a function of d) the starburst luminosity,
e) the AGN luminosity, and f) the sum of both. The gray diagonal lines
indicate ratios $\dot{E}_{\mathrm{tot}}/L_{\mathrm{*,AGN,TOTAL}}=0.2$, 1, and 5\%.
Filled circles indicate the values obtained by ignoring the (uncertain)
low velocity ($<200$ \kms) components, while open circles include them.
Galaxies are labeled as in Fig.~\ref{ener}.
}   
\label{pdotglb}
\end{figure*}

\begin{figure*}
\begin{center}
\includegraphics[angle=0,scale=.68]{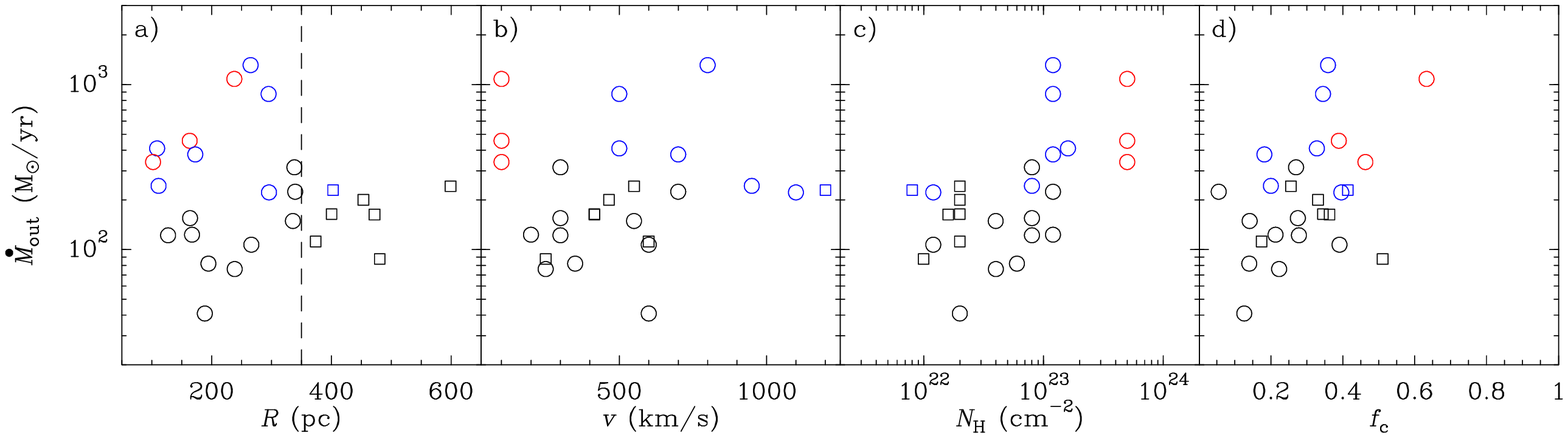}
\end{center}
\caption{The mass outflow rate, $\dot{M}_{\mathrm{out}}$, of the individual
  components involved in the best fit models, are
  shown as a function of the radius $R$, velocity $v$, column density
  $N_{\mathrm{H}}$, and covering factor $f_c$. Red symbols indicate
  CORE components; otherwise they indicate ENVELOPE components (see
  Fig.~\ref{modeltypes}). Blue symbols indicate components with high
  $\dot{P}_{\mathrm{out}}>1.0\times10^{36}$ dyn.
  a) The dashed vertical line marks the value we use for the
  classification of compact ($R<350$\,pc) and extended ($R>350$\,pc)
  components. 
  b-d) Circles and squares indicate compact and extended components,
  respectively. 
}   
\label{mdot}
\end{figure*}

\subsection{Overall continua and energetics} \label{modener}

The predictions for the galaxy continua
are compared with the observed SEDs in Fig.~\ref{fitscont}, showing that the
total predicted continuum in most sources (in red) is similar to the observed
SED in the transition from mid- to far-IR wavelengths (i.e., from 25 to
$\sim50$ $\mu$m). This total predicted continuum also includes the warm
components with non-outflowing (or slowly outflowing) OH, so that the good
fits of the modeled continua indicate that OH is an excellent tracer of the
warm regions in ULIRGs  with and without outflowing gas. The close agreement
we find in most sources, however,
fails in IRAS~09022$-$3615 and IRAS~13120$-$5453, where OH65 is not 
detected. In most sources, the far-IR emission at $\gtrsim60$ $\mu$m
is underpredicted, indicating the presence of cold dust unrelated to OH.

The mass outflow rate and momentum and energy fluxes, $\dot{M}_{\mathrm{out}}$,
$\dot{P}_{\mathrm{out}}$, and $\dot{E}_{\mathrm{out}}$, associated with a
given component depend on 4 parameters (eqs.~\ref{mdot_out}, \ref{pdot_out},
and \ref{edot_out}): the size $R$, the column density $N_{\mathrm{H}}$, the
gas velocity $v$, and the covering factor $f_c$. The dependence of
$\dot{M}_{\mathrm{out}}$ on all these  
parameters is shown in Fig.~\ref{mdot} for all individual components
involved in the best fit models of our sample. The individual components can
be classified according to their predicted spatial extent. Components with
$R<350$\,pc are considered ``compact'' and are indicated with circles
in Fig.~\ref{mdot}, while those with $R>350$\,pc are considered
``extended'' and shown with squares. With typical
radii of $\sim100$\,pc, all CORE components are compact, as well as 
most ENVELOPE models. The small radii in these models are a direct
consequence of the need to match the outflowing gas observed in high-lying
OH84 and OH65 doublets, which require high $T_{\mathrm{dust}}\gtrsim70$ K for
the associated continuum source to pump the excited OH levels, and hence 
small sizes to avoid overestimating the continuum emission. Conversely, the
extended components reproduce the observed absorption and emission in the
ground OH119 and OH79 doublets at velocities where no contribution is seen
from the high-lying lines, with effective $T_{\mathrm{dust}}\sim55$ K.

The highest mass outflow rates $\dot{M}_{\mathrm{out}}\gtrsim300$
\Msun\,yr$^{-1}$ are 
mostly found in compact components, and can exhibit both low and high gas
velocities, but they consistently show high column densities
($N_{\mathrm{H}}\gtrsim10^{23}$ \cmd). 
From eqs.~(\ref{mdot_out}), (\ref{pdot_out}), and (\ref{edot_out}), we
  express
\begin{eqnarray}
\dot{M}_{\mathrm{out}} & = & 320\times \frac{f_c}{0.3}
\times \frac{R}{150\,\mathrm{pc}}
\times \frac{N_{\mathrm{H}}}{10^{23}\,\mathrm{cm^{-2}}} 
\times \frac{v}{500\,\mathrm{km\,s^{-1}}}, \nonumber \\
\dot{P}_{\mathrm{out}} & = & 10^{36}\times \frac{f_c}{0.3}
\times \frac{R}{150\,\mathrm{pc}}
\times \frac{N_{\mathrm{H}}}{10^{23}\,\mathrm{cm^{-2}}}      \nonumber \\
& \times & \left(\frac{v}{500\,\mathrm{km\,s^{-1}}}\right)^2, \nonumber \\
\dot{E}_{\mathrm{out}} & = & 2.5 \times 10^{43}\times \frac{f_c}{0.3}
\times \frac{R}{150\,\mathrm{pc}}
\times \frac{N_{\mathrm{H}}}{10^{23}\,\mathrm{cm^{-2}}}      \nonumber \\
& \times & \left(\frac{v}{500\,\mathrm{km\,s^{-1}}}\right)^3, 
\label{dotmphys}
\end{eqnarray}
in $\mathrm{M_{\odot}/yr}$, dyn, and erg/s, respectively. 
The observational requirement for high column densities,
$N_{\mathrm{OH}}\gtrsim2\times10^{17}$ \cmd, comes from the absorption in the
high OH84 and OH65 doublets. Very high values of $\dot{M}_{\mathrm{out}}$
are also associated with moderate expansion
velocities of the CORE components, although they are uncertain due
to the dependence on the adopted redshift. The components showing the
highest momentum fluxes ($\dot{P}_{\mathrm{out}}>10^{36}$ dyn, blue
symbols in Fig.~\ref{mdot}) are found in IRAS~03158+4227, IRAS~08572+3915,
Mrk~231, and IRAS~20100$-$4156, and are associated with either high columns 
or extreme velocities ($>1000$ \kms). In spite of their higher radii and
overall high outflow velocities, most extended components have 
moderate $\dot{M}_{\mathrm{out}}<300$ $\mathrm{M_{\odot}/yr}$, because
$N_{\mathrm{OH}}\lesssim5\times10^{16}$ 
\cmd\ are enough to account for the ground-state OH doublets. 

The total $\dot{M}_{\mathrm{tot}}$ estimates (Fig.~\ref{ener}) show hints of a
positive correlation with $L_{\mathrm{IR}}$. Most sources with 
$L_{\mathrm{IR}}\lesssim2\times10^{12}$ \Lsun\ have
  $\dot{M}_{\mathrm{tot}}\sim200$ \Msun/yr, with $\dot{M}_{\mathrm{tot}}$
  increasing sharply for $L_{\mathrm{IR}}>2\times10^{12}$ \Lsun. 
We note that 
IRAS~08572+3915 is not necessarily an outlier, because its luminosity may be
significantly higher than the apparent one according to the model by
\cite{efs14}. $\dot{M}_{\mathrm{tot}}$ appears also to be weakly correlated
with $L_{\mathrm{AGN}}$ and $L_{\mathrm{*}}$, the latter indicating loading
factors of $1-10$. 

With the exception of IRAS~14348$-$1447, the low-velocity components 
transport relatively low momentum in comparison with the high-velocity
  components. 
{\it Starburst99} models \citep{lei99} predict that a 40 Myr old continuous
  starburst can supply a maximum momentum \citep[see also][]{vei05,hec15} of
  $\dot{P}_*\sim3.5L_*/c$ (including radiation pressure at a level of
  $L_*/c$), which is marked as a shaded rectangle in Fig.~\ref{ener}. 
Our inferred total momentum flux measurements are
    $\dot{P}_{\mathrm{tot}}\gtrsim3L_*/c$ (Fig.~\ref{ener}), 
indicating that starbursts alone are most likely unable to drive 
the outflows via momentum-conserving winds, but may play an important
  role in some sources. The momentum supplied by the AGN 
may approach $\dot{P}_{\mathrm{AGN}}\sim2L_{\mathrm{AGN}}/c$ (treating both
radiation pressure on dust grains and AGN inner winds as
$L_{\mathrm{AGN}}/c$ each), also below the inferred total values. 
  Nevertheless, Fig.~\ref{pdotglb}abc shows the absolute values of
  $\dot{P}_{\mathrm{tot}}$ as a function of $\dot{P}_*$, 
  $\dot{P}_{\mathrm{AGN}}$, and $\dot{P}_*+\dot{P}_{\mathrm{AGN}}$, indicating
a remarkable similarity between $\dot{P}_{\mathrm{tot}}$ and the combined 
momentum rates supplied by the starburst and the AGN in most galaxies.
The outflows in IRAS~03158+4227, IRAS~08572+3915, Mrk~231, and
IRAS~20100$-$4156 (Fig.~\ref{pdotglb}) seem to
require significant momentum boosts.
These rough estimates are refined in \S\ref{dynmod}, where 
consideration of the covering factors, the extended nature of the starburst,
the potential well, and the momentum boost due to radiation trapping is
made, and in which we show that momentum-driven winds may indeed be
responsible for some outflow components with the combined effect of the AGN
(in most cases dominant) and the starburst.

It is estimated that supernovae and stellar winds can provide a mechanical
luminosity of up to $\sim1.8$\% of the starburst luminosity
\citep{lei99,vei05,har14}, and only a fraction of the mechanical power will go
into bulk motion of the ISM \citep[1/4 according to][]{wea77}.
Energy-conserving winds from the starburst are then probably unable to
drive the molecular outflows in Mrk~231, IRAS~08572+3915, 
IRAS~03158+4227, and IRAS~20100$-$4156, where mechanical
luminosities of $>1$\% of $L_*$ are inferred (Fig.~\ref{ener}). For the rest
of the sources, an expanding energy-conserving bubble generated by
supernovae could in principle account for the observed outflows, with
coupling efficiencies of $\sim10-50$\%. We also refine these estimates in
\S\ref{econs}, where we find that only some low-velocity components can
be driven in this way. Energy-conserving bubbles generated by AGN winds are
believed to be able to supply a power of up to $\sim5$\% of
$L_{\mathrm{AGN}}$ \citep[e.g.,][]{kin15}, from which a fraction of $\sim1/2$
may go into bulk motion of the ISM \citep{fau12}. 
Fig.~\ref{ener} shows that the 
energy flux is $\lesssim5$\% for all sources, including the strong
  AGNs Mrk~231, IRAS~08572+3915, IRAS~03158+4227, and IRAS~20100$-$4156, in
general agreement with theoretical predictions. We show in \S\ref{econs} that
{\it partially} energy-conserving flows driven by the AGN are generally
required for the high-velocity components. 

 The most powerful outflows (Fig.~\ref{pdotglb}def) are found in
   IRAS~03158+4227 (belonging to a 
  widely separated pair), IRAS~08572+3915 and IRAS~20100$-$4156 (mergers with
  projected separation of $\approx5.5$ kpc), and Mrk~231 (post-merger), 
  apparently unrelated with merging stage. They are mostly associated with
  high AGN luminosities (Fig.~\ref{pdotglb}e). They are best identified by
  high-velocity absorption in the high-lying OH65 doublet, and represent
  $\sim20$\% of all local ULIRGs observed in OH65. This suggests that
  stochastic and episodic strong-AGN feedback events occur throughout the
    merger process.

\begin{deluxetable*}{lcccccccccc}
\tabletypesize{\scriptsize}
\tablecaption{Physical parameters of individual outflowing components}
\tablewidth{0pt}
\tablehead{
\colhead{Galaxy} & 
\colhead{T} &
\colhead{$V$} &
\colhead{$R_1$} &
\colhead{$R$} &
\colhead{$N_{\mathrm{H}}$} &
\colhead{$f_c$} &
\colhead{$M_{\mathrm{out}}$} &
\colhead{$\dot{M}_{\mathrm{out}}$} &
\colhead{$\dot{P}_{\mathrm{out}}$} &
\colhead{$\dot{E}_{\mathrm{out}}$} \\
\colhead{name} & 
\colhead{} & 
\colhead{(km/s)} & 
\colhead{(pc)} & 
\colhead{(pc)} & 
\colhead{($10^{21}$ \cmd)} & 
\colhead{} & 
\colhead{($10^6$ \Msun)} &
\colhead{(\Msun/yr)} & 
\colhead{($10^{34}$ dyn)}  &
\colhead{($10^{41}$ erg/s)}  \\
\colhead{(1)} & 
\colhead{(2)} & 
\colhead{(3)} & 
\colhead{(4)} &
\colhead{(5)} &
\colhead{(6)} &
\colhead{(7)} &
\colhead{(8)} &
\colhead{(9)} &
\colhead{(10)} &
\colhead{(11)} 
}
\startdata
 IRAS 03158+4227  & E &  800 & 160 & 270 & 120 & 0.36 &  420 & 1300 &  660 & 2700  \\
                  & E & 1200 & 260 & 400 &   8 & 0.41 &   75 &  230 &  170 & 1000  \\
 IRAS 05189-2524  & E &  200 &  77 & 170 & 120 & 0.21 &  100 &  120 &   16 &   16  \\
                  & E &  550 & 130 & 340 &  40 & 0.14 &   89 &  150 &   52 &  140  \\
 IRAS 08572+3915  & E &  500 &  62 & 110 & 160 & 0.33 &   88 &  410 &  130 &  330  \\
                  & E &  950 &  50 & 110 &  80 & 0.20 &   28 &  240 &  150 &  700  \\
 IRAS 10565+2448  & E &  250 & 340 & 480 &  10 & 0.51 &  170 &   88 &   14 &   17  \\
                  & E &  420 & 240 & 400 &  20 & 0.34 &  160 &  160 &   43 &   90  \\
 Mrk 231          & C &  100 & 100 & 160 & 500 & 0.39 &  730 &  460 &   29 &   14  \\
                  & E &  550 & 300 & 600 &  20 & 0.26 &  260 &  240 &   84 &  230  \\
                  & E &  700 &  74 & 170 & 120 & 0.18 &   91 &  380 &  170 &  590  \\
 IRAS 13120-5453  & E &  420 & 280 & 470 &  16 & 0.36 &  180 &  160 &   43 &   89  \\
 Mrk 273          & E &  300 &  67 & 130 &  80 & 0.28 &   51 &  120 &   23 &   35  \\
                  & E &  700 &  79 & 340 & 120 & 0.05 &  110 &  220 &   99 &  350  \\
 IRAS 14348-1447  & C &  100 & 190 & 240 & 500 & 0.63 & 2500 & 1100 &   69 &   34  \\
                  & E &  600 & 160 & 370 &  20 & 0.17 &   68 &  110 &   43 &  130  \\
                  & E &  300 & 180 & 340 &  80 & 0.27 &  350 &  310 &   60 &   90  \\
 IRAS 14378-3651  & E &  600 & 170 & 270 &  12 & 0.39 &   47 &  110 &   41 &  120  \\
                  & E &  250 & 110 & 240 &  40 & 0.22 &   71 &   76 &   12 &   15  \\
 IRAS 20100-4156  & E &  500 & 170 & 300 & 120 & 0.35 &  510 &  880 &  280 &  690  \\
                  & E & 1100 & 190 & 300 &  12 & 0.40 &   58 &  220 &  150 &  850  \\
 IRAS 20551-4250  & E &  300 &  86 & 160 &  80 & 0.27 &   83 &  160 &   30 &   44  \\
                  & E &  600 &  67 & 190 &  20 & 0.13 &   13 &   41 &   16 &   47  \\
 IRAS 23365+3604  & C &  100 &  70 & 100 & 500 & 0.46 &  340 &  340 &   22 &   11  \\
                  & E &  350 &  73 & 190 &  60 & 0.14 &   45 &   82 &   18 &   32  \\
                  & E &  470 & 260 & 450 &  20 & 0.33 &  190 &  200 &   59 &  140  
\enddata
\label{tbl-2}
\tablecomments{(1) Galaxy name; (2) Type of model; C: CORE, E: ENVELOPE;
  (3) Average velocity; (4) Minimum radius; (5) Adopted radius, after
  correcting for the covering factor $f_c$; (6) Hydrogen column density, adopting an
  abundance $X(\mathrm{OH})=2.5\times10^{-6}$ relative to H nuclei; (7)
  Covering factor of the far-IR continuum by the outflowing OH, as defined in
  \S\ref{energetics}; (8) Outflowing gas mass; (9) Mass outflow rate;  
  (10) Momentum flux; (11) Energy flux. }
\end{deluxetable*}

\begin{deluxetable*}{lccccccc}
\tabletypesize{\scriptsize}
\tablecaption{Total outflow energetics inferred from models of OH}
\tablewidth{0pt}
\tablehead{
\colhead{Galaxy} & 
\colhead{$M_{\mathrm{tot}}$ (HV)} &
\colhead{$M_{\mathrm{tot}}$ (T)} &
\colhead{$\dot{M}_{\mathrm{tot}}$ (HV)} &
\colhead{$\dot{M}_{\mathrm{tot}}$ (T)} &
\colhead{$\dot{P}_{\mathrm{tot}}$ (HV)} &
\colhead{$\dot{P}_{\mathrm{tot}}$ (T)} &
\colhead{$\dot{E}_{\mathrm{tot}}$ (T)} \\
\colhead{name} & 
\colhead{($10^7$ \Msun)} &
\colhead{($10^7$ \Msun)} & 
\colhead{(\Msun/yr)} & 
\colhead{(\Msun/yr)} &
\colhead{($10^{34}$ dyn)}  &
\colhead{($10^{34}$ dyn)}  &
\colhead{($10^{42}$ erg/s)}  \\
\colhead{(1)} & 
\colhead{(2)} & 
\colhead{(3)} & 
\colhead{(4)} &
\colhead{(5)} &
\colhead{(6)} &
\colhead{(7)} &
\colhead{(8)} 
}
\startdata
 IRAS 03158+4227  & $  50_{- 24}^{+  9}$ & $  50_{- 24}^{+120}$ & $1500_{-280}^{+340}$ & $1500_{-280}^{+830}$ & $ 840_{-140}^{+290}$ & $ 840_{-140}^{+290}$ & $ 370_{- 81}^{+170}$ \\
 IRAS 05189-2524  & $  19_{- 11}^{+ 27}$ &  & $ 270_{-130}^{+ 22}$ &  & $  67_{- 29}^{+ 13}$ &  & $  16_{-  7}^{+  4}$ \\
 IRAS 08572+3915  & $  12_{-  8}^{+  5}$ & $  12_{-  7}^{+ 33}$ & $ 650_{-340}^{+230}$ & $ 650_{-320}^{+390}$ & $ 280_{-110}^{+170}$ & $ 280_{- 99}^{+170}$ & $ 100_{- 33}^{+100}$ \\
 IRAS 10565+2448  & $  32_{- 10}^{+ 13}$ & $  32_{- 10}^{+ 28}$ & $ 250_{- 81}^{+ 25}$ & $ 250_{- 81}^{+150}$ & $  57_{- 21}^{+ 13}$ & $  57_{- 20}^{+ 13}$ & $  11_{-  4}^{+  4}$ \\
 Mrk 231          & $  35_{- 14}^{+  1}$ & $ 110_{-  8}^{+ 20}$ & $ 620_{-260}^{+ 25}$ & $1100_{-150}^{+ 41}$ & $ 250_{-110}^{+ 36}$ & $ 280_{- 97}^{+ 38}$ & $  83_{- 26}^{+ 23}$ \\
 IRAS 13120-5453  & $  18_{-  4}^{+  4}$ &  & $ 160_{- 33}^{+ 34}$ &  & $  43_{-  9}^{+  9}$ &  & $   9_{-  2}^{+  2}$ \\
 Mrk 273          & $  16_{-  8}^{+  2}$ & $  16_{-  8}^{+ 60}$ & $ 350_{-150}^{+ 35}$ & $ 350_{-130}^{+490}$ & $ 120_{- 63}^{+ 10}$ & $ 120_{- 63}^{+ 44}$ & $  38_{- 26}^{+ 10}$ \\
 IRAS 14348-1447  & $  42_{- 29}^{+ 71}$ & $ 290_{- 35}^{+ 97}$ & $ 430_{-280}^{+180}$ & $1500_{-250}^{+270}$ & $ 100_{- 63}^{+ 97}$ & $ 170_{- 48}^{+ 97}$ & $  25_{- 12}^{+ 32}$ \\
 IRAS 14378-3651  & $  12_{-  1}^{+ 13}$ &  & $ 180_{- 33}^{+180}$ &  & $  53_{- 14}^{+ 37}$ &  & $  14_{-  5}^{+  6}$ \\
 IRAS 20100-4156  & $  56_{- 41}^{+ 22}$ &  & $1100_{-620}^{+830}$ &  & $ 430_{-250}^{+620}$ &  & $ 150_{- 80}^{+320}$ \\
 IRAS 20551-4250  & $  10_{-  5}^{+ 11}$ &  & $ 200_{-100}^{+210}$ &  & $  45_{- 21}^{+ 81}$ &  & $   9_{-  4}^{+ 29}$ \\
 IRAS 23365+3604  & $  24_{- 13}^{+ 26}$ & $  58_{- 47}^{+ 73}$ & $ 280_{-160}^{+260}$ & $ 620_{-430}^{+490}$ & $  77_{- 43}^{+ 90}$ & $  99_{- 48}^{+110}$ & $  18_{-  8}^{+ 45}$ 
\enddata
\label{tbl-3}
\tablecomments{(1) Galaxy name; 
  (2)-(3) Outflow gas mass;  
  (4)-(5) Mass outflow rate; 
  (6)-(7) Momentum flux;
  (8) Energy flux. HV refers to the values calculated for only the
  high-velocity ($>200$ \kms) component(s), and the (T) columns include
    the values of the low-velocity
  components (only tabulated when the value or error is different from
  that of the corresponding HV column). The low-velocity components
  contribution to the the energy flux is negligible, and only the total value
  is given.} 
\end{deluxetable*}

\begin{figure*}
\begin{center}
\includegraphics[angle=0,scale=.58]{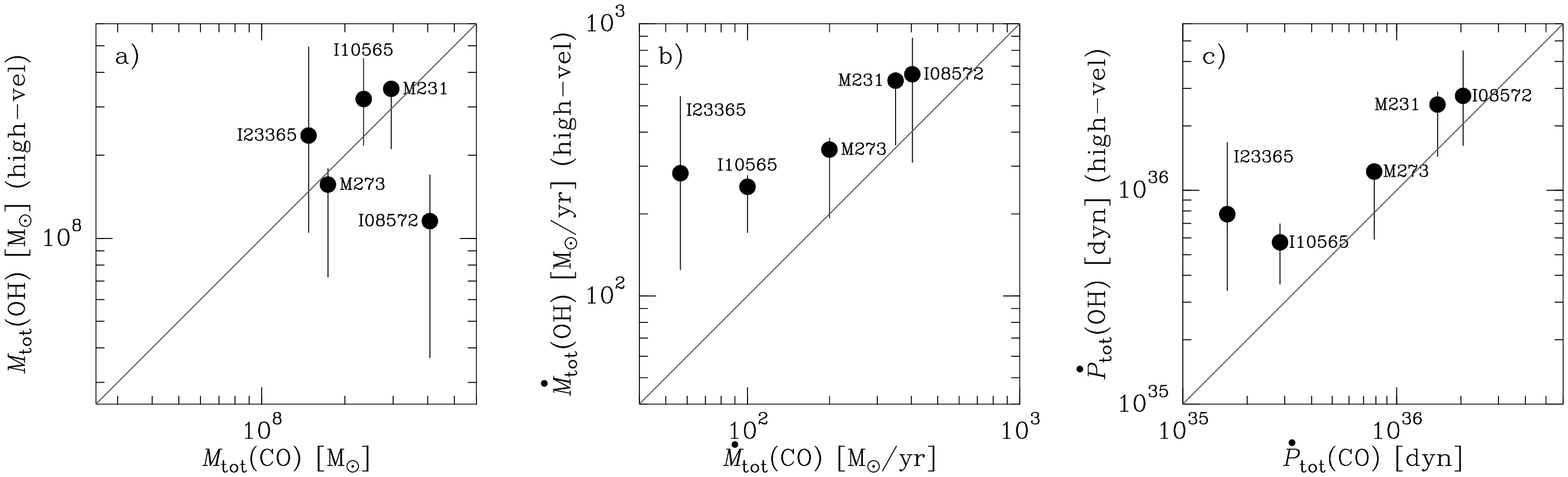}
\end{center}
\caption{Comparison between the total properties of the outflows,
  $M_{\mathrm{tot}}$, $\dot{M}_{\mathrm{tot}}$, and $\dot{P}_{\mathrm{tot}}$,
  derived from OH and from CO (1-0) \citep{cic14} for the sources that have
  been analyzed in both species. For OH, only the high-velocity ($>200$ \kms)
  components are considered. The values of $\dot{M}_{\mathrm{tot}}$ and
  $\dot{P}_{\mathrm{tot}}$ in \cite{cic14} have been divided by 3 to match our
  definitions in eqs.~(\ref{mdot_out}) and (\ref{pdot_out}).
}   
\label{ohcomod}
\end{figure*}

\begin{figure*}
\begin{center}
\includegraphics[angle=0,scale=.61]{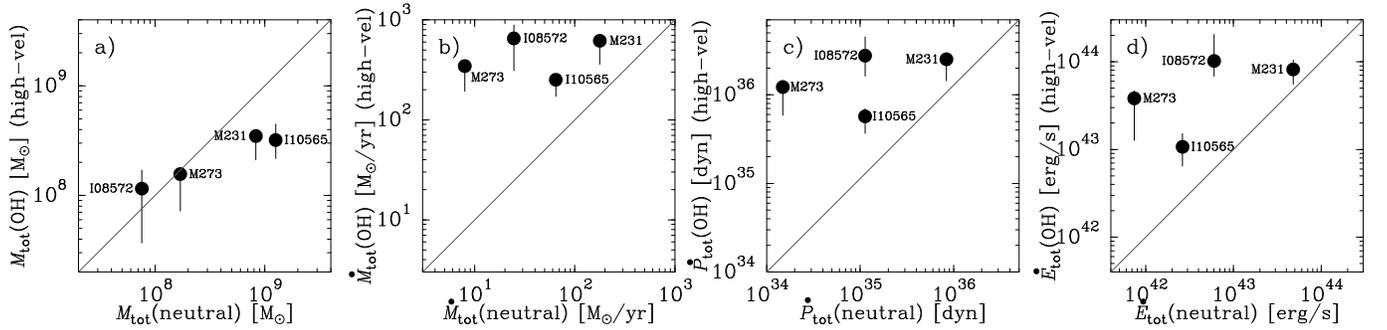}
\end{center}
\caption{Comparison between the properties of the outflows,
  $M_{\mathrm{tot}}$, $\dot{M}_{\mathrm{tot}}$, $\dot{P}_{\mathrm{tot}}$, and
  $\dot{E}_{\mathrm{tot}}$, derived from OH and from Na  {\sc i} D
  \citep{rup13a} for the sources that 
  have been analyzed in both species. For OH, only the high-velocity ($>200$
  \kms) components are considered. 
}   
\label{ohnamod}
\end{figure*}

\subsection{Comparison with other tracers} \label{mdotohco}

\begin{figure}
\begin{center}
\includegraphics[angle=0,scale=.5]{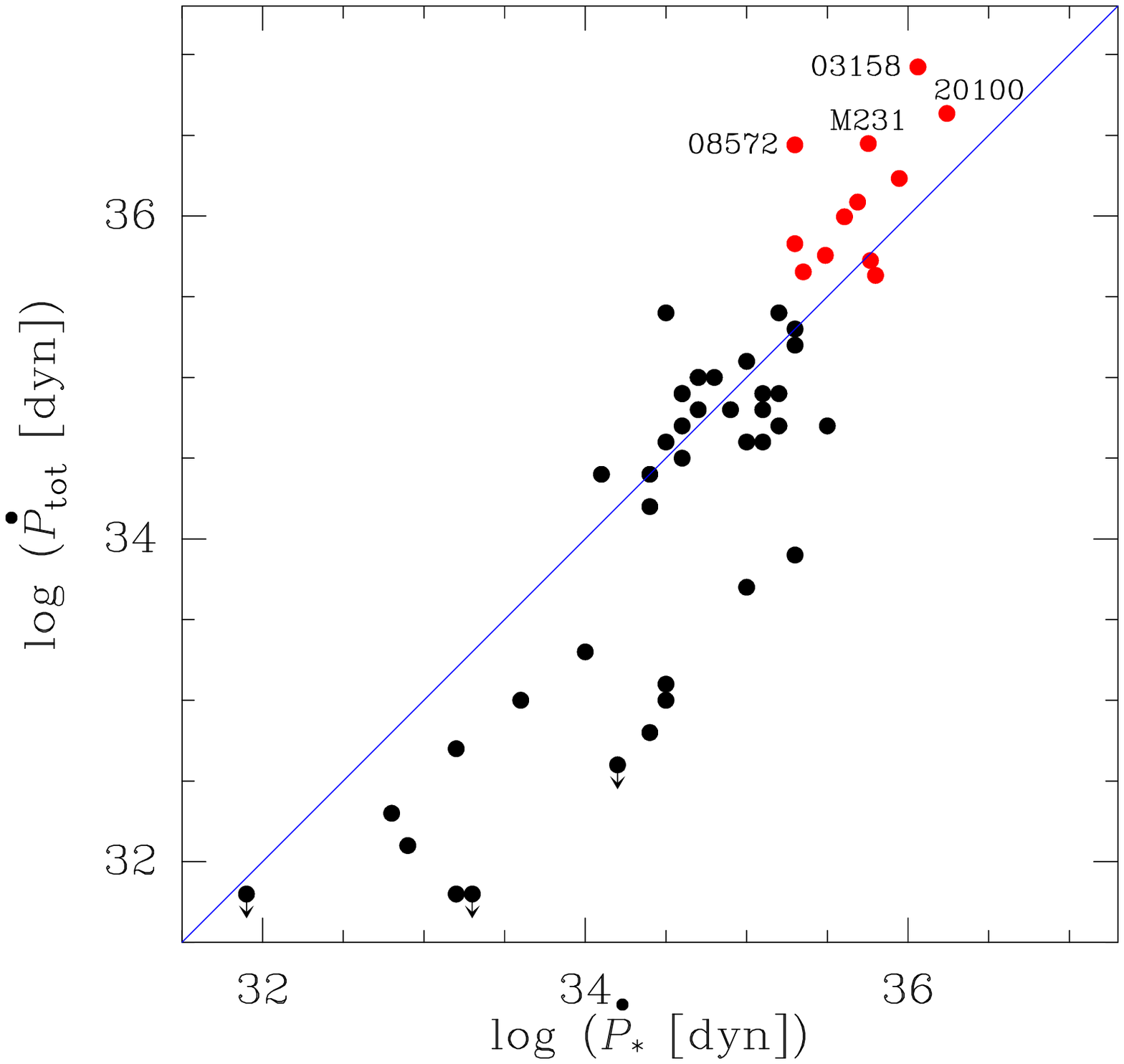}
\end{center}
\caption{The total momentum fluxes ($\dot{P}_{\mathrm{tot}}$) of ULIRGs (red
  symbols, inferred from OH) and of starburst galaxies \citep[black symbols,
    inferred from ionized UV lines in][]{hec15}, are plotted as a function of
  the momentum rate supplied by the starburst 
  ($\dot{P}_{\mathrm{*}}=3.5\times L_*/c$). The blue line indicates equal
  values in both axis. 
}   
\label{pdotglbstar}
\end{figure}

Our calibration of the energetics from OH observations is based on
$(i)$ the reliability of the present models, and in particular of the inferred
outflow sizes that are not directly observed but inferred from the OH
excitation and the observed and predicted continuum; $(ii)$ our adopted OH
abundance relative to H nuclei, $X_{\mathrm{OH}}=2.5\times10^{-6}$ (GA14,
GA15), which in turn is based on both observations and chemical models. These
assumptions are worth checking by comparing our values of $M_{\mathrm{tot}}$,
$\dot{M}_{\mathrm{tot}}$, and $\dot{P}_{\mathrm{tot}}$ (eqs.~\ref{par_glob})
with the values inferred from totally  
independent measurements of CO for the 5 sources where both OH and CO have
been observed and analyzed: Mrk~231, IRAS~08572+3915, Mrk~273,
IRAS~10565+2448, and IRAS~23365+3604 \citep{cic12,cic14}. Since CO is not
sensitive to low-velocity outflows due to contamination by the core of the
line, the values we include here from OH exclude the low-velocity ($<200$
\kms) components. On the other hand, we divide the values of
$\dot{M}_{\mathrm{OF}}$ and $\dot{P}_{\mathrm{OF}}$ in \cite{cic12,cic14} by
a factor of 3 to match our definitions given in eqs.~(\ref{mdot_out}) and
(\ref{pdot_out}). 


The outflowing masses, $M_{\mathrm{out}}$, are compared in
Fig.~\ref{ohcomod}a. The agreement is within a factor of 2 in all
sources except IRAS~08572+3915, where $M_{\mathrm{out}}$(OH) is a factor of
$\sim4$ lower than $M_{\mathrm{out}}$(CO). Since
$M_{\mathrm{out}}(\mathrm{OH})\propto R^2$ but 
$M_{\mathrm{out}}(\mathrm{CO})$ is independent of $R$, the discrepancy is a
consequence of the very compact outflow ($\sim100$\,pc) inferred for OH:
both components (moderate and high velocities) are very excited, with OH84 as
strong as OH119, as well as absorption in OH65
(Fig.~\ref{ohblue}). Since the observed absorption is dominated by the high
amounts of outflowing nuclear OH, and the source of far-IR absorption is
expected to be compact, OH is not sensitive to the more extended outflowing
gas in this source. On the redshifted side of the OH spectrum, in addition,
IRAS~08572+3915 shows an emission feature in OH84 
(and marginally also in OH65) at positive velocities
that is not fitted (see Fig.~\ref{ohspec}) 
and does not appear to belong
to the same components we model in Fig.~\ref{fitsoh84}. Nevertheless, the
values of $\dot{M}_{\mathrm{tot}}$ and $\dot{P}_{\mathrm{tot}}$ inferred from
both species (Fig.~\ref{ohcomod}bc) are similar in this source, because
the flow time scale, $t_{\mathrm{flow}}\sim R/v$, is correspondingly higher
in CO.  

In all other sample sources, $\dot{M}_{\mathrm{tot}}$ and
$\dot{P}_{\mathrm{tot}}$ are somewhat higher for OH than for CO. The largest
discrepancies are found for IRAS~10565+2448 and IRAS~23365+3604, where we
infer higher values by factors of $\sim2$ and $\sim4$ for OH,
respectively. These discrepancies are a consequence of the extended emission
estimated in CO, $R>1$ kpc \citep{cic14}, while we infer $R\sim0.4-0.5$ kpc
and $R\sim0.2-0.45$ kpc for OH in IRAS~10565+2448 and IRAS~23365+3604,
respectively. High-resolution CO observations in these sources will check
whether there is a decrease in $\dot{M}$ with increasing distance to the
center, as found in Mrk~231 \citep[GA14,][]{fer15}. Nevertheless, the
difference in the OH and CO spatial distributions may be real, as the maximum
velocity observed in CO is lower than in OH in both sources (Fig.~\ref{ohco}).

It is also interesting to compare the energetics inferred from OH with those
derived from Na {\sc i} D \citep{rup13a} for the four sources that 
have been analyzed in both species. \cite{rup13a} used the same expressions
for the mass, momentum, and energy flux as we use in this study
(eqs.~\ref{mof}, \ref{mdot_out}, \ref{pdot_out}, \ref{edot_out}).
The comparisons are shown in Fig.~\ref{ohnamod}, where the comparable values
of the outflow masses inferred from the two species is remarkable. This is 
surprising because $M_{\mathrm{out}}\propto R^2 N_{\mathrm{H}}$; while our
inferred values of $R$ are $<1$ kpc, the radii observed by \cite{rup13a}
are $2-5$ kpc, but the much higher columns inferred from OH compensate for
this, and thus result in similar total masses. The mass, momentum, and energy
fluxes are then much higher for the molecular gas. 

{\it The similar outflow masses found for OH and for the more spatially
  extended CO and Na {\sc i} D suggests that a significant fraction of the
total outflow mass is loaded at small radii (a few $\times100$ pc), and
the outflows are less efficient in evacuating the gas initially located in the
extended kpc-scale disks}.

The total momentum fluxes ($\dot{P}_{\mathrm{tot}}$) inferred for ULIRGs are 
compared in Fig.~\ref{pdotglbstar}
with the analogous quantities inferred by \cite{hec15} from observations of UV
ionized lines in a sample of starburst galaxies. Most of the composite ULIRGs 
lie close to the line $\dot{P}_{\mathrm{tot}}=\dot{P}_{\mathrm{*}}$ that is also 
a good fit for a number of pure starbursts, but usually lie above it 
probably due to the AGN contribution. The comparison appears to
indicate that nuclear starbursts also contribute to the molecular feedback
in most ULIRGs, but are not enough to drive the observed
outflows alone in most sources. As a corollary, if our estimate of 
the relative contributions of the AGN and the starburst is sufficiently
accurate, it appears that feedback
from various combinations of the AGN and the nuclear starburst regulate the
star formation and the growth of the SMBH in gas-rich mergers. 
However, the 4 outlier ULIRGs labeled in Fig.~\ref{pdotglbstar}
appear to require very strong AGN contribution, likely in form of
energy-conserving winds.  
Comparison of the inferred molecular outflow properties with those of  
the ionized outflow phase in starburst galaxies
\citep{hec15} is further explored in Appendix~\ref{heck15}.

Overall, the comparison of the energetics inferred from OH and CO, especially
the mass outflow rate and the momentum flux, as well as the comparison
  between the observed SEDs and the model predictions in Fig.~\ref{fitscont},
  are satisfactory. We now push onward to the analysis of the molecular
outflows by considering the individual components of our fits in the next
section, as different components in a given source have different associated
energetics.

\section{Discussion} \label{discussion}

In our initial analysis of the momentum deposition rates supplied by the
AGN and the starburst in \S\ref{modener} (Fig.~\ref{pdotglb}), we ignore the
clumpiness of the molecular outflows, the momentum boost due to
trapping of infrared radiation, and the finite extension of the starbursts, as
well as the gravitational potential well in the central regions of the
galaxies. We attempt to overcome these limitations with a simple model that
accounts for these aspects, in the framework of momentum-driven and
energy-driven outflows.

\subsection{Momentum-driven outflows}
\label{dynmod}

We attempt to evaluate the impact of radiation pressure on dust grains
from both the AGN and the compact, but not point-like, central starburst,
as well as the role of momentum-deposition by the AGN winds, stellar
winds, and supernovae in launching the molecular outflows, using a simple
analytical model of momentum-driven flows based on recent work in the
literature \citep{mur05,tho15,ish15,hec15}. 
As pointed out by \cite{ste16}, the 
observed momentum fluxes are the result of the time-averaged forces acting on
the outflowing gas. It is therefore necessary to integrate the equation of
motion beginning from the time that the outflowing shell is close to
the energy source (and presumably covering a large fraction of the
$4\pi$\,sr), taking into account the change of the covering factor with
increasing radius. The dynamics of the gas are given by 
\begin{equation}
\frac{dP_{\mathrm{mod}}}{dt}= -\frac{M_{\mathrm{out}}v_{\mathrm{cir}}^2}{r}+
\dot{P}_*(r) + \dot{P}_{\mathrm{AGN}}(r),
\label{dyn}
\end{equation}
where $P_{\mathrm{mod}}$ is the (modeled) momentum of the outflow.
$v_{\mathrm{cir}}$ is the ``circular'' 
velocity characterizing the potential well of an isothermal
sphere \citep{hec15}. We follow the approach of \cite{mur05}
and use $v_{\mathrm{cir}}^2=2\sigma^2$ for the sources where $\sigma$, the
stellar velocity dispersion tabulated in \cite{das06} and \cite{gen01}, 
is available. Mrk~231 has an unusually low value of
$\sigma$ \citep{tac02} and we adopt $v_{\mathrm{cir}}=240$ 
\kms\ to account for the dynamical mass of $6.7\times10^9$ \Msun\ within
$r=500$\,pc \citep{dav04}.  
For the rest of the sources we adopt $v_{\mathrm{cir}}=205$, $150$, $220$,
  and $260$  \kms\ in IRAS~08572+3915, IRAS~10565+2448,
IRAS~03158+4227, and IRAS~20100$-$4156, respectively, based on our fits to 
the OH65 doublet, and $v_{\mathrm{cir}}=226$ \kms\ in
 IRAS~13120$-$5453 based on recent molecular observations with ALMA
 \citep{pri16}.  
Our $v_{\mathrm{cir}}$ values are not corrected for
rotation and flattening and are thus lower limits; nevertheless, the
high $M_{\mathrm{out}}$ values in the nuclear regions of ULIRGs
(Table~\ref{tbl-2}) suggest gas that is emanating from the rotating
structures where the bulk of the nuclear gas mass is stored, with
significant rotational support.

$\dot{P}_*(r)$ and $\dot{P}_{\mathrm{AGN}}(r)$ are the rates of momentum
deposition onto the ISM at radial position $r$ due to the starburst and the
AGN, respectively. For the starburst,
\begin{equation}
\dot{P}_*(r)=(1+\tau_{\mathrm{IR}}+\tau_{\mathrm{*,w}})f_c(r) 
\frac{L_{*,\mathrm{encl}}(r)}{c},
\label{pdotstar}
\end{equation}
where $\tau_{\mathrm{IR}}$ is the infrared opacity of the outflowing shell
and accounts for the momentum boost due to absorption of re-radiated IR
emission \citep{tho15}. Although $\tau_{\mathrm{IR}}$ is expected to vary with
$r$, we avoid overestimating its effect \citep[see][]{rot12}
by simply taking a characteristic constant value of 
$\mu m_H k_{\mathrm{IR}} N_{\mathrm{H}}$, where an absorption-mass coefficient
of $k_{\mathrm{IR}}=5$ cm$^2$/g of gas is adopted \citep{tho15,ish15}. 
$\tau_{\mathrm{*,w}}$ describes the contribution of stellar
winds and supernovae remnants, for which we adopt $\tau_{\mathrm{*,w}}=2.5$
based on Starburst99 \citep{lei99,vei05,hec15}.
$f_c(r)$ is the covering factor of the starburst feedback by the outflowing
OH, for which we adopt two approaches: $(i)$ a ``mixed'' shell-cloud
approach, in which a shell with $f_c=1$ is 
used for $r\leq R_1$ (see eq.~\ref{mshell}), while $f_c=(R_1/r)^2$ for 
$r>R_1$ due to shell breakup and geometrical dilution of the cloud
ensemble. We thus neglect here the expansion of clouds included in
\cite{tho15} because the high-density molecular tracers HCN and HCO$^+$ 
  are detected in the outflow of Mrk~231 at large distances from the center
\citep{aal12,aal15,lin16}. $(ii)$ An ``aligned momenta'' approach is also
  considered in which $f_c=1$ at all radii, reflecting the
  possibility that the AGN and starburst momenta are preferentially released
  in the same directions as the clumpy, outflowing OH.  
$L_{*,\mathrm{encl}}(r)$ is the starburst luminosity enclosed within radius
$r$. While in Mrk~231, \cite{dav04} fitted the spatial distribution of the
nuclear starburst intensity with an exponential $I(r)=I_0\,\exp\{-r/r_d\}$,
where $r_d=150-200$\,pc, we have approximated the corresponding enclosed
luminosity as a linear function of $r$,
\begin{equation}
L_{*,\mathrm{encl}}(r) = \frac{r}{3r_d} L_*, \,\, r\leq3\,r_d
\end{equation}
where $L_*$ is the total starburst luminosity, and applied this expression to
all galaxies with $r_d=150$\,pc. We thus assume a very compact starburst to
avoid underestimating its momentum deposition\footnote{The
  implied stellar luminosity to mass ratio within $3r_d$ is 
$L_{*,\mathrm{encl}}(r)/M_{*,\mathrm{encl}}(r)=G\,L_*/(3r_dv_{\mathrm{cir}}^2)$,
which gives $240$ \Lsun/\Msun\ for $L_{*}=10^{12}$ \Lsun\ and 
$v_{\mathrm{cir}}=200$ \kms.}. Since in eq.~(\ref{dyn}) we also neglect 
the inward force exerted by the stars formed outside $r$, the role of the
nuclear starburst in launching the outflows is most likely not underestimated.

The momentum deposition due to AGN feedback is described as
\begin{equation}
\dot{P}_{\mathrm{AGN}}(r)=(1+\tau_{\mathrm{IR}}+\tau_{\mathrm{AGN,w}})f_c(r) 
\frac{L_{\mathrm{AGN}}}{c},
\label{pdotagn}
\end{equation}
with $\tau_{\mathrm{AGN,w}}=1$ \citep[e.g.,][]{zub12}. We implicitely assume
that, after being scattered by electrons in a Compton-thick medium, most
photons peak in the UV and are absorbed by dust -thus contributing at least
twice (for $f_c=1$) to the momentum deposition onto the ISM.
Comparison of this simple momentum-driven model with observations enables us
to evaluate the role of energy-driven flows
\citep[e.g.,][]{buh12,zub12,fau12,tom15,fer15,ste16} in the OH outflows, which
is considered below in \S\ref{econs}.  

In the following we ignore different terms in eqs.~(\ref{pdotstar}) and
(\ref{pdotagn}) to isolate the relative roles of radiation pressure on dust
grains and winds. The equation of motion is then integrated from the
``launching'' radius ($R_0$, see below) to the ``observed'' radius ($R$,
Table~\ref{tbl-2}), by 
assuming a constant outflowing gas mass, so that the first member in
eq.~(\ref{dyn}) is $0.5 M_{\mathrm{out}}dv^2/dr$. 
By defining the modeled momentum flux in the same way as for the
observations, i.e. $\dot{P}_{\mathrm{mod}}=M_{\mathrm{out}}v^2/R$ 
(eq.~\ref{pdot_out}), the model predictions can be directly compared with our
values for the observed momentum rates inferred for the individual components,
$\dot{P}_{\mathrm{out}}$ (\S\ref{energetics}, Table~\ref{tbl-2}).  
We also rely on the AGN fractions
($\alpha_{\mathrm{AGN}}=L_{\mathrm{AGN}}/L_{\mathrm{bol}}$, Table~\ref{tbl-1})
given in \cite{vei13}, some of them being uncertain due to high extinction
even at far-IR wavelengths (GA15).

\subsubsection{Can radiation pressure on dust grains drive the outflows?}
\label{radpress}

\begin{figure*}
\begin{center}
\includegraphics[angle=0,scale=.6]{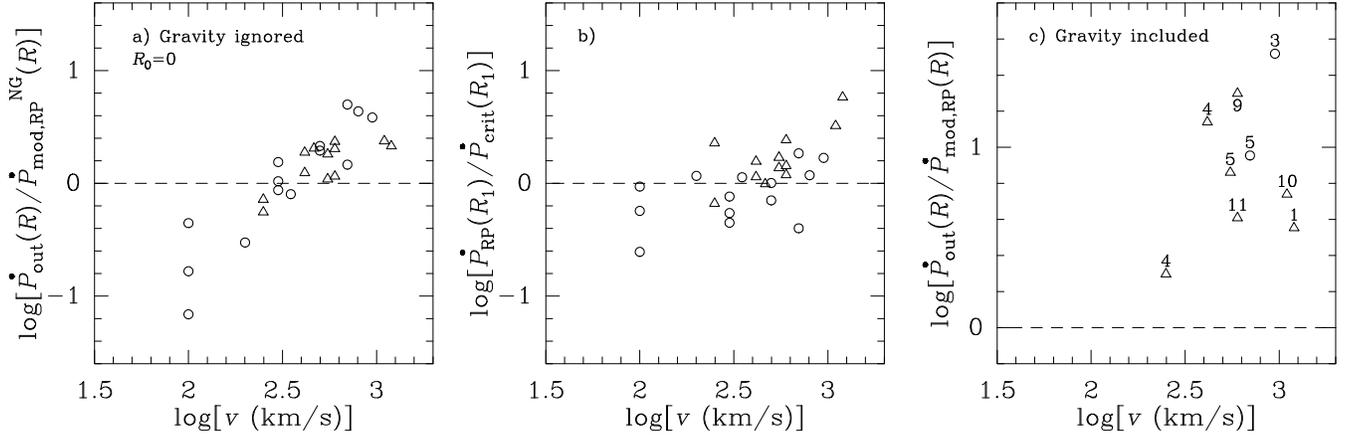}
\end{center}
\caption{Evaluation of the possibility that radiation pressure from the AGN
  and the starburst can drive the observed molecular outflows 
  with the shell-cloud approach 
  (\S\ref{radpress}). Circles and triangles indicate components with
  $\tau_{\mathrm{IR}}\ge0.5$ and $<0.5$, respectively, thus distinguishing
  components with significant momentum boost owing to radiation trapping.
  a) Ratio of the observed momentum 
  rates for the individual components, $\dot{P}_{\mathrm{out}}$, and the
  values predicted by eq.~(\ref{pdotrpng}) which ignores the effect of the
  potential well. b) Ratio of the rates of momentum deposition due to
  radiation pressure, $\dot{P}_{\mathrm{RP}}$, to the critical momentum
  flux $\dot{P}_{\mathrm{crit}}$ that needs to
  be overcome in order to launch the outflow. Both values are evaluated at
  position $R_1$ where the ratio is expected to be maximum. c) Ratio of the
  observed momentum rates to the values predicted by eq.~(\ref{pdotrpg})
  that include the gravitational term (only those components that satisfy
  $\dot{P}_{\mathrm{RP}}(R_1)>\dot{P}_{\mathrm{crit}}(R_1)$ have solutions).
  Galaxies are labeled as in Fig.~\ref{ener}.
}   
\label{testradp}
\end{figure*}

First we ignore the terms proportional to $\tau_{\mathrm{*,w}}$ and 
$\tau_{\mathrm{AGN,w}}$ in eqs.~(\ref{pdotstar}) and (\ref{pdotagn}),
thus taking into account only radiation pressure from both
the AGN and the starburst. By also neglecting the gravitational term
(superscript NG) and assuming that the outflow is launched at very small
radii (taking $R_0=0$), a strong upper limit is found for the momentum flux
that can be attained due to radiation pressure in the shell-cloud
  approach:  
\begin{eqnarray}
\dot{P}_{\mathrm{mod,RP}}^{\mathrm{NG}}(R) & = & 
\frac{R_1}{R} 
\dot{P}_*(R_1) 
\left[1+2\ln\left(\frac{R}{R_1}\right)\right] \nonumber \\
& + & 
\frac{2R_1}{R} 
 \dot{P}_{\mathrm{AGN}}(R_1) \left(2-\frac{R_1}{R}\right),
\label{pdotrpng}
\end{eqnarray}
where $\dot{P}_*(R_1)=(1+\tau_{\mathrm{IR}})L_*R_1/(3cr_d)$ and
$\dot{P}_{\mathrm{AGN}}(R_1)=(1+\tau_{\mathrm{IR}})L_{\mathrm{AGN}}/c$
only include the radiative terms. We compare in Fig.~\ref{testradp}a the
observed momentum fluxes $\dot{P}_{\mathrm{out}}$ with the strong  
upper limits of eq.~(\ref{pdotrpng}), showing 
that the observed values are above the limiting values for most
  high-velocity components. 

By now including the gravitational term, we obtain the critical
momentum flux
$\dot{P}_{\mathrm{crit}}(r)=M_{\mathrm{out}}v_{\mathrm{cir}}^2/r$ that must
be overcome at some position $r$ by $\dot{P}_*(r)+\dot{P}_{\mathrm{AGN}}(r)$
in order to allow the outflow launching to proceed
\citep{hec15,tho15}. In the shell-cloud approach, this condition should
be met at least at $r=R_1$, where 
$(\dot{P}_*(r)+\dot{P}_{\mathrm{AGN}}(r))/\dot{P}_{\mathrm{crit}}(r)$
is highest, and thus we compare in Fig.~\ref{testradp}b
$\dot{P}_{\mathrm{RP}}(R_1)\equiv\dot{P}_*(R_1)+\dot{P}_{\mathrm{AGN}}(R_1)$ and
$\dot{P}_{\mathrm{crit}}(R_1)$. Interestingly,
both values are found to be similar in most sources within the uncertainties,
indicating that, together with rotation, radiation pressure (dominated
by the AGN in most cases) is able to support the structures
against gravity particularly in the direction of the rotation axis
\citep[see also][and GA15]{mur05}. This support is further 
explored in Appendix~\ref{radp}.

For the model components that satisfy
$\dot{P}_{\mathrm{RP}}(R_1)>\dot{P}_{\mathrm{crit}}(R_1)$,
we calculate the expected momentum flux at the observed position $R$:
\begin{eqnarray}
\dot{P}_{\mathrm{mod,RP}}(R) & = & 
-\frac{2R_1}{R} \dot{P}_{\mathrm{crit}}(R_1) \ln\frac{R}{R_0}
\nonumber \\
& + &
\frac{R_1}{R} \dot{P}_*(R_1) 
\left[1-\left(\frac{R_0}{R_1}\right)^2+2\ln\frac{R}{R_1}\right] 
\nonumber \\
& + &
\frac{2R_1}{R} \dot{P}_{\mathrm{AGN}}(R_1)
\left[2-\frac{R_0}{R_1}-\frac{R_1}{R}\right] 
\label{pdotrpg}
\end{eqnarray}
where $R_0$ is the launching radius
($\dot{P}_{\mathrm{RP}}(R_0)=\dot{P}_{\mathrm{crit}}(R_0)$). 
The resulting values are compared with the observed momentum rates in 
Fig.~\ref{testradp}c, showing that no outflowing component 
can be explained with this model. 
We conclude that radiation pressure alone can provide (partial) support
against gravitation in the vertical direction \citep[see also][]{tho05}, but
can hardly drive the observed outflows. 

\subsubsection{Inclusion of winds}
\label{winds}

\begin{figure*}
\begin{center}
\includegraphics[angle=0,scale=.6]{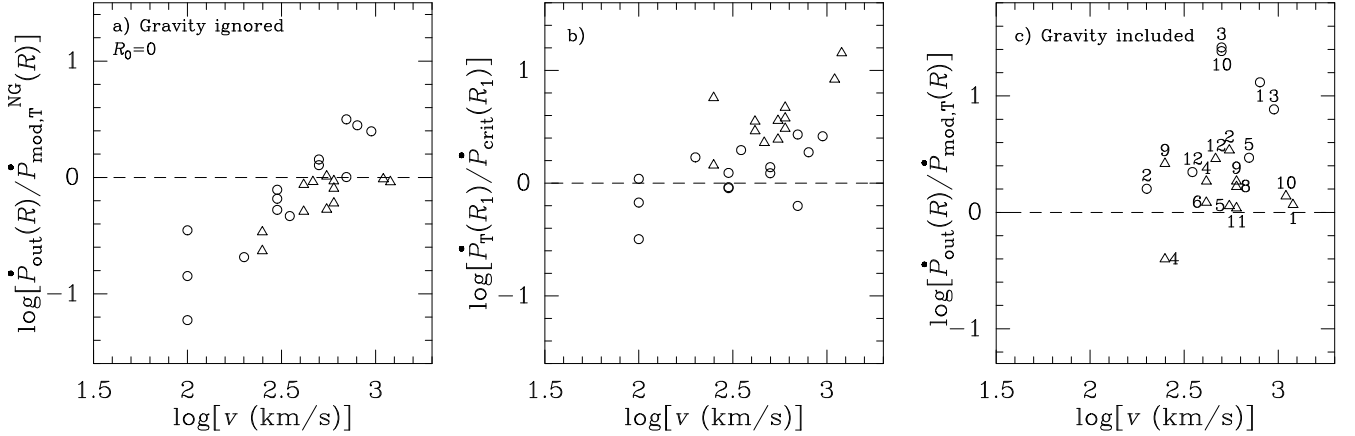}
\end{center}
\caption{Evaluation of the possibility that momentum-driven winds and
  radiation pressure from the AGN and the starburst can drive the observed
  molecular outflows with the shell-cloud approach
  (\S\ref{winds}). Circles and triangles indicate 
  components with $\tau_{\mathrm{IR}}\ge0.5$ and $<0.5$, respectively, thus
  indicating the components with significant momentum boost owing to radiation
  trapping. 
  a) Ratio of the observed momentum
  rates for the individual components, $\dot{P}_{\mathrm{out}}$, and the
  predicted values when the potential well is ignored. b)
  Ratio of the rates of momentum deposition due to radiation 
  pressure and winds, $\dot{P}_{\mathrm{T}}$, and the critical
  momentum flux $\dot{P}_{\mathrm{crit}}$ that must be overcome in order to
  launch the outflow. Both values are evaluated at 
  position $R_1$ where the ratio is expected to be maximum. c) Ratio of the
  observed momentum rates and the values predicted by 
  including the gravitational term (only those components that satisfy
  $\dot{P}_{\mathrm{T}}(R_1)>\dot{P}_{\mathrm{crit}}(R_1)$ have solutions).
Galaxies are labeled as in Fig.~\ref{ener}.
}   
\label{testradpw}
\end{figure*}

We now include the terms proportional to $\tau_{\mathrm{*,w}}$ and 
$\tau_{\mathrm{AGN,w}}$ in eqs.~(\ref{pdotstar}) and (\ref{pdotagn}), and
follow the same three steps as in \S\ref{radpress}. 
In the shell-cloud approach, the relevant equations
have the same forms as eqs.~(\ref{pdotrpng}) and
(\ref{pdotrpg}). Ignoring the gravitational term, Fig.~\ref{testradpw}a shows 
that the observed momentum fluxes are mostly lower than or similar to the
predictions.  This scenario may describe gas 
that is rotationally supported and radially pushed, flowing close to the
equatorial plane of the disk. Because of this extra support, molecular
outflows coplanar with the disk can be roughly
accounted for via momentum deposition by the 
AGN and the starburst, with no need for energy-conserving phases.

\begin{figure}
\begin{center}
\includegraphics[angle=0,scale=.9]{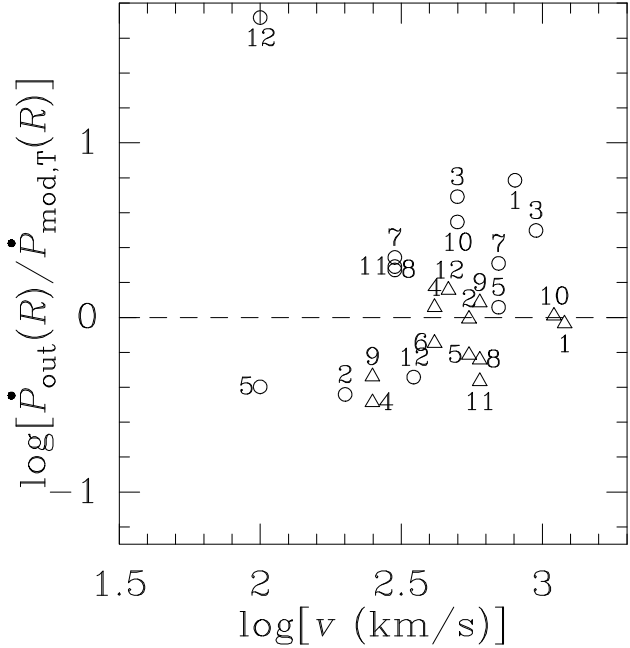}
\end{center}
\caption{Same as Fig.~\ref{testradpw}c but for the aligned-momenta
    approach in which $fc=1$ is assumed at all radii. The potential well is
    included in these calculations. Galaxies are labeled as
    in Fig.~\ref{ener}. 
}
\label{testradpwfc1}
\end{figure}

If rotational support is unimportant (e.g. if the outflow is driven in
  the polar direction), the total momentum deposited by the AGN
and the starburst,
$\dot{P}_{\mathrm{T}}(R_1)\equiv\dot{P}_*(R_1)+\dot{P}_{\mathrm{AGN}}(R_1)$,
must be higher than the critical value $\dot{P}_{\mathrm{crit}}(R_1)$ needed
to launch the outflows.  Fig.~\ref{testradpw}b compares
$\dot{P}_{\mathrm{T}}(R_1)$ to $\dot{P}_{\mathrm{crit}}(R_1)$, showing that
the potential well can be overcome for most outflow components once winds 
are included. The calculated momentum fluxes 
at the effective observed position $R$ are usually lower than
those inferred from observations (Fig.~\ref{testradpw}c), 
but at least one velocity component in 
IRAS~03158+4227, IRAS~05189$-$2524, IRAS~14348$-$1447,
IRAS~14378$-$3651, Mrk~231, IRAS~20100$-$4156, IRAS~20551$-$4250, 
IRAS~23365+3604, and possibly the two components in IRAS~10565+2448, 
appear to be outflows with little momentum boost 
$(\lesssim3)$.

The aligned-momenta approach, simulating the scenario in which
winds from the AGN and starburst are released in the same directions as the
molecular clumps, provides a much more efficient coupling between the released
and outflow momenta. 
As shown in Fig.~\ref{testradpwfc1}, most modeled and
observed momentum fluxes are similar in this scenario. 
The potential well is included in these calculations.
Among the high-velocity
components, only the components with high column densities in 
IRAS~03158+4227, IRAS~08572+3915, and IRAS~20100$-$4156 have momentum boosts
higher than 3. The momentum fluxes of the 100 km/s, very massive components in
IRAS~14348$-$1447 and IRAS~23365+3604 cannot be explained in this way; results
for these components are very sensitive to $v_{\mathrm{cir}}$ 
(see Figs.~\ref{testradp}a and \ref{testradpw}a).

\subsection{Energy-driven outflows}
\label{econs}

\begin{figure*}
\begin{center}
\includegraphics[angle=0,scale=.8]{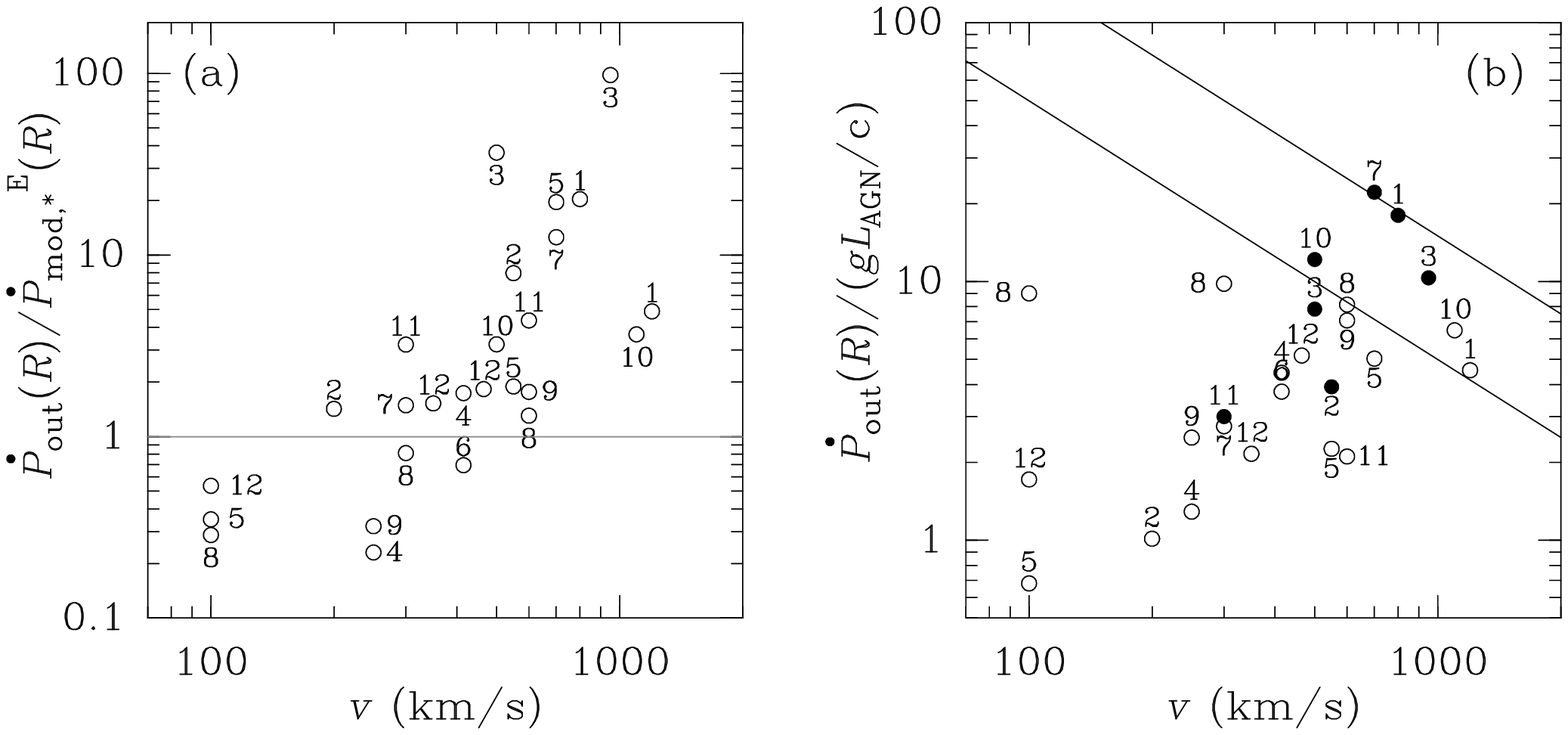}
\end{center}
\caption{Predictions for energy-driven outflows, which use a
    shell-cloud approach and assume that gravity and
  radiation pressure balance and thus are both ignored (\S\ref{econs}). a)
  Ratio of the observed momentum rates for the individual components,
  $\dot{P}_{\mathrm{out}}$, and the predicted maximum values for starburst
  winds (eq.~\ref{pdoteconsstar}). b) Ratio of the observed momentum rates and
  $gL_{\mathrm{AGN}}/c$, where $g$ is of order unity
  (eq.~\ref{pdoteconsagn}). In energy-driven flows, this ratio is expected
  to be $\beta_{\mathrm{out,AGN}} \, v_{\mathrm{in}}/v$, where
  $v_{\mathrm{in}}$ is the velocity of the inner AGN wind and
  $\beta_{\mathrm{out,AGN}}$ is the fraction of the injected energy that goes
  into bulk motion of the ISM gas. The two solid lines show the expected
  relationship for $v_{\mathrm{in}}=0.1c$ and $0.033c$, with
  $\beta_{\mathrm{out,AGN}}=1/2$. Filled symbols indicate components
    that, within a factor of 3, can neither be explained through
    momentum-driven by the starburst and AGN, or through energy-driven by the
    starburst, and thus require an enery-conserving phase driven by the AGN.
Galaxies are labeled as in Fig.~\ref{ener}.
}   
\label{pboost}
\end{figure*}

Both starburst and AGN winds can create an adiabatic hot bubble that would
drive a vigorous outflow. Since radiation pressure, together with rotation,
can support the structures against gravity (\S\ref{radpress}), here 
we drop both the gravitational and radiation pressure terms, and use for
simplicity  $R_0=0$. Starting with starburst winds (both stellar winds and 
supernovae), the energy injection in the outflowing ISM is
\begin{equation}
\frac{dE_{\mathrm{mod}}}{dt}=\beta_{\mathrm{out,*}} f_c(r) K_w L_{*,\mathrm{encl}}(r),
\label{econsstar}
\end{equation}
where $K_wL_*$ is the power of the winds, and
$\beta_{\mathrm{out,*}}$ is the fraction of this energy that goes into bulk
motion of the shocked ISM. We adopt $K_w=0.02$ \citep{lei99,vei05,har14} and
$\beta_{\mathrm{out,*}}=1/4$ \citep{wea77}. Integrating eq.~(\ref{econsstar})
gives for the momentum flux in the shell-cloud approach
\begin{equation}
\dot{P}_{\mathrm{mod,*}}^{\mathrm{E}}=
\frac{\beta_{\mathrm{out,*}} K_w L_* R_1}{2r_dv} 
\times\frac{R_1}{R}\left(1+2\ln\frac{R}{R_1}\right),
\label{pdoteconsstar}
\end{equation}
where $v$ is the velocity of the outflowing gas. Figure~\ref{pboost}a compares
the observed values of $\dot{P}_{\mathrm{out}}$ with the values given by 
eq.~(\ref{pdoteconsstar}) for all individual components. Some
components could be explained by feedback from a compact starburst,
including the $\sim100$ \kms\ outflows. However, most high-velocity
components show momentum fluxes that exceed the calculated upper limit.
The reason for this behavior is that the momentum boost, relative to the
momentum-driven flows, is given by 
$\beta_{\mathrm{out,*}}\times3000\,\mathrm{km\,s^{-1}}/v$ ($\leq7.5$ in all
cases), and thus little boost is expected for the high-velocity components due
to the moderate velocities of the starburst winds (a few $\times10^3$
\kms). This result suggests that bubbles generated by AGN winds are required
for the high-velocity components.

For the AGN winds, we solve
\begin{equation}
\frac{dE_{\mathrm{mod}}}{dt}=\beta_{\mathrm{out,AGN}} f_c(r) \dot{E}_{\mathrm{AGN}},
\label{econsagn}
\end{equation}
where $\beta_{\mathrm{out,AGN}}$ is the fraction of the injected energy
($\dot{E}_{\mathrm{AGN}}$) that goes into bulk motion of the shocked
ISM. Following \cite{fau12} and \cite{ste16}, we adopt
$\beta_{\mathrm{out,AGN}}=1/2$ and integrate eq.~(\ref{econsagn}) to give
\begin{equation}
\dot{P}_{\mathrm{mod,AGN}}^{\mathrm{E}}=g(R) \frac{L_{\mathrm{AGN}}}{c} 
\beta_{\mathrm{out,AGN}} \, \frac{v_{\mathrm{in}}}{v},
\label{pdoteconsagn}
\end{equation}
where $g(R)\equiv3R_1/(2R)\times(2-R_1/R)$ is of order unity, and the momentum
of the inner wind is written as $L_{\mathrm{AGN}}/c$ and has a velocity 
$v_{\mathrm{in}}$.  Fig.~\ref{pboost}b shows 
$\dot{P}_{\mathrm{out}}/(g\,L_{\mathrm{AGN}}/c)$ together with 
the solid lines indicative of the $\dot{P}\propto v^{-1}$ relationship
expected for an energy-conserving outflow
\citep[e.g.,][]{fau12,zub12,ste16,tom15} with 
assumed nuclear wind speeds of $v_{\mathrm{in}} = 0.1c$ and $0.033c$.
The required momentum boosts are found $\lesssim10$ in most
sources and $\lesssim20$ in all cases, in general consistent
with $v_{\mathrm{in}}\lesssim 0.1c$.  
We also find that the expected trend of increasing momentum boost with
  decreasing velocity no longer holds in this regime of relatively low
  velocities ($100-1000$ \kms).

\subsection{Interpretation of these models}
\label{interpret}

\cite{ste16} argue that the pressure of the hot
shocked wind can be constrained by emission line ratios measured in the colder
line-emitting gas.  They showed that in quasars 
radiation pressure is probably dominating at all scales, with an upper limit
$P_{\mathrm{hot}}/P_{\mathrm{rad}}\leq6$ for the spatial domain we sample with
OH. \cite{ste16} also indicated that high momentum boosts could only be
obtained in buried quasars where the shocked wind was tightly confined, and 
the high momentum boosts reported for quasars in the literature could thus
only be obtained early in their evolution, during the buried stage. 
This conclusion is consistent with \cite{zub14}, who found that an
energy-conserving bubble with escape channels for the hot gas cools
adiabatically and hence is less efficient in driving the ISM, leading to
small momentum boosts that approach the momentum-conserving scenario.

GA15 argued that OH65 is an excellent tool to probe this buried stage.  All
sources in the present sample except IRAS~09022$-$3615, IRAS~13120$-$5453, 
and IRAS~10565+2448 show strong OH65 features.  Fig.~\ref{oh65v84} shows
that the highest OH velocities are indeed found in sources with strong 
OH65 features, consistent with the model described by \cite{zub14} and
\cite{ste16}. However, 
some sources with very strong OH65 absorption only show low-velocity outflows,
or even lack evidence of outflows (region III in Fig.~\ref{oh65v84}),
suggesting 
that they are still in an early stage of this process, and the columns
around some extremely obscured sources are so high that they cannot 
(yet) be accelerated to high velocities and dispersed. 
Therefore, perhaps it is not
surprising that high-velocity molecular outflows are found in
moderately (but not extremely) obscured 
sources, in which the shocked wind is moderately confined but finds paths
with columns $\lesssim10^{23}$ \cmd\ that can be efficiently accelerated. Once
these paths are found and the wind emerges, its pressure will decrease and
moderate momentum boosts are expected, as found here in most ULIRGs. 

Our results for the momentum boost are, however, sensitive to the way in
which the gravitational potential well and the geometrical dilution of the
outflowing gas are treated (Figs.~\ref{testradpw}a,c and \ref{testradpwfc1}).
If the AGN and starburst winds are not isotropic but restricted to the solid
angles of the outflowing OH (the aligned momentum approach discussed in
Section~\ref{winds}), no geometrical loss of momentum is expected and 
most outflows approach the momentum-conserving scenario. On the other hand,
rotationally supported structures (i.e., ignoring the potential well,
\S\ref{winds}) are obviously more easily launched and accelerated, 
and a momentum-conserving approach would be enough to account for 
most of the inferred momenta (Fig.~\ref{testradpw}a). 
However, this only applies to the outflow components that are
  launched close to the equatorial plane of the nuclear disk or torus around 
the AGN.  Because we require high columns to account for the molecular
outflows, which are observed against strong sources of far-IR radiation (and
hence against regions with high columns), at least some low-velocity 
components can indeed be associated with ``equatorial'' flows.
This may not be the case for most high-velocity components, as recent CO
  interferometric observations show that the high velocity gas is not coplanar
  with the rotating disk in NGC~1614 and IRAS~17208$-$0014 \citep{gar15} and
  in Mrk~231 \citep{fer15}.
Nevertheless, the OH outflows are wide-angle, and the possibility that
  some low-velocity components flow close  
to the equatorial plane in a momentum-driven phase is again 
consistent with the theoretical approach by \cite{zub14}, who found that
in the directions of higher resistance, the ISM is mainly affected by the
incident momentum of the inner outflow, while the outflow energy tends to
  escape along leaky paths.  This makes the deposition of AGN energy in
the ISM (feedback) less efficient, potentially explaining the
  large masses that SMBHs attain and, likewise, the
  $M_{\mathrm{BH}}-\sigma$ relation \citep{zub14,kin15}.

In some ULIRGs, energy-conserving phases with momentum boosts in the
  range $\sim3-20$ are most likely required to explain some outflowing 
components (Figs.~\ref{testradpw}c, \ref{testradpwfc1}, and
  \ref{pboost}b). The corresponding sources are IRAS~03158+4227, 
IRAS~08572+3915, IRAS~20100$-$4156, and possibly Mrk~231 and Mrk~273,
although the momentum boost in the last two objects is small if $f_c=1$
  is assumed (Fig.~\ref{testradpwfc1}). Most of these galaxies have strong
  absorption in OH65 at blueshifted velocities. Our model for Mrk~273,  
however, is relatively uncertain because of the weakness of the
high-velocity ($\sim700$ \kms) spectral feature in OH119, OH79, and
  OH84, which suggests a jet-driven collimated outflow with energetics
unrelated to the source luminosity. Momentum boosts of
  $<20$ are lower than the values obtained in
  theoretical approaches to energy-conserving outflows
  for an inner wind velocity of $v_{\mathrm{in}}=0.1c$
  \citep[e.g.,][]{fau12,kin15}. Since this may be indicative of only partial
  covering and confinement of the hot shocked gas \citep{ste16} even in
  optically thick ULIRGs, we denote this scenario as 
  ``{\it partially} energy-conserving outflows''.

\subsection{Depletion time and flow timescales}
\label{sectime}

\begin{figure}
\begin{center}
\includegraphics[angle=0,scale=.55]{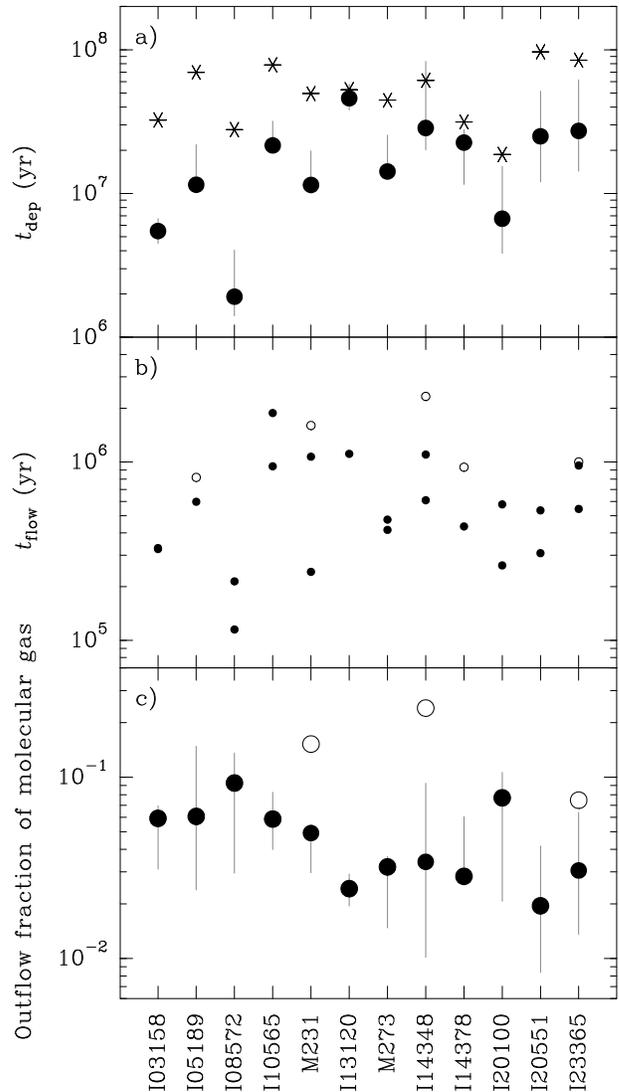}
\end{center}
\caption{a) Outflow depletion time (filled circles) calculated from the
  total molecular mass ($M_{\mathrm{H2}}$) of the galaxy and the total values
  of the mass outflow rates,
  $t_{\mathrm{dep}}=M_{\mathrm{H2}}/\dot{M}_{\mathrm{tot}}$.   
Most $M_{\mathrm{H2}}$ values have been inferred from the CO (1-0)
  emission, (with references in GA15 and A. Gowardhan et al., in prep.); in
  the case of IRAS~13120$-$5453, $M_{\mathrm{H2}}=7.5\times10^9$ \Msun\ in the
  central 1 kpc diameter region is
  estimated from the continuum associated with the H$_2$O submillimeter
  emission \citep{pri16}. The values of $\dot{M}_{\mathrm{tot}}$
  (eq.~\ref{par_glob}) only include the high-velocity components ($>200$
  \kms). Starred symbols
  show the gas consumption time scales, $M_{\mathrm{H2}}/\mathrm{SFR}$, where
  $\mathrm{SFR}(\mathrm{M}_{\odot}/\mathrm{yr})=10^{-10}L_*(\mathrm{L_{\odot}})$. 
  b) Flowing times for the individual outflowing components, calculated as 
  $t_{\mathrm{flow}}=r/v$. Open and filled squares indicate low ($<300$ \kms)
  and high ($>300$ \kms) velocity components, respectively. c) Fraction of the
  total molecular gas of the galaxy involved in the observed
  outflows. Open and filled squares include/do not include low ($<200$ \kms)
  velocity components, respectively. Errorbars in panels a and c only
    include uncertainties in $\dot{M}_{\mathrm{tot}}$ and $M_{\mathrm{tot}}$,
    respectively. 
} 
\label{time}
\end{figure}

\begin{figure}
\begin{center}
\includegraphics[angle=0,scale=.55]{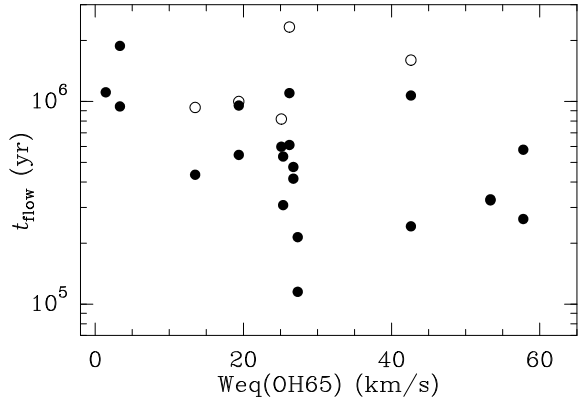}
\end{center}
\caption{Flow times for the individual outflow components (see
  Fig.~\ref{time}) as a function of the equivalent width of the OH65 doublet
  (between $-200$ and $+200$ \kms, GA15). Open and filled squares indicate low
  ($<300$ \kms) and high ($>300$ \kms) velocity components,
  respectively. Recent outflow events are found in active sources as
  diagnosed by the absorption in the high-lying OH65 doublet. 
} 
\label{tflow}
\end{figure}

Fig.~\ref{time}a shows the depletion time scales ($t_{\mathrm{dep}}$) calculated 
from the ratios of total molecular gas masses ($M_{\mathrm{H2}}$) to total
mass loss rates $\dot{M}_{\mathrm{tot}}$. 
The ratios lie mostly between $10^7$ and $10^8$ yr, except in the powerful
IRAS~03158+4227, IRAS~08572+3915, and IRAS~20100$-$4156. 
These timescales are significantly shorter than the gas consumption time
scales due to star formation \citep[$t_{\mathrm{con}}$, starred symbols in
  Fig.~\ref{time}a, see also][]{eva02}, and they are much shorter than the
strangulation time scales of local galaxies \citep{pen15}.
The values of $M_{\mathrm{H2}}$ used to estimate both 
$t_{\mathrm{dep}}$ and $t_{\mathrm{con}}$, as inferred in most cases
from simple scaling values of CO 1-0 luminosities (GA15), should be considered
with caution because the fraction of the CO population in the lowest
rotational levels is tiny; values of the mass-to-CO luminosity ratio a factor
of $\sim2$  higher are not excluded \citep{sol97}. 
In addition, the mass stored in the neutral (atomic) and ionized gas
components is not taken into account. 
The gas masses may be some factor higher than measured in
CO, but an upper limit of $t_{\mathrm{dep}}\lesssim 10^8$ yr can be
established within uncertainties. 
The $t_{\mathrm{con}}/t_{\mathrm{dep}}=1.1-15$ ratio is however independent of
gas mass, and indicates that the outflows mainly limit the amount of gas
  that will be converted into stars, i.e., the outflows are mainly
  responsible for the quenching of star formation. 
The flow time scales, defined as the time required for the
outflows to arrive at the current (inferred) radial position, are shown in
Fig.~\ref{time}b for the individual components, indicating that they are much
shorter than $t_{\mathrm{dep}}$. Some low-velocity components
($<300$ \kms, open symbols) have $t_{\mathrm{flow}}$ comparable to that of the
high-velocity components in the same source, suggesting that they may be a
consequence of the same AGN ``event''. The fraction of the total
molecular mass carried by the high-velocity outflows (Fig.~\ref{time}c)
is $\lesssim10$\%, so that the current ``cycle'' would have to be repeated
a dozen times to deplete the galaxy's molecular gas. 
We show in Fig.~\ref{tflow}, however, that sources with weak absorption in
OH65 have $t_{\mathrm{flow}}\gtrsim1$ Myr, while shorter $t_{\mathrm{flow}}$ are
only found in sources with strong absorption in OH65, i.e., the 
regions that are expected to represent the most buried
and active stage of nuclear starburst-AGN co-evolution
(GA15). The two sources with weak OH65 absorption, IRAS~13120$-$5453 and
IRAS~10565+2448 (see also Fig.~\ref{oh65v84}), show outflowing signatures that
may be a vestige of past strong nuclear activity, which has subsided over time
as a result of the outflowing activity. This 
suggests that the molecular outflows are efficient at evacuating the
nuclear regions of ULIRGs, but less efficient at expelling the gas
located at larger distances.

\subsection{Negative feedback: the relative roles of the AGN and the
  starburst} 
\label{negfeed}

\begin{figure*}
\begin{center}
\includegraphics[angle=0,scale=.6]{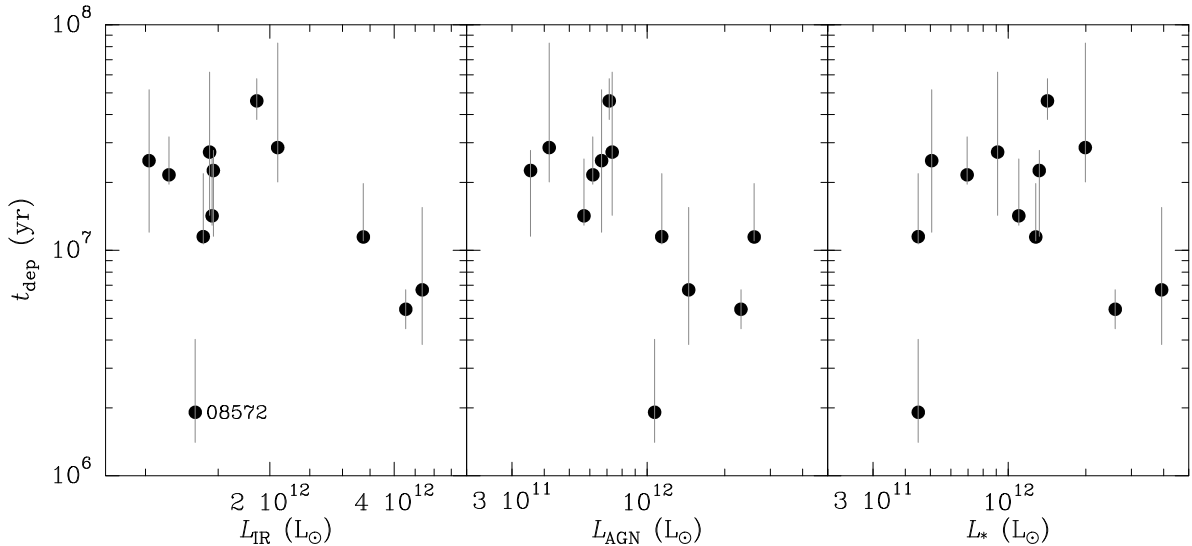}
\end{center}
\caption{Depletion timescales (Fig.~\ref{time}a) as a function of 
$L_{\mathrm{IR}}$, $L_{\mathrm{AGN}}$, and $L_{\mathrm{*}}$. The source with 
$t_{\mathrm{dep}}\sim2\times10^6$ yr is IRAS~08572+3915, for which
  $L_{\mathrm{AGN}}$ (and hence $L_{\mathrm{IR}}$) is probably underestimated
  \citep{efs14}. The best anti-correlation is found between $t_{\mathrm{dep}}$
  and $L_{\mathrm{AGN}}$ (correlation coefficient of $-0.59$, and
  $-0.74$ if IRAS~08572+3915 is excluded), indicating that
  the feedback from the AGN dominates the quenching of star formation at least
  for $L_{\mathrm{AGN}}\gtrsim10^{12}$ \Lsun.
} 
\label{tdeplum}
\end{figure*}

The outflow depletion timescale is plotted as a function of
  $L_{\mathrm{IR}}$, $L_{\mathrm{AGN}}$, and $L_{\mathrm{*}}$ in
  Fig.~\ref{tdeplum}. Since $t_{\mathrm{dep}}$ is sensitive to both the
  cumulative effect (the $M_{\mathrm{H2}}$ reservoir decreases as the
  molecular gas is expelled) and the present effect (through
  $\dot{M}_{\mathrm{tot}}$) of the negative feedback, 
  Fig.~\ref{tdeplum} allows us to check the relative roles of the AGN
  and the starburst in quenching the star formation in the sources. The middle
  and right-hand panels indicate that $t_{\mathrm{dep}}$ is better correlated
  with $L_{\mathrm{AGN}}$ than with $L_{\mathrm{*}}$, indicating that the AGN
  is primarily responsible for the short $t_{\mathrm{dep}}\lesssim10^7$ yr found 
  in IRAS~08572+3915, IRAS~03158+4227, IRAS~20100$-$4156, Mrk~231,
  IRAS~05189$-$2524, and most probably Mrk~273. The
  $t_{\mathrm{dep}}-L_{\mathrm{AGN}}$ correlation would be even tighter if
  IRAS~08572+3915 had an AGN luminosity well in excess of $10^{12}$ \Lsun\ as
  was argued by \cite{efs14}. The $L_{\mathrm{AGN}}$ threshold for AGN-dominated
  feedback, $\sim10^{12}$ \Lsun, is similar to that found by \cite{vei13}, 
  $10^{11.8}$ \Lsun, from the analysis of the outflow velocities inferred
  from the OH119 doublet in a larger sample of (U)LIRGs.

For the rest of the sources, the starburst may have a significant role in
  the observed feedback, but the relative roles of the AGN and the starburst
  are more uncertain. According to the models developed in \S\ref{dynmod}
  and \S\ref{econs}, and taking at face value the
  $\alpha_{\mathrm{AGN}}$ values in Table~\ref{tbl-1},
  the AGN appears to dominate in IRAS~20551$-$4250 and IRAS~23365+3604,
  and the starburst may have a comparable contribution 
  in IRAS~10565+2448 and a dominant role in IRAS~13120$-$5453,
  IRAS~14348$-$1447, and IRAS~14378$-$3651. However, the main
  caveats for these assignments are 
  $(i)$ the extended outflow components may be the relic of an epoch when the
  AGN was very active, but has since then subsided; 
  $(ii)$ $L_{\mathrm{AGN}}$ may be underestimated in sources with strong OH65
  absorption, as in IRAS~14348$-$1447; $(iii)$ the presence of high-velocity
  wings in the OH119 doublet in all these sources (with velocities approaching
  or exceeding 800 km/s, see Fig.~\ref{fitsoh119}) may be difficult to explain
  through SNe explosions \citep{mart16}, perhaps indicating the dominance of
  the AGN in driving at least this extremely high velocity molecular gas.

\section{Conclusions} \label{conclusions}

\subsection{Observational results}

The main observational results of our analysis of the line intensities and
profiles of four OH doublets (at 119, 79, 84, and 65\,$\mu$m) in a sample of 
14 (U)LIRGs observed with {\it Herschel}/PACS are: \\
$\bullet$ In ULIRGs in which a P-Cygni profile or a high-velocity blueshifted
absorption wing is detected in the ground-state OH119 transition,
we confirm that the more optically thin OH79 transition also shows clear
indication of outflowing gas in its line shape, though at lower
velocities \citep[see also][]{stu11}. \\   
$\bullet$ OH119 is optically thick even in the line wings, and its strength
provides a measure of the covering factor of the 119 $\mu$m continuum by the
outflowing OH. The overall covering factor ranges from $0.1$ to $>0.5$
(\S\ref{cov}). \\ 
$\bullet$ The highest outflowing velocities arise in sources with strong
absorption in the high-lying OH65 doublet at central velocities
(\S\ref{oh65of}), indicating that the buried nuclear regions of ULIRGs
are most efficient in launching and accelerating the molecular outflows. \\
$\bullet$ From the velocity shifts observed in OH65 and their correspondence
with those of OH79, we find evidence for low-velocity expansion of the nuclear
regions in several ULIRGs, mostly in those sources that also present P-Cygni
line shapes in OH119. The peak absorption velocity in the high-lying lines is
usually closer to systemic than in the ground-state OH119 and OH79 doublets
(\S\ref{oh65vshift}), indicating increasing column densities for lower outflow
velocities. \\  
$\bullet$ Spatially compact and extended outflows are best distinguished by
the OH84/OH119 ratio at blueshifted velocities (\S\ref{comext}).  \\  

\subsection{Modeling results}

We have modeled the observed OH doublet profiles in 14 ULIRGs
through $\chi^2$ minimization of combinations of single-component models
(\S\ref{libr}). The models yield satisfactory fits to the observed OH doublets
in 12 sources, while closely matching the observed SEDs in the
transition from mid- to far-IR wavelengths ($25-50$ $\mu$m,
Figs.~\ref{fitsoh119}-\ref{fitscont}). The models enable us to estimate the
energetics associated with the 
individual components and with each source globally
(\S\ref{energetics}). \\   
$\bullet$ Excluding the low-velocity ($\sim100$ \kms) components,
the total mass outflow rates range from $\sim150$  to $\sim1500$
\Msun\,yr$^{-1}$ (Table~\ref{tbl-3}), the mass loading 
factors are $1.5-8$ (Fig.~\ref{ener}), the momentum fluxes are
$(2-20)\times L_{\mathrm{IR}}/c$, and the energy fluxes are 
$(0.1-3)$\% of $L_{\mathrm{IR}}$. The corresponding momentum boosts are
$<10$ if the momentum rates due to radiation pressure, winds, and
  supernovae supplied by the AGN ($2L_{\mathrm{AGN}}/c$)
and the starburst ($3.5L_{\mathrm{*}}/c$) are
combined (\S\ref{modener}, Fig.~\ref{pdotglb}). These
values may be significantly higher in some sources (e.g., IRAS~14348$-$1447,
IRAS~23365+3604, Mrk~231) if the low-velocity components are included. \\  
$\bullet$ The most powerful outflows, 
with energy fluxes of $\sim10^{10.5-11}$ \Lsun\ (Fig.~\ref{pdotglb}f),
  are found in IRAS~03158+4227 (belonging to a
  widely separated pair), IRAS~08572+3915 and IRAS~20100$-$4156 (mergers with
  projected separation of $\approx5.5$ kpc), and Mrk~231 (post-merger), 
  apparently uncorrelated with the merging stage. They are best
  identified by 
  high-velocity absorption in the high-lying OH65 doublet, and represent
  $\sim20$\% of all local ULIRGs observed in OH65. This suggests stochastic
  episodic strong-AGN feedback events throughout the ULIRG phase of the
  merging process. \\    
$\bullet$ Overall, the energetics inferred from OH are consistent with 
those derived from CO. The similar outflow masses found for OH and for the
more spatially extended CO and Na {\sc i} suggests that a significant fraction
of the total outflow mass is loaded at small radii (a few $\times100$ pc),
and the outflows are less efficient at evacuating the gas initially located in
the extended kpc-scale disks. \\    
$\bullet$ We use a simple analytical model for momentum-driven flows 
to check whether radiation pressure due to the AGN and starburst, and
momentum-conserving winds, can account for the observed OH outflows
(\S\ref{dynmod}). Our primary adopted geometry is a combination of a
shell for small radii, and a collection of clouds at larger radii with
  decreasing covering factor relative to the released momenta, but we also
  check the results obtained with no geometrical loss of input momentum. 
The model also roughly simulates the extended
nature of the starbursts and the momentum boost due to radiation
trapping. We find that radiation pressure may partially support the gas
against gravity (\S\ref{radpress} and App.~\ref{radp}), but is unable to drive
the outflows. Inclusion of momentum-conserving winds (\S\ref{winds}) could
explain some of the outflowing components, in particular if
the gravitational potential well is assumed to be balanced by
rotation or the input momentum is not geometrically lost. However,
partially energy-conserving phases are most likely required for some 
high-velocity  components with high columns (IRAS~03158+4227,
IRAS~08572+3915, IRAS~20100$-$4156 and possibly Mrk~231). \\  
$\bullet$ Energy-conserving flows are treated separately for the starburst 
and AGN (\S\ref{econs}). Using limiting values for feedback from a 
nuclear starburst, some low-velocity components can be accounted for. 
Winds from the AGN with energy-conserving phases are required
to explain some of the highest-velocity OH outflows, with momentum boosts
relative to $L_{\mathrm{AGN}}/c$ that, for the individual components, do not
exceed a factor $\sim20$ and are mostly $\lesssim10$. These values are
mostly lower than the maximum momentum boosts predicted by theoretical models
for energy-driven outflows. For the
$<1000$ \kms\ outflows probed by OH, the momentum flux carried by
the highest velocity components is usually higher than that carried by lower
velocity components. Our results indicate that energy-driven and
momentum-driven outflows may coexist in the same sources. \\
$\bullet$ Depletion time scales generally lie between $10^7$ and $10^8$ yr,
except in the cases of IRAS~03158+4227, IRAS~08572+3915, and
IRAS~20100$-$4156, where $t_{\mathrm{dep}}$ is somewhat shorter
(Fig.~\ref{time}). These estimates are comparable to but significantly shorter
than the gas consumption time scales. The flow time scales 
($\sim R/v$) are much shorter, $\sim 10^5-10^6$ yr, and also favor recent or
ongoing nuclear outflow activity in highly excited and buried sources as
probed by the absorption in the high-lying OH65 doublet. \\
$\bullet$ We find an anticorrelation between $t_{\mathrm{dep}}$ and
  $L_{\mathrm{AGN}}$, indicating that the AGN is primarily responsible for the
  short $t_{\mathrm{dep}}\lesssim10^7$ yr found in IRAS~08572+3915,
  IRAS~03158+4227, IRAS~20100$-$4156, Mrk~231, IRAS~05189$-$2524, and most
  probably Mrk~273. For the rest of the sources the relative roles of the AGN
  and the starburst are more uncertain, but the AGN either dominates or has a
  significant role in the observed negative feedback.

\acknowledgments

We thank Avani Gowardhan for providing us with the CO fluxes of 
IRAS~03158+4227 and IRAS~20100$-$4156 prior to the publication.
E.GA thanks the hospitality of the Harvard-Smithsonian
Center for Astrophysics, where part of the present study was carried on.
We would also like to acknowledge the contribution made by Will Lewitus who
carried out some early data analysis on these outflows while a student intern
at NRL. PACS has been developed by a consortium of institutes
led by MPE (Germany) and including UVIE (Austria); KU Leuven, CSL, IMEC
(Belgium); CEA, LAM (France); MPIA (Germany); 
INAFIFSI/OAA/OAP/OAT, LENS, SISSA (Italy); IAC (Spain). This development
has been supported by the funding agencies BMVIT (Austria), ESA-PRODEX
(Belgium), CEA/CNES (France), DLR (Germany), ASI/INAF (Italy), and
CICYT/MCYT (Spain). E.GA is a Research Associate at the Harvard-Smithsonian
Center for Astrophysics, and thanks the Spanish 
Ministerio de Econom\'{\i}a y Competitividad for support under projects
FIS2012-39162-C06-01 and  ESP2015-65597-C4-1-R, and NASA grant ADAP
NNX15AE56G. Basic research in IR astronomy at NRL is funded by 
the US ONR; J.F. also acknowledges support from NHSC/JPL subcontracts 139807
and 1456609. H.S. acknowledges support from NHSC/JPL subcontract 1483848.
S.V. thanks NASA for partial support of this research through
NASA grant ADAP NNX16AF24G. 
This research has made use of NASA's Astrophysics Data System
(ADS) and of GILDAS software (http://www.iram.fr/IRAMFR/GILDAS).

{\it Facilities:} \facility{{\it Herschel Space Observatory} (PACS)}.

\appendix

\section{A. Individual sources}
\label{indsour}

{\it\underline{Mrk 231}}: 
Our 3-component model fit is consistent with the more 
detailed, 4-component model reported in GA14: we fit a ``quiescent''
component (QC) that accounts for the high-lying absorption at systemic
velocities, a compact component with gas outflowing up to 
$>1300$ \kms\ (HVC), 
and an extended component (EC) with intermediate velocities. The
HVC and EC have associated $\dot{M}\sim400$ and $\sim250$ \Msun\,yr$^{-1}$,
respectively. These values are roughly consistent with the inferred values
from CO (2-1) by \cite{fer15} once their values are corrected to match our
definition. Tentatively, we find in the current work that the QC is also
expanding at $\sim100$ \kms\ and with $\sim450$ \Msun\,yr$^{-1}$, matching the
velocity of peak absorption in the OH84 doublet though not that of the OH65
doublet. On the other hand, we find that the EC could be more compact
than adopted in GA14 \citep[which was based on interferometric CO 1-0 emission
in][]{cic12}; a good match to the line profiles is also obtained with
$R\sim400$ pc.

{\it\underline{IRAS 08572+3915}}: This is an AGN-dominated
double-nucleus source with CO detected only in the NW nucleus \citep{eva02},
and with no trace of PAH emission at 3.3 $\mu$m \citep{ima08}.
It shows strong absorption in the blue
wings of OH119, OH79, and OH84 out to $-1000$ \kms, and somewhat more
tentatively in OH65. At blueshifted velocities, OH84 is
almost as strong as OH119 (Fig.~\ref{ohblue}), indicating a compact
outflow. There are also hints of the presence of a non-outflowing (i.e.,
peaking at systemic velocities) component, but it is 
spectrally diluted by the prominent wings. It is the only source in our sample
that shows a P-Cygni profile in the highly excited OH65 doublet. 
Our best-fit model for the blueshifted wing requires 2 components for the
outflow, though both are very compact with similar $r\sim110$\,pc. The
lower velocity ($200-800$ \kms) component, with
$N_{\mathrm{H}}\sim1.5\times10^{23}$ \cmd, and the high velocity 
($800-1100$ \kms) component, with $N_{\mathrm{H}}\sim8\times10^{22}$ \cmd,
both have similar $\dot{P}_{\mathrm{out}}\sim(1.3-1.5)\times10^{36}$ dyn, about 3
times the momentum supplied by the dominant AGN and the starburst
($\sim5\times10^{35}$ dyn). However, \cite{efs14} argue that this Sy 2 galaxy
has a much larger AGN luminosity, but the radiation mainly escapes along the
axis of the torus away from our line-of-sight. If this is the case, and the
AGN has a luminosity of 
$\sim10^{13}$ \Lsun\ \citep{efs14}, it would supply a momentum rate of
$2.5\times10^{36}$ dyn which is just the sum of the momentum fluxes estimated
for the two OH outflowing components. 

With $\sim100$\,pc resolution, \cite{rup13b} have resolved the outflow in
H$_2$ ro-vibrational lines, showing a blueshifted wing similar to OH119 with
extent up to 400\,pc and centered at $\sim200$\,pc from the center. This is
more extended than the radius inferred for OH, which is perhaps consistent
with the high extinction of the nuclear source at far-IR wavelengths.
Nevertheless, the OH appears to be missing the more extended component of the
outflow. \cite{cic14} indeed measured a radius of $\sim0.8$ kpc based on CO
emission but, in contrast with other sources, OH does not show 
evidence for extended emission in the blueshifted wing. This may be due to the
compactness of the far-IR continuum. OH84 shows a prominent P-Cygni
profile (see Fig.~\ref{ohspec}), but the redshifted emission (at velocities
between 1000 and 1600 km/s relative to the blue component, very likely
extended) does not appear to 
belong to the component responsible for the blue absorption wing. This
redshifted feature appears to have its counterpart in CO (1-0) \citep{cic14},
and has not been modeled.  

\cite{shi13} have detected the outflow in the ro-vibrational lines of CO at
$\sim5$ $\mu$m, up to blueshifted velocities of $-400$ \kms, and derived a
column of $N_{\mathrm{H}}\sim5\times10^{22}$ \cmd. Since the 5 $\mu$m
continuum emission that CO is absorbing is expected to arise from a very warm,
small region close to the AGN \citep{shi13}, CO and OH probably probe
different regions, but the two tracers indicate high columns of outflowing gas
close to the central engine.

{\it\underline{Mrk 273}}: 
This is a double system with the northern nucleus containing
most of the molecular mass and probably also an AGN that accounts for roughly 
$1/3-1/2$ of the total luminosity \citep{rup13a}.
Mrk~273 is similar in OH79 to IRAS~05189$-$2524. In addition, both sources 
(and also IRAS~20551$-$4250) show an OH119 absorption trough of only
$\sim10$\% of the continuum, and OH79 correspondingly shows weak absorption
and emission in comparison with other ULIRGs with similar luminosity such as
IRAS~23365+3604, but extending to higher velocities. This indicates a
relatively low covering factor of the continuum by the outflow, perhaps
associated with the biconical shape of the outflowing gas \citep{rup13a}. At
systemic velocities, the high-lying OH84 and 
OH65 doublets show strong absorption, most likely associated with the very
compact molecular core detected by \cite{dow98}. Mrk~273 shows two well
defined velocity components on the blue side of the OH profiles, both detected
in OH84, suggesting a compact molecular outflow. \cite{cic14} indeed found
the shortest radius in Mrk~273, $\sim0.55$ kpc, among the ULIRG outflows
observed in CO. The $\sim300$ \kms\ component, detected in both OH84 and OH65,
appears to be extremely compact $\sim130$\,pc and has high columns
($N_{\mathrm{H}}\sim8\times10^{22}$ \cmd), yielding
$\dot{P}_{\mathrm{out}}\sim2\times10^{35}$ dyn. If the northern nucleus
harbours an embedded AGN with $1/3-1/2$ of the total luminosity, the above
momentum flux is $1-2$ times that provided by 
the AGN. While suggesting a moderate momentum boost, overcoming the deep 
potential well is the main drawback to explaining this outflow component
through a momentum-driven approach.  For the high-velocity ($\sim700$
\kms) component, which is detected in OH84 with similar strength as OH119, we
infer $N_{\mathrm{H}}\sim1\times10^{23}$ \cmd\ and $r\sim350$\,pc, giving
$\dot{P}_{\mathrm{out}}\sim10^{36}$ dyn, i.e., nearly twice the momentum rate
supplied by the starburst and the AGN combined. A high momentum boost is
required (Fig.~\ref{pboost}), though we cautiously note the weakness of the
OH119, OH79 and OH84 high-velocity wings, which may suggests a jet-driven
collimated outflow component.

{\it\underline{IRAS 20551$-$4250}}: This is an advanced merger with a single
nucleus that shows evidence for both a deeply embedded AGN and a starburst
from hard X-rays to mid-IR emission \citep{fra03,haa11,gen98,san12}. The
relative contribution from the AGN and the starburst based on mid-IR line
diagnostics may be misleading, as [Ne {\sc ii}] and [O {\sc iv}] are not
detected but the equivalent width of the PAH 6.2 $\mu$m feature is low,
suggesting high extinction \citep{far07}. 
The strong OH65 absorption indeed indicates that the
source is optically thick even at far-IR wavelengths (GA15). \cite{san12}
estimated a 26\% contribution by the AGN, while \cite{vei13} estimated 57\%
from the $f_{15}/f_{30}$ flux density ratio. OH65 shows a blueshifted
line wing up to $-400$ \kms, while the rest of the doublets show absorption up
to $-800$ \kms. These are best matched with two outflowing components, that
provide a total momentum flux of $\sim4.5\times10^{35}$ dyn, 15\% in excess of
the combined momentum rate supplied by the AGN and the starburst.

{\it\underline{IRAS 23365+3604}}: This LINER-type ULIRG shows no evidence of a
double nucleus \citep{dow98}, and the tidal tails observed in the optical
\citep{mur96} suggest a post-coalescence, late-stage merger \citep{martin16}.
About half of its luminosity is associated
with an AGN \citep{vei13}. Besides OH119 and OH79, OH84 also shows a P-Cygni
profile in this source, and the OH65 peak absorption is blueshifted by
$\approx100$ \kms\ relative to [C {\sc ii}] 157$\mu$m. OH119 shows blue
absorption up to $-1000$ \kms\ and possibly more, but the other doublets only
show absorption up to $-500$ \kms. Our best fit model
involves three outflowing components, with the CORE component outflowing at
$\sim100$ \kms\ ($\dot{M}\sim340$ M$_{\odot}$\,yr$^{-1}$). The other two
components (green and magenta curves in Figs.~\ref{fitsoh119}-\ref{fitsoh65})
are probably not independent. They combine to give a total 
$\dot{P}_{\mathrm{out}}\sim7.5\times10^{35}$ dyn, $\sim4\times$ the momentum
rate supplied by the AGN. If a compact starburst is considered, the outflow in
IRAS~23365+3604 requires little momentum boost for both components 
($\lesssim3$).

{\it\underline{IRAS 14348$-$1447}}: No evidence for broad-line emission or
very high ionization lines in the optical and near-IR is found in this
LINER-type ULIRG \citep{vei97}, suggesting either a weak AGN contribution to
the luminosity or an extremely buried AGN. In the far-IR, the strong
absorption found in OH65 (GA15, Fig.~\ref{fitsoh65}) strengthens the
latter possibility. It is a very massive merger, with projected
nuclear separation of $\sim5$ kpc, and with the two CO $1-0$ components 
associated with the stellar nuclei \citep{eva00}. The southwest (SW) nucleus
is the most luminous in both CO \citep{eva00} and in the near-IR
\citep{sco00}, and the OH profiles suggest that that it is the source of the
outflow because its redshift, $z_{\mathrm{SW}}=0.0827$ \citep{eva00},
is in good agreement with the center of the OH119 P-Cygni profile 
(Fig.~\ref{fitsoh119}. We have used in Figs~\ref{fitsoh119}-\ref{fitsoh65}
$z_{\mathrm{CII}}=0.08257$, Table~\ref{tbl-1}, but a better fit to OH119 and
OH79 is obtained with $z_{\mathrm{SW}}$). A very high velocity wing, in excess
of 1000 km/s, is observed in OH119. An interesting characteristic of this
source is the blueshift of $\approx100$ \kms\ observed in the OH65 absorption
feature (Figs.~\ref{ciioh65prof} and \ref{fitsoh65}). Although the northeast
(NE) nucleus has a redshift that coincides with that of the OH65 feature, the
close correspondence between the OH65 and the OH79 absorptions 
strongly suggests that OH65 is also produced toward the SW nucleus, i.e., the
nucleus is expanding. The associated momentum flux, $\sim7\times10^{35}$ dyn,
is of the same order as the momentum rate that can be supplied by the dominant
energy source, whether starburst or AGN, and can be explained through
a momentum-driven approach if it is rotationally supported. Still, some
non-negligible contribution of the NE nucleus may be present and the above
estimates are upper limits. The other, higher velocity components together
account for $\sim10^{36}$ dyn.

{\it\underline{IRAS 05189$-$2524}}: This is an AGN-dominated ULIRG, with 
$\alpha_{\mathrm{AGN}}$ ranging from 0.6 to 1 using the six diagnostics in
\cite{vei09}. It has a single identifiable nucleus, and is proposed to be a
late-stage merger system due to the observed tidal tails in the optical
\citep{san88}. Its optical classification is Sy 2, but it shows broad
($\sim2000$ \kms) emission near-IR lines (Pa$\alpha$, Pa$\beta$, He)
and is thus classified as Sy 1 in the near-IR \citep{sev00}.
\cite{spo09} found from blueshifted [Ne {\sc iii}] and [Ne {\sc v}]
  mid-IR emission lines detected in this and other ULIRGs that the gas
  ionization increase with increasing blueshift. 
While the ground-state OH119 and OH79 profiles show well defined P-Cygni line
shapes, the excited OH84 and OH65 profiles show strong 
absorption at systemic velocities (GA15). At moderate blueshifted velocities
($\sim-200$ \kms), OH79 has an outflowing counterpart in both OH84 and OH65
(Fig.~\ref{ciioh65prof}), indicating a compact component with
$N_{\mathrm{H}}\sim10^{23}$ \cmd\ and momentum flux 
($\dot{P}_{\mathrm{out}}\sim1.5\times10^{35}$ dyn) somewhat lower than that
provided by the AGN ($\sim3\times10^{35}$ dyn), suggesting a momentum driven
shell (Figs.~\ref{testradpw}c and \ref{pboost}b). 
OH119 shows a flat absorption profile
between $-200$ and $-600$ \kms\ at $\approx10$\% level, indicating that the
high-velocity gas, not detected in OH84 or OH65, is more extended and
with a low covering factor. 
With  $r\sim340$\,pc, the high-velocity component can be roughly
explained with $N_{\mathrm{H}}\sim4\times10^{22}$ \cmd, 
yielding $\dot{P}_{\mathrm{out}}\sim5\times10^{35}$ dyn and thus
requiring moderate momentum boost. Although the OH79 absorption at
$v<-600$ \kms\ is not quite reproduced, this is the AGN-dominated ULIRG with
the lowest momentum boost (Fig.~\ref{ener}).

{\it\underline{IRAS 14378$-$3651}}: 
\cite{duc97} found no
obvious optical companion to this distorted galaxy, which the authors
suggest is a final stage merger.  This is consistent with
the OH observations: the prominent outflow probed by the OH119, OH79, and 
OH84 doublets appears to have expelled a significant fraction of the nuclear
molecular gas, leaving a relatively weak OH65 absorption (GA15,
Fig.~\ref{fitsoh65}).  A moderate (250 \kms) 
velocity component, mostly responsible for the broad and blueshifted OH84
absorption, and a high velocity component give a total momentum flux of
$\sim5\times10^{35}$ dyn, which is of the same order as the momentum rate
supplied by the (dominant) starburst. This suggests at most moderate
momentum boost ($\lesssim3$).

{\it\underline{IRAS 10565+2448}}: In the optical ($r$-band) this galaxy
appears to be a triple system \citep{mur96}, but CO is only detected in the  
western nucleus \citep{dow98} which dominates the luminosity of the system 
\citep{shi10,rup13a}. The AGN luminosity may be significantly
lower than that of the starburst \citep{vei09,vei13,rup13a} since the value of
$\alpha_{\mathrm{AGN}}=0.47$ is higher than that obtained from other
diagnostics in \cite{vei09}. 
The source shows prominent OH119 and OH79 P-Cygni 
profiles with blue absorption trough and red emission strength similar to or
stronger than AGN-dominated sources, although with lower velocity extent
(Figs.~\ref{fitsoh119} and \ref{fitsoh79}). The ground-state OH lines thus
indicate outflow parameters for the extended component of the outflow that are
similar to AGN-dominated ULIRGs. Our model fits two 
outflow components with $N_{\mathrm{H}}\sim(1-2)\times10^{22}$ \cmd\ and
$r\sim400-500$\,pc each, yielding $\dot{P}_{\mathrm{tot}}\sim5.5\times10^{35}$
dyn. This is $\sim1.2$ times the momentum supplied by the starburst and AGN
combined, suggesting a momentum-conserving molecular outflow. The high total 
outflowing molecular mass of $\sim3\times10^8$ \Msun\ still falls short of 
the outflowing atomic gas of $\sim10^9$ \Msun\ inferred from Na {\sc I} D by
\cite{rup13a} , which extends out to at least $\sim4$ kpc from the western 
nucleus \citep{shi10}. 
In spite of the prominent extended outflow inferred from the
ground-state OH119 and OH79 doublets, the source shows weak OH65 and OH84
absorption, indicative of relatively low columns or dust temperatures in
the central region (GA15). This may be consistent with the model by
\cite{dow98}, who fitted an {\it unfilled} disk (i.e., with a central gap) to
the CO $1-0$ interferometric data.
In addition, the weak OH84 appears to be globally
shifted to the blue (Fig.~\ref{fitsoh84}), 
indicating that a significant part of the excited, nuclear region is
outflowing, and that the source recently rid itself of most gas in the
nuclear region. Together, these clues appear to indicate that, in
spite of the high luminosity of the source, the nuclear outflowing activity in
IRAS 10565+2448 has subsided in the last $\sim1$ Myr. This is consistent with
the fact that the most obscured regions are found spatially coincident with
the highest velocity outflow, rather than with the western nucleus
\citep{rup13a}.

{\it\underline{IRAS 13120$-$5453}}: This ULIRG is classified as a post-merger
with a single nucleus \citep{sti13}. Its luminosity is dominated by the
starburst, though it has a Compton-thick AGN \citep{iwa11,ten15}. The
source is not detected in OH65 (GA15), indicating that the material
responsible for the AGN obscuration is too compact to have any significant
contribution to the far-IR. 
Our fit to the OH doublets consists of a central velocity component,
best identified in OH84 (Fig.~\ref{fitsoh84}), and an outflow component with
moderate column density ($\sim1.5\times10^{22}$ \cmd) and probably
extended. It gives an estimated momentum flux of $\sim4\times10^{35}$
dyn, which is significantly lower than the predicted momentum rate supplied by
the starburst and the AGN. This suggests a momentum-conserving flow. A weak
but apparent blueshifted wing in OH119 up to 
velocities of $\sim-1500$ \kms\ (Fig.~\ref{fitsoh119}) is not modeled, 
and could reveal the feedback by the buried AGN.

{\it\underline{IRAS 19297$-$0406}}: This is a multiple colliding system
dominated by star formation. Though the source is undetected in OH79, it
still shows a prominent absorption in OH65 at systemic velocities. The
energetics of the outflow are very uncertain, but the model is consistent with
an outflowing 
column of $N_{\mathrm{H}}\sim10^{22}$ \cmd, $M_{\mathrm{out}}\sim4\times10^7$
\Msun, $\dot{M}\sim100$ \Msun/yr, and $\dot{P}\sim3\times10^{35}$ dyn. This is
significantly lower than the momentum rate that a galaxy with
$L_{\mathrm{IR}}\sim2.5\times10^{12}$ \Lsun\ can provide, suggesting that only
a single component of the multiple system is involved in the outflow.

{\it\underline{IRAS 09022$-$3615}}: Optically classified as a star-forming
galaxy \citep{lee11}, the red near-IR continuum in the $L$-band suggests a
buried AGN in this ULIRG \citep{lee12}. 
The OH65 absorption in this source is weak but detected, 
and shows a slight blueshift relative to the systemic
velocity. This is the only source in our sample where the emission feature in
OH119 is stronger than the absorption feature, thus involving important
geometrical effects or significant collisional excitation. Our model is thus
relatively uncertain, though it involves $\dot{M}\sim130$ \Msun/yr and
$\dot{P}\sim1\times10^{35}$ dyn. As in the case of IRAS 19297$-$0406, this is
significantly lower than the momentum rate that the galaxy could supply
based on its luminosity. However, the energetics could be significantly
underestimated if the weak absorption is due to low far-IR continuum
brightness, rather than to low column densities. 

{\it\underline{IRAS 03158$+$4227}}: This galaxy has a companion (projected
separation of $\sim50$ kpc) with no apparent sign of interaction in the
$r-$band \citep{mur96}, but \cite{meu01} found a curved tail in the companion
that matches the expected result of a binary interaction. The optical
spectrum shows no indication of an AGN \citep{meu01}, and
\cite{ris00} find no evidence for an AGN from hard X-rays activity. 
However, this is (together with IRAS~20100$-$4156) the most extreme source in
our sample. At systemic velocities the OH65 absorption is very deep 
with nearly flat absorption between the two doublet components
(Fig.~\ref{ohspec}). This indicates very high extinction in the far-IR,
capable of obscuring any AGN signature at shorter wavelengths. On the other
hand, the outflow is observed out to $\gtrsim1500$ \kms\ in OH119 as
emphasized by \cite{spo13}. In OH79 the observed P-Cygni profile has a
blueshifted wing apparently extending up to $-1700$ \kms, though it may be
contaminated by H$_2$O primarily at velocities more blueshifted than $-1300$
\kms. The OH84 absorption trough is deep, 10\% at $-800$ \kms, and a very
prominent OH65 absorption is seen up to $-1000$ \kms. Our simple 3-component
model approach cannot account for all spectroscopic details in the wide
velocity range depicted by the OH doublets. Even neglecting the high-velocity
wing in OH79, the OH65 absorption at central velocities is underpredicted and
OH84 is slightly overpredicted over the full velocity range. Nevertheless, the
model captures the presence of a compact, massive ($4\times10^8$ \Msun), and
excited component flowing at high velocities ($300-1300$ \kms), with an
estimated mass outflow rate of $1300$ \Msun/yr and a momentum flux of
$\sim7\times10^{36}$ dyn. These astonishing values cannot be understood in
terms of the momentum rate supplied by a galaxy with even such high luminosity
as IRAS~03158$+$4227, and a very high momentum boost is required. This is
probably an ``exploding'' quasar, with enormous masses of gas suddenly
accelerated to high velocities, and a nuclear energy conserving phase within
this extremely buried source is required to understand the involved
energetics. In addition, another outflowing component is required to account
for the extremely high-velocity OH wings. Our model for this latter component
is more uncertain, but probably does not dominate the energetics of the
source and can be explained through a momentum conserving phase
  (Fig.~\ref{testradpw}c).  

{\it\underline{IRAS 20100$-$4156}}: From optical and near-IR images,
\cite{duc97} classify this galaxy as a close interacting pair of disk galaxies
(projected separation of $\approx5.4$ kpc),
with a starburst or LINER-type spectrum. A possible third component of the
merger was identified by \cite{bus02} with NICMOS. The galaxy shows strong
megamaser emission in OH at centimeter wavelengths \citep{har16}, with a
plateau of OH emission whose 
blue wing traces the velocities of the absorption found in the far-IR
OH doublets. Searches in X-rays have identified the possible presence of a
buried AGN \citep{fra03}. 
The OH doublets indicate a galaxy with extreme properties. The OH65
absorption is very deep and broad, with little decrease of the absorption
at wavelengths between the two doublet components. Absorption in OH119 is
observed out to $-1300$ \kms, and to $-1000$ \kms\ in the other 3 doublets.  
Similar to IRAS~03158+4227, our model involves a compact, massive
($5\times10^8$ \Msun), and excited component flowing at velocities up to 800
\kms\ with a momentum flux of $\sim3\times10^{36}$ dyn. It most likely
requires a high momentum boost. A second, even
higher speed component with one order of magnitude lower mass accounts for 
$\sim1.5\times10^{36}$ dyn, and could be explained via a momentum-driven
outflow. The OH119 absorption is underestimated by our model on the red side
of the profile, suggesting additional, foreground absorption. The very high
velocities in this source suggest that the AGN contribution to the luminosity
is underestimated.


\section{B. Comparison with ionized lines in starburst galaxies}
\label{heck15}

\begin{figure*}
\begin{center}
\includegraphics[angle=0,scale=.7]{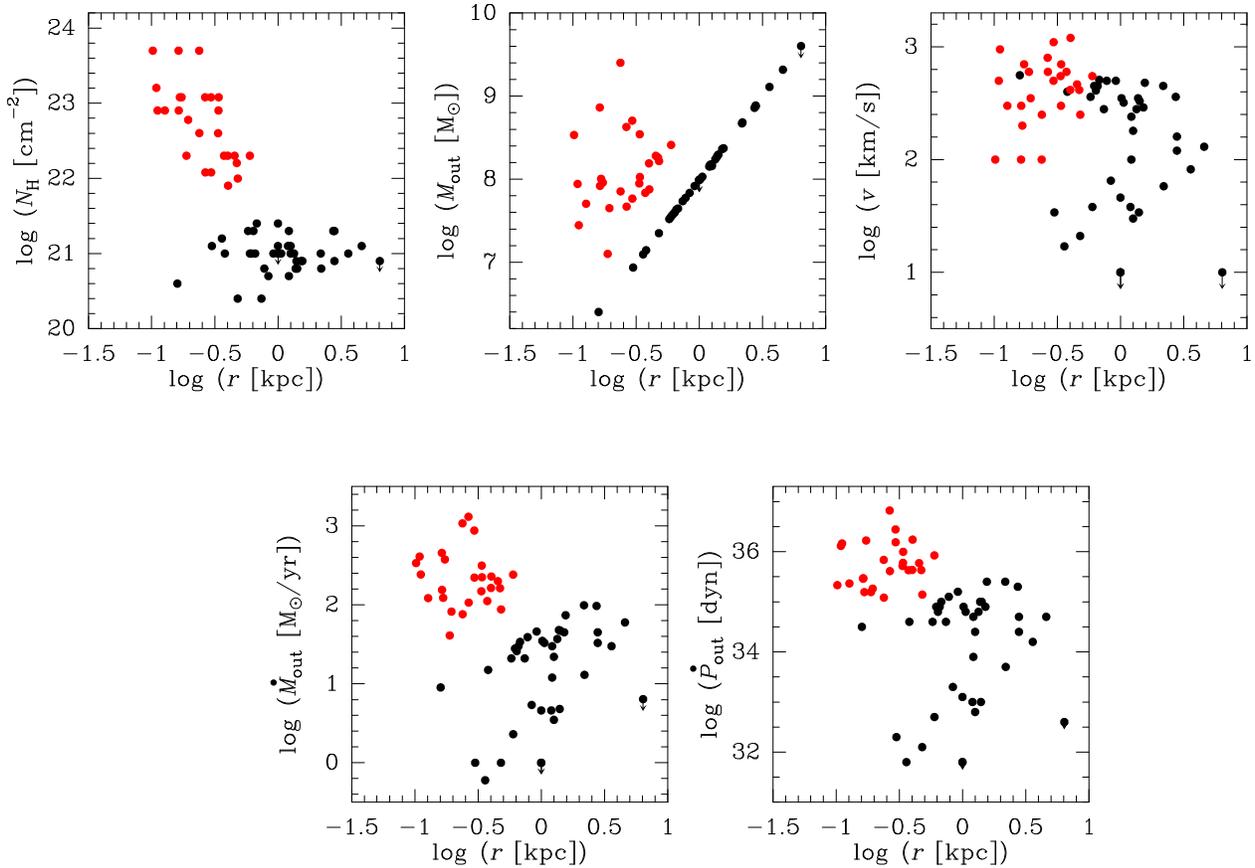}
\end{center}
\caption{Comparison of the outflow properties of local starburst
  galaxies inferred by \cite{hec15} (black symbols) with those of ULIRGs as
  inferred from OH (individual components, in red) in this work. 
  The column density 
  ($N_{\mathrm{H}}$), outflowing mass ($M_{\mathrm{out}}$), velocity ($v$),
  mass outflow rate ($\dot{M}_{\mathrm{out}}$), and momentum flux
  ($\dot{P}_{\mathrm{out}}$), are shown as a function of the estimated radius
  ($r$). 
}   
\label{heck1}
\end{figure*}

We compare here the outflow properties of local starburst galaxies inferred
by \cite{hec15} and of ULIRGs as derived in this work. Since these properties
are derived in basically the same way (i.e., eqs~\ref{mshell}, \ref{mdot_out},
and \ref{pdot_out}), while the starburst galaxies are analyzed in lines of
ions, the comparison is relevant to contrast different phases of the outflow
and to infer the main differences between pure starburst galaxies and ULIRGs
where the AGN plays an important role. 
Figure~\ref{heck1} shows the outflow column
density, mass, velocity, mass outflow rate, and momentum flux, as a function
of the outflow radius, for the starburst galaxies (black squares) and the
ULIRGs (individual components, in red).

A quantitative difference between the two phases is found in their column
densities, which are moderate in the ionized outflow phase of starburst
galaxies, $\sim10^{21}$ \cmd, but are much larger, $\gtrsim10^{22}$ \cmd, for
the molecular outflows in ULIRGs. 
The outflow size, which is adopted as twice the
size of the starburst in \cite{hec15}, is also larger than the extents we have
derived for the OH outflows. The outflowing gas mass (eq.~\ref{mshell}) is
then found roughly similar in both types of sources and phases. 
The derived mass outflow rates and
momentum fluxes are significantly higher for the OH outflows.

\section{C. Radiation pressure support of the highly excited nuclear
  structures} 
\label{radp}

\cite{tho15} consider two types of structures that are subject
to radiation pressure: a spherical shell and a cloud ensemble. We show here
that the pressure exerted by the radiation responsible for the {\it observed}
molecular excitation in ULIRGs is close to that required for support against
gravity, but probably not enough to overcome gravity and drive an outflow
under typical physical conditions in the nuclear regions of
ULIRGs. Starting with a shell at initial position $r_0$ and IR optical depth
$\tau_{\mathrm{IR}}>>1$, the Eddington luminosity to gas mass ratio is 
\citep[from eq. 8 by][]{tho15}
\begin{equation}
\frac{L_{\mathrm{EDD}}}{M_{\mathrm{gas}}}= 
\frac{4\pi \, c \, G}{f_g \, k_{\mathrm{IR}}} =
1.76\times10^4 
\times \left( \frac{0.15}{f_g} \right) 
\times \left( \frac{5\,\mathrm{cm^2/g}}{k_{\mathrm{IR}}} \right) 
\, L_{\odot}/M_{\odot},
\label{eq:leddmgas_sh}
\end{equation}
where $M_{\mathrm{gas}}$ denotes the gas mass within $r_0$, $f_g$ is the gas
mass fraction in that region, and $k_{\mathrm{IR}}$ (dependent on
$T_{\mathrm{dust}}$) is the IR opacity. \cite{com13} estimate that 
the average $f_g$ in local galaxies is $\lesssim0.10$ though with high
dispersion, and \cite{dow98} derived $f_g\sim0.17$ in the nuclear
regions of ULIRGs. We adopt a reference value of $f_g=0.15$. 

\begin{figure*}
\begin{center}
\includegraphics[angle=0,scale=.6]{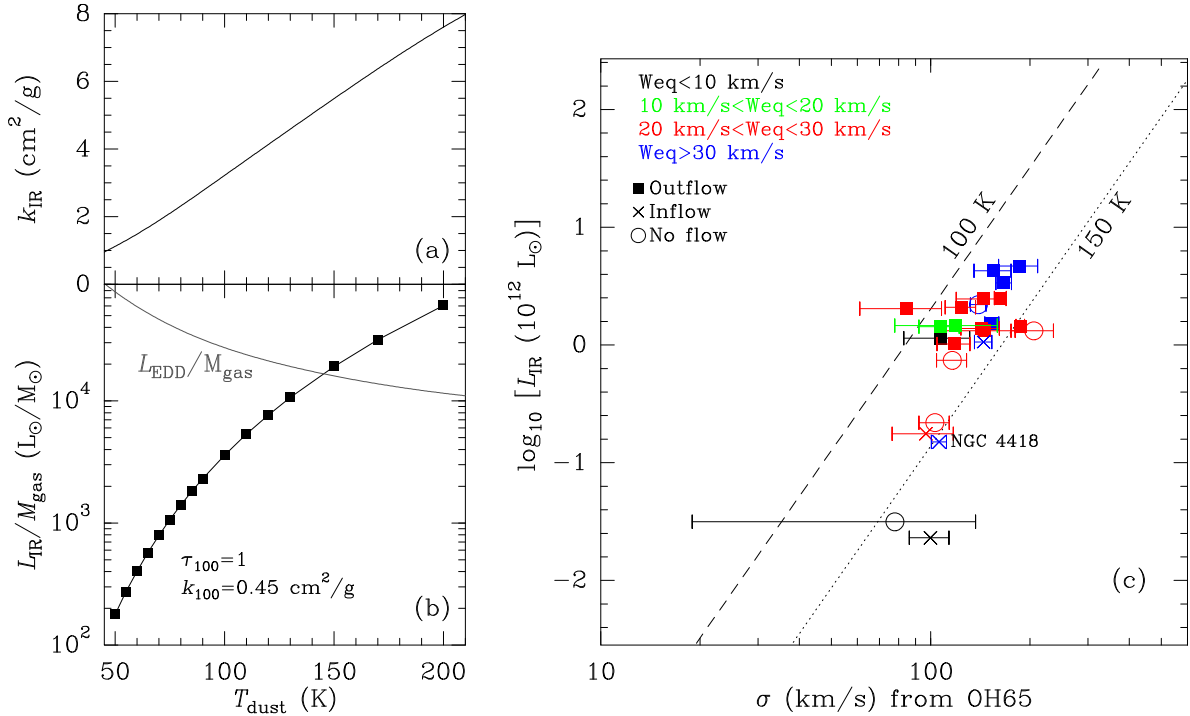}
\end{center}
\caption{Radiation pressure support in (U)LIRGs with measurable absorption in
  the high-lying OH65 doublet. a) The Planck-averaged mass-absorption
  coefficient ($k_{\mathrm{IR}}$) as a function of the dust temperature
  ($T_{\mathrm{dust}}$). b) The gray curve shows
  $L_{\mathrm{EDD}}/M_{\mathrm{gas}}$ as predicted by
  eq.~(\ref{eq:leddmgas_sh}) with a gas fraction of $f_{\mathrm{g}}=0.15$ and
  the $k_{\mathrm{IR}}$ profile of panel a. The solid line (squares) shows the
  calculated $L_{\mathrm{IR}}/M_{\mathrm{gas}}$ for a continuum optical depth
  at 100 $\mu$m of unity (GA15). For these optically thick sources we expect
  radiation pressure support for $T_{\mathrm{dust}}\sim140$ K. c) Infrared
  luminosities ($L_{\mathrm{IR}}$) as a function of the velocity dispersion 
  ($\sigma$) for all (U)LIRGs where OH65 is detected (GA15). $\sigma$ is
  calculated from Gaussian fits to the OH65 doublet. The dashed and dotted
  lines indicate the Eddington luminosities $L_{\mathrm{EDD}}$ predicted by
  eq.~(\ref{eq:ledd_sh}) for $T_{\mathrm{dust}}=100$ and 150 K, respectively.
}   
\label{radp_supp}
\end{figure*}

Figure~\ref{radp_supp}a shows the Planck-averaged opacity $k_{\mathrm{IR}}$ as
a function of $T_{\mathrm{dust}}$, based on the mass-absorption coefficient of
dust given in \cite{gon14b} and a nominal gas-to-dust ratio by mass of
$g/d=100$. The reference value given by \cite{tho15}, $k_{\mathrm{IR}}=5$
cm$^2$/g of gas, is obtained for $T_{\mathrm{dust}}=140$ K. Based on this
$k_{\mathrm{IR}}$ profile and on eq.~(\ref{eq:leddmgas_sh}), we plot in
Fig.~\ref{radp_supp}b $L_{\mathrm{EDD}}/M_{\mathrm{gas}}$ (gray curve) as a
function of $T_{\mathrm{dust}}$. We also show in Fig.~Fig.~\ref{radp_supp}b
the calculated values of $L_{\mathrm{IR}}/M_{\mathrm{gas}}$ (squares, solid
line) for a spherical source with $\tau_{100}=1$, the minimum value we inferred
from radiative transfer models to the OH65 doublet (GA15). To calculate 
$M_{\mathrm{gas}}$ from $\tau_{100}$, we adopt an absorption coefficient of
$0.45$ cm$^2$/g of gas at 100 $\mu$m \citep[based again on][]{gon14b}
and $g/d=100$.

The figure indicates that $T_{\mathrm{dust}}\approx140$ K is required for
radiation pressure support against gravity for $\tau_{100}=1$. 
For $\tau_{100}>1$, the $L_{\mathrm{IR}}/M_{\mathrm{gas}}$ values would be
lower because $L_{\mathrm{IR}}$ would remain nearly the same owing to the fact
that the far-IR emission is optically thick; hence, higher
$T_{\mathrm{dust}}$ would be required. 

In the far-IR, the values of $T_{\mathrm{dust}}$ as
directly inferred from the OH and H$_2$O excitation in buried nuclei
\citep[GA12, GA14, GA15,][]{fal15} are $100-150$ K, suggesting that these
nuclear regions are close to the regime of radiation pressure support against
gravity in the vertical direction \citep{tho05},
but may be insufficient to drive an outflow. The most excited
source in our original sample is the LIRG NGC~4418 where
$T_{\mathrm{dust}}\sim150$ K is inferred (GA12), but no molecular outflow is
detected in this source. 
We emphasize that in these sources that are optically thick in the
  far-IR, higher $T_{\mathrm{dust}}$ are expected to be prevalent
  within the cocoon of dust due to radiation trapping, but the relevant values
  for radiation pressure support at the scales probed by the far-IR are the
  surface values where the material is illuminated from basically one side.

An alternative way to illustrate this result is by substituting
$M_{\mathrm{gas}}=2\sigma^2 f_g r_0/G$ in eq.~(\ref{eq:leddmgas_sh}),
as is appropriate for an isothermal sphere, and 
$r_0\approx (L_{\mathrm{EDD}} / 4\pi \sigma_{\mathrm{SB}}T_{\mathrm{dust}}^4)^{1/2}$, 
where $\sigma_{\mathrm{SB}}$ is the Stefan-Boltzman constant, yielding
\begin{equation}
L_{\mathrm{EDD}}= 
\frac{16\pi \,
  c^2}{k_{\mathrm{IR}}^2\sigma_{\mathrm{SB}}\,T_{\mathrm{dust}}^4} \,\sigma^4= 
1.34\times10^{13}
\times \left( \frac{100 \, \mathrm{K}}{T_{\mathrm{dust}}} \right)^4 
\times \left( \frac{5\,\mathrm{cm^2/g}}{k_{\mathrm{IR}}} \right)^2
 \times \left( \frac{\sigma}{200\,\mathrm{km/s}} \right)^4
\, L_{\odot}.
\label{eq:ledd_sh}
\end{equation}
Figure~\ref{radp_supp}c shows the IR luminosity as a function of the velocity
dispersion $\sigma$, which is here measured from Gaussian fits to the 
high-lying OH65 doublet to filter the nuclear regions, for all
(U)LIRGs where OH65 is detected (GA15). The OH65 doublet is in part broadened
due to radial motions but $\sigma$ is not corrected for inclination effects
and thus the intrinsic values are higher than in Fig.~\ref{radp_supp}c; in
addition, the $L_{\mathrm{IR}}$ values are also upper limits because a
fraction of the 
luminosity most likely arise from more extended regions. The lines indicate
the $L_{\mathrm{EDD}}$ values for $T_{\mathrm{dust}}=100$ and 150 K as given by
eq.~(\ref{eq:ledd_sh}). The results are roughly consistent with 
Fig.~\ref{radp_supp}b, i.e., radiation pressure support in the vertical
direction of the region sampled
by OH65 is expected for $T_{\mathrm{dust}}=100-150$ K, coinciding with the
highest values of $T_{\mathrm{dust}}$ inferred from the high-lying OH and
H$_2$O lines in some sources, but this is insufficient to drive an outflow. 

Using now the cloud approach in \cite[][their eq. 38]{tho15}, the Eddington
luminosity to gas mass ratio is
\begin{equation}
\frac{L_{\mathrm{EDD}}}{M_{\mathrm{gas}}}= 
\frac{2\pi \, c \, G \, \mu \, m_{\mathrm{H}} \, N_{\mathrm{H}}}{f_g} =
1.03\times10^4 
\times \left( \frac{0.15}{f_g} \right) 
\times \left( \frac{N_{\mathrm{H}}}{10^{23}\,\mathrm{cm^{-2}}} \right) 
\, L_{\odot}/M_{\odot},
\label{eq:leddmgas_cl}
\end{equation}
where $N_{\mathrm{H}}$ is the column density of the cloud, which is assumed to
have $\tau_{\mathrm{IR}}\gtrsim1$. Writing again $M_{\mathrm{gas}}$ in terms
of $\sigma$ and $T_{\mathrm{dust}}$, we find
\begin{equation}
L_{\mathrm{EDD}}= 
\frac{4\pi \, c^2 \mu^2 \, m_{\mathrm{H}}^2 \, 
N_{\mathrm{H}}^2}{\sigma_{\mathrm{SB}}\,T_{\mathrm{dust}}^4}\,\sigma^4 =
4.60\times10^{12}
\times \left( \frac{100 \, \mathrm{K}}{T_{\mathrm{dust}}} \right)^4
\times \left( \frac{N_{\mathrm{H}}}{10^{23}\,\mathrm{cm^{-2}}} \right)^2 
\times \left( \frac{\sigma}{200\,\mathrm{km/s}} \right)^4
\, L_{\odot}.
\label{eq:ledd_cl}
\end{equation}
We have assumed here that the impinging radiation field on the cloud is UV
dominated \citep{tho15}, but we note that the outflows are observed in
absorption lines against a strong far-IR field and thus the radiation has been
already reprocessed and reemitted at far-IR wavelengths. Hence, 
eqs.~(\ref{eq:leddmgas_cl}) and (\ref{eq:ledd_cl}) probably
underestimate $L_{\mathrm{EDD}}/M_{\mathrm{gas}}$ and $L_{\mathrm{EDD}}$ by a
factor of $\sim2$.

\section{D. Velocity field} \label{vfield}

In GA14, we favored a decelerating velocity field in Mrk~231 based on three
arguments: $(i)$ OH84 and OH65 showed absorption at extreme velocities
($\lesssim-1000$ \kms), indicating the presence of high-velocity gas close to
the nuclear source of far-IR radiation, while the extended outflowing
component probed by OH119 and OH79 indicated expanding velocities
significantly lower, up to 
$\approx900$ \kms. $(ii)$ CO (1-0) and (2-1) emission in the line wings,
tracing an extended outflow, was found up to $\approx800$ \kms\ \citep{cic12},
significantly lower than OH. $(iii)$ In our non-clumpy spherical models,
  the line profiles are better fit with decelerating flows.

We examine here whether the decelerating flows are also applicable
to the other ULIRGs of our sample.
Fig.~\ref{vpaoh} indicates that, in all sample sources except
Mrk~273 and IRAS~08572+3915, the ground-state OH119 and OH79 doublets peak at
more blueshifted velocities than the excited OH84 and OH65. In addition,
Fig.~\ref{ohblue} shows that, in all sources except IRAS~03158+4227,
Mrk~231, IRAS~08572+3915, and IRAS~20100$-$4156 (and IRAS~09022$-$3615,
where OH65 is not detected), 
OH119 shows absorption well in excess of $500$ \kms\ but the OH65 absorption
is restricted to expanding velocities $v<500$ \kms. Since the very excited
OH65 is formed very close to the nuclear source of strong far-IR emission, and
OH119 is generally generated further from it, does this observational evidence
indicate that the gas is accelerated as it moves away from the nuclear region?

Not necessarily. Besides high \tdust,
significant absorption of OH65 requires high columns of gas. If the gas column
per unit of velocity decreases with increasing outflowing velocity, it is
expected that the high-velocity gas will not be detected in OH65 due to
insufficient columns, even if a significant fraction of it is located close to
the nuclear region. In addition, absorption and reemission in the ground-state
  doublets over large volumes can bring the intensity close to zero at
  central velocities, thus shifting the velocity of the peak absorption to
  high velocities, while this effect cannot happen in the excited lines
  (\S\ref{oh65vshift}). 

For instance, based on our models
of IRAS~14378$-$3651 (see Table~\ref{tbl-2}), the high-velocity
component has a moderate 
column of $N_{\mathrm{OH}}=3\times10^{16}$ \cmd\ and is not expected to
generate any OH65 absorption even though it is
located only $200-300$\,pc from the center. 
On the other hand, in IRAS~20551$-$4250 and
IRAS~10565+2448, both the high- and low-velocity components 
have similar spatial distributions. Although our model components also include
accelerated flows, the best fits are mostly found for decelerating or constant
velocity fields (except in two components), though this may be an artificial
effect of our non-clumpy approach. 

Our favored interpretation for the apparent higher spatial extent of the
high-velocity relative to the low-velocity gas in our sample galaxies is
not necessarily that the gas is smoothly accelerated from the
central region, but that the different doublets trace components with
different columns that have been accelerated to different velocities. The
escape velocity is given by 
$v_{\mathrm{esc}}\approx 200 \, \sqrt{M_9/R_{200}}$ \kms, where 
$M_9=M/10^9\,\mathrm{M_{\odot}}$ and $R_{200}=R/200\,\mathrm{pc}$, so that a
significant fraction of the low-velocity gas may not escape 
the central potential well of the host galaxy, while the high-velocity gas
will escape and be observed at larger distances.


\begin{thebibliography}{}
  \bibitem[Aalto et al.(2012)]{aal12} Aalto, S., Garc\'{\i}a-Burillo, S.,
    Muller, S., Winters, J. M., van der Werf, P., Henkel, C., 
    Costagliola, F., \& Neri, R. 2012, A\&A, 537, A44
  \bibitem[Aalto et al.(2015)]{aal15} Aalto, S., Garc\'{\i}a-Burillo, S.,
    Muller, S., Winters, J. M., Gonz\'alez-Alfonso, E., van der Werf, P.,
    Henkel, C., Costagliola, F., \& Neri, R. 2015, A\&A, 574, A85 
  \bibitem[Adams et al.(2001)]{ada01} Adams, F. C., Graff, D. S., \&
    Richstone, D. O. 2001, \apj, 551, L31
  \bibitem[Alatalo et al.(2011)]{ala11} Alatalo, K., Blitz, L., Young, L. M.,
    et al. 2011, \apj, 735, 88
  \bibitem[Alatalo et al.(2015)]{ala15} Alatalo, K., Lacy, M., Lanz,
    L., et al. 2015, \apj, 798, 31 
  \bibitem[Arav et al.(2013)]{ara13} Arav, N., Borguet, B., Chamberlain, C.,
    Edmonds, D., Danforth, C. 2013, MNRAS, 436, 3286
  \bibitem[Baldry et al.(2004)]{bal04} Baldry, I. K., Glazebrook, K.,
    Brinkmann, J., Ivezi\'c, Z., Lupton, R. H., Nichol, R. C., \& Szalay,
    A. S. 2004, ApJ, 600, 681
  \bibitem[Beifiori et al.(2012)]{bei12} Beifiori, A., Courteau, S., Corsini,
    E. M., \& Zhu, Y. 2012, MNRAS, 419, 2497
  \bibitem[Bolatto et al.(2013)]{bol13} Bolatto, A. D., Warren, S. R.; Leroy,
    A. K., et al. 2013, Nature, 499, 450
  \bibitem[Borguet et al.(2013)]{bor13} Borguet, B. C. J., Arav, N., Edmonds,
    D., Chamberlain, C., \& Benn, C. 2013, \apj, 762, 49
  \bibitem[Burkert \& Silk(2001)]{bur01} Burkert, A., \& Silk, J. 2001, \apj,
    554, L151
  \bibitem[Bushouse et al.(2002)]{bus02} Bushouse, H. A., Borne, K. D.,
    Colina, L., et al. 2002, \apjs, 138, 1
  \bibitem[Cicone et al.(2012)]{cic12} Cicone, C., Feruglio, C., Maiolino, R.,
    et al. 2012, A\&A, 543, A99
  \bibitem[Cicone et al.(2014)]{cic14} Cicone, C., Maiolino, R., Sturm, E., et
    al. 2014, A\&A, 562, A21
  \bibitem[Combes et al.(2013)]{com13} Combes, F., Garc\'{\i}a-Burillo, S.,
    Braine, J., et al. 2013, A\&A, 550, A41
  \bibitem[Costagliola et al.(2013)]{cos13} Costagliola, F., Aalto, S.,
    Sakamoto, K., Mart\'{\i}n, S., Beswick, R., Muller, S., Kl\"ockner,
    H.-R. 2013, A\&A, 556, A66
  \bibitem[Dasyra et al.(2006)]{das06} Dasyra, K. M., Tacconi, L. J., Davies,
    R. I., et al. 2006, \apj, 651, 835
  \bibitem[Davies et al.(2004)]{dav04} Davies, R. I., Tacconi, L. J., \&
    Genzel, R. 2004, \apj, 613, 781
  \bibitem[DeBuhr et al.(2012)]{buh12} DeBuhr, J., Quataert, E., \& Ma,
    C.-P. 2012, MNRAS, 420, 2221
  \bibitem[Dekel \& Birnboim(2006)]{dek06} Dekel, A., \& Birnboim, Y. 2006,
    MNRAS, 368, 2
  \bibitem[di Matteo et al.(2005)]{mat05} di Matteo, T., Springel, V., \&
    Hernquist, L. 2005, Nature, 433, 604
  \bibitem[D\'{\i}az-Santos et al.(2016)]{dia16} D\'{\i}az-Santos, T., Assef,
    R. J., Blain, A. W., Tsai, C.-W., Aravena, M., Eisenhardt, P., Wu, J.,
    Stern, D., \& Bridge, C. 2016, \apj, 816, L6
  \bibitem[Downes et al.(1993)]{dow93} Downes, D., Solomon, P. M., \& Radford,
    S. J. E. 1993, \apj, 414, L13
  \bibitem[Downes \&Solomon(1998)]{dow98} Downes, D., \& Solomon, P. M. 1998,
    \apj, 507, 615
  \bibitem[Duc et al.(1997)]{duc97} Duc, P.-A., Mirabel, I. F., \& Maza,
    J. 1997, A\&AS, 124, 533 
  \bibitem[Efstathiou et al.(2014)]{efs14} Efstathiou, A., Pearson, C.,
    Farrah, D., et al. 2014, MNRAS, 437, L16
  \bibitem[Evans et al.(2000)]{eva00} Evans, A. S., Surace, J. A., \&
    Mazzarella, J. M. 2000, \apj, 529, L85
  \bibitem[Evans et al.(2002)]{eva02} Evans, A. S., Mazzarella, J. M., Surace,
    J. A., \& Sanders, D. B. 2002, \apj, 580, 749
  \bibitem[Fabian \& Iwasawa(1999)]{fab99b} Fabian, A. C., \& Iwasawa,
    K. 1999, MNRAS, 303, L34
  \bibitem[Fabian(1999)]{fab99} Fabian, A. C. 1999, MNRAS, 308, L39
  \bibitem[Fabian(2012)]{fab12} Fabian, A. C. 2012, ARA\&A, 50, 455
  \bibitem[Falstad et al.(2015)]{fal15} Falstad, N., Gonz\'alez-Alfonso, E.,
    Aalto, S., van der Werf, P. P., Fischer, J., Veilleux, S., Mel\'endez, M.,
    Farrah, D., \& Smith, H. A. 2015, A\&A, 580, A52  
  \bibitem[Falstad et al.(2017)]{fal17} Falstad, N., Gonz\'alez-Alfonso, E.,
    Aalto, S., \& Fischer, J. 2017, A\&A, 597, A105
  \bibitem[Farrah et al.(2007)]{far07} Farrah, D., Bernard-Salas, J., Spoon,
    H. W. W., et al. 2007, \apj, 667, 149  
  \bibitem[Farrah et al.(2012)]{far12} Farrah, D., Urrutia, T., Lacy, M., et
    al. 2012, \apj, 745, 178
  \bibitem[Farrah et al.(2013)]{far13} Farrah, D., Lebouteiller, V., Spoon,
    H. W. W., et al. 2013, \apj, 776, 38 
  \bibitem[Faucher-Gigu\`ere \& Quataert(2012)]{fau12} Faucher-Gigu\`ere,
    C.-A., \& Quataert, E. 2012, MNRAS, 425, 605
  \bibitem[Ferrarese \& Merritt(2000)]{fer00} Ferrarese, L., \& Merritt,
    D. 2000, ApJ, 539, L9
  \bibitem[Ferrarese \& Ford(2005)]{fer05} Ferrarese, L., \& Ford, H. 2005,
    Space Science Reviews, 116, 523 
  \bibitem[Feruglio et al.(2010)]{fer10} Feruglio, C., Maiolino, R.,
    Piconcelli, E., Menci, N., Aussel, H., Lamastra, A., \& Fiore, F.
    2010, A\&A, 518, L155 
  \bibitem[Feruglio et al.(2015)]{fer15} Feruglio, C., Fiore, F., Carniani,
    S., et al. 2015, A\&A, 583, A99 
  \bibitem[Fischer et al.(2010)]{fis10} Fischer, J., Sturm, E.,
    Gonz\'alez-Alfonso, et al. 2010, A\&A, 518, L41
  \bibitem[Franceschini et al.(2003)]{fra03} Franceschini, A., Braito, V.,
    Persic, M., et al. 2003, MNRAS, 343, 1181
  \bibitem[Garc\'{\i}a-Burillo et al.(2015)]{gar15} Garc\'{\i}a-Burillo, S.,
    Combes, F., Usero, A., et al. 2015, A\&A, 580, A35
  \bibitem[Garc\'{\i}a-Mar\'{\i}n et al.(2006)]{gar06} 
    Garc\'{\i}a-Mar\'{\i}n, M., Colina, L., Arribas, S., Alonso-Herrero, A.,
    \& Mediavilla, E. 2006, \apj, 650, 850
  \bibitem[Gebhardt et al.(2000)]{geb00} Gebhardt, K., Bender, R., Bower, G.,
    et al. 2000, ApJ, 539, L13
  \bibitem[Genzel et al.(1998)]{gen98} Genzel, R., Lutz, D.; Sturm, E., et
    al. 1998, ApJ, 498, 579 
  \bibitem[Genzel et al.(2001)]{gen01} Genzel, R., Tacconi, L. J., Rigopoulou,
    D., Lutz, D., \& Tecza, M. 2001, \apj, 563, 527
  \bibitem[Goicoechea \& Cernicharo(2002)]{goi02} Goicoechea, J. R. \&
    Cernicharo, J. 2002, \apj, 576, L77
  \bibitem[Goicoechea et al.(2006)]{goi06} Goicoechea, J. R., Cernicharo, J.,
    Lerate, M. R., Daniel, F., Barlow, M. J., Swinyard, B. M., Lim, T. L.,
    Viti, S., \& Yates, J. 2006, \apj, 641, L49
  \bibitem[Goicoechea et al.(2011)]{goi11} Goicoechea, J. R., Joblin, C.,
    Contursi, A., Bern\'e, O., Cernicharo, J., Gerin, M., Le Bourlot, J.,
    Bergin, E. A., Bell, T. A., \& R\"ollig, M. 2011, A\&A, 530, L16
  \bibitem[Gonz\'alez-Alfonso \& Cernicharo(1997)]{gon97} Gonz\'alez-Alfonso,
    E., \& Cernicharo, J. 1997, A\&A, 322, 938
  \bibitem[Gonz\'alez-Alfonso \& Cernicharo(1999)]{gon99} Gonz\'alez-Alfonso,
    E., \& Cernicharo, J. 1999, \apj, 525, 845
  \bibitem[Gonz\'alez-Alfonso et al.(2012)]{gon12} Gonz\'alez-Alfonso,
    E., Fischer, J., Graci\'a-Carpio, J., et al. 2012, A\&A, 541, A4
  \bibitem[Gonz\'alez-Alfonso et al.(2013)]{gon13} Gonz\'alez-Alfonso,
    E., Fischer, J., Bruderer, S., et al. 2013, A\&A, 550, A25
  \bibitem[Gonz\'alez-Alfonso et al.(2014a)]{gon14a} Gonz\'alez-Alfonso,
    E., Fischer, J., Graci\'a-Carpio, J., et al. 2014a, A\&A, 561, A27 (GA14)
  \bibitem[Gonz\'alez-Alfonso et al.(2014b)]{gon14b} Gonz\'alez-Alfonso,
    E., Fischer, J., Aalto, S., \& Falstad, N. 2014b, A\&A, 567, A91 
  \bibitem[Gonz\'alez-Alfonso et al.(2015)]{gon15} Gonz\'alez-Alfonso,
    E., Fischer, J., Sturm, E., et al. 2015, \apj, 800, 69 
  \bibitem[Graci\'a-Carpio et al.(2011)]{gra11} Graci\'a-Carpio, J., Sturm,
    E., Hailey-Dunsheath, S., et al. 2011, \apj, 728, L7
  \bibitem[Graham et al.(2001)]{gra01} Graham, A. W., Erwin, P., Caon, N., \&
    Trujillo, I. 2001, \apj, 563, L11
  \bibitem[Graham \& Driver(2007)]{gra07} Graham, A. W. \& Driver, S. P. 2007,
    \apj, 655, 77
  \bibitem[Guillard et al.(2015)]{gui15} Guillard, P., Boulanger, F., Lehnert,
    M. D., Pineau des For\^ets, G., Combes, F., Falgarone, E., \&
    Bernard-Salas, J. 2015, A\&A, 574, A32
  \bibitem[Haan et al.(2011)]{haa11}  Haan, S., Surace, J. A., Armus, L., et
    al. 2011, \aj, 141, 100 
  \bibitem[Harrison et al.(2014)]{har14} Harrison, C. M., Alexander, D. M.,
    Mullaney, J. R., \& Swinbank, A. M. 2014, MNRAS, 441, 3306
  \bibitem[Harvey-Smith et al.(2016)]{har16} Harvey-Smith, L., Allison, J. R.,
    Green, J. A., et al. 2016, MNRAS, 460, 2180
  \bibitem[Heckman et al.(1990)]{hec90} Heckman, T. M., Armus, L., \& Miley,
    G. K. 1990, ApJS, 74, 833
  \bibitem[Heckman et al.(2015)]{hec15} Heckman, T. M., Alexandroff, R. M.,
    Borthakur, S., Overzier, R., \& Leitherer, C. 2015, \apj, 809, 147
  \bibitem[Hopkins et al.(2006a)]{hop06a} Hopkins, P. F.; Hernquist, L., Cox,
    T. J., Robertson, B., \& Springel, V. 2006a, ApJS, 163, 50
  \bibitem[Hopkins et al.(2006b)]{hop06b} Hopkins, P. F.; Hernquist, L., Cox,
    T. J., di Matteo, T., Robertson, B., \& Springel, V. 2006b, ApJS, 163, 1
  \bibitem[Hopkins et al.(2009)]{hop09} Hopkins, P. F., Cox, T. J., Younger,
    J. D., \& Hernquist, L. 2009, \apj, 691, 1168
  \bibitem[Imanishi et al.(2008)]{ima08} Imanishi, M., Nakagawa, T., Ohyama,
    Y., Shirahata, M., Wada, T., Onaka, T., \& Oi, N. 2008, PASJ, 60, 489 
  \bibitem[Imanishi et al.(2016)]{ima16} Imanishi, M., Nakanishi, K., \&
    Izumi, T. 2016, AJ, 152, 218 
  \bibitem[Ishibashi \& Fabian(2015)]{ish15} Ishibashi, W., \& Fabian,
    A. C. 2015, MNRAS, 451, 93
  \bibitem[Iwasawa et al.(2011)]{iwa11} Iwasawa, K., Sanders, D. B., Teng,
    S. H., et al. 2011, A\&A, 529, A106
  \bibitem[Jahnke \& Macci\`o(2011)]{jah11} Jahnke, K., \& Macci\`o,
    A. V. 2011, \apj, 734, 92
  \bibitem[Janssen et al.(2016)]{jan16} Janssen, A. W., Christopher, N.,
    Sturm, E., et al. 2016, \apj, 822, 43
  \bibitem[Khochfar \& Ostriker(2008)]{kho08} Khochfar, S., \&  Ostriker,
    J. 2008, \apj, 680, 54
  \bibitem[Kim et al.(1998)]{kim98} Kim, D.-C., Veilleux, S., \& Sanders,
    D. B. 1998, \apj, 508, 627
  \bibitem[King(2003)]{kin03} King, A. 2003, \apj, 596, L27 
  \bibitem[King(2005)]{kin05} King, A. 2005, \apj, 635, L121 
  \bibitem[King \& Pounds(2015)]{kin15} King, A., \& Pounds, K. 2015, ARA\&A,
    53, 115
  \bibitem[Lang et al.(2014)]{lan14} Lang, P., Wuyts, S., Somerville, R. S.,
    et al. 2014, \apj, 788, 11
  \bibitem[Lee et al.(2011)]{lee11} Lee, J. C., Hwang, H. S., Lee, M. G., Kim,
    M., Kim, S. C. 2011, MNRAS, 414, 702
  \bibitem[Lee et al.(2012)]{lee12} Lee, J. C., Hwang, H. S., Lee, M. G., Kim,
    M., Lee, J. H. 2012, \apj, 756, 95
  \bibitem[Leitherer et al.(1999)]{lei99} Leitherer, C., Schaerer, D.,
    Goldader, J. D., Delgado, R. M., Robert, C., Kune, D. F., de Mello, D. F.,
    Devost, D., Heckman, T. M. ApJS, 123, 3
  \bibitem[Lindberg et al.(2016)]{lin16} Lindberg, J. E., Aalto, S., Muller,
    S., et al. 2016, A\&A, 587, A15
  \bibitem[L\'{\i}pari et al.(2005)]{lip05} L\'{\i}pari, S., Terlevich, R.,
    Zheng, W., Garc\'{\i}a-Lorenzo, B., S\'anchez, S. F., \& Bergmann,
    M. 2005, MNRAS, 360, 416
  \bibitem[L\'{\i}pari et al.(2009)]{lip09} L\'{\i}pari, S., S\'anchez, S. F.,
    Bergmann, M., et al. 2009, MNRAS, 392, 1295
  \bibitem[Magorrian et al.(1998)]{mag98} Magorrian, J., Tremaine, S.,
    Richstone, D. et al. 1998, AJ, 115, 2285
  \bibitem[Maiolino et al.(2012)]{mai12} Maiolino, R., Gallerani, S., Neri,
    R., et al. 2012, MNRAS, 425, 66
  \bibitem[Marconi \& Hunt(2003)]{mar03} Marconi, A., \& Hunt, L. K. 2003,
    \apj, 589, L21
  \bibitem[Mart\'{\i}n et al.(2016)]{mar16} Mart\'{\i}n, S., Aalto, S.;
    Sakamoto, K., et al. 2016, A\&A, 590, A25
  \bibitem[Martin \& Soto(2016)]{martin16} Martin, C. L., \& Soto, K. 2016,
    \apj, 819, 49
  \bibitem[Martizzi et al.(2016)]{mart16} Martizzi, D., Fielding, D.,
    Faucher-Gigu\`ere, C.-A., \& Quataert, E. 2016, MNRAS, 459, 2311
  \bibitem[Meijerink et al.(2011)]{mei11} Meijerink, R., Spaans, M., Loenen,
    A. F., \& van der Werf, P. P. 2011, A\&A, 525, A119
  \bibitem[Meusinger et al.(2001)]{meu01} Meusinger, H., Stecklum, B., Theis,
    C., \& Brunzendorf, J. 2001, A\&A, 379, 845
  \bibitem[Murphy et al.(1996)]{mur96} Murphy, T. W. Jr., Armus, L., Matthews,
    K., Soifer, B. T., Mazzarella, J. M., Shupe, D. L., Strauss, M. A., \&
    Neugebauer, G. 1996, \aj, 111, 1025
  \bibitem[Murray et al.(2005)]{mur05} Murray, N., Quataert, E., \& Thompson,
    T. A. 2005, ApJ, 618, 569
  \bibitem[Nardini et al.(2010)]{nar10} Nardini, E., Risaliti, G., Watabe, Y.,
    Salvati, M., \& Sani, E. 2010, MNRAS, 405, 2505
  \bibitem[Peng(2007)]{pen07} Peng, C. Y. 2007, \apj, 671, 1098
  \bibitem[Peng et al.(2015)]{pen15} Peng, Y., Maiolino, R., \& Cochrane, R.
    2015, Nature, 521, 192
  \bibitem[Pereira-Santaella et al.(2016)]{per16} Pereira-Santaella, M.,
    Colina, L., Garc\'{\i}a-Burillo, S., et al. 2016, A\&A, 594, A81
  \bibitem[Pilbratt et al.(2010)]{pil10} Pilbratt, G. L.; Riedinger, J. R.;
    Passvogel, T., et al. 2010, A\&A, 518, L1
  \bibitem[Poglitsch et al.(2010)]{pog10} Poglitsch, A., Waelkens, C., Geis,
    N., et al. 2010, A\&A, 518, L2
  \bibitem[Privon et al.(2016)]{pri16} Privon, G. C., Aalto, S., Falstad, N.,
    et al. 2016, \apj, in press (arXiv:1612.04401)
  \bibitem[Risaliti et al.(2000)]{ris00} Risaliti, G., Gilli, R., Maiolino,
    R., \& Salvati, M. 2000, A\&A, 357, 13
  \bibitem[Rodr\'{\i}guez Zaur\'{\i}n et al.(2013)]{rod13} Rodr\'{\i}guez
    Zaur\'{\i}n, J., Tadhunter, C. N., Rose, M., \& Holt, J. 2013, MNRAS, 432,
    138 
  \bibitem[Roth et al.(2012)]{rot12} Roth, N., Kasen, D., Hopkins, P. F., \&
    Quataert, E. 2012, \apj, 759, 36 
  \bibitem[Rupke et al.(2002)]{rup02} Rupke, D. S., Veilleux, S., \&
    Sanders, D. B. 2002, \apj, 570,588
  \bibitem[Rupke et al.(2005a)]{rup05a} Rupke, D. S., Veilleux, S., \&
    Sanders, D. B. 2005a, \apj, 632, 751
  \bibitem[Rupke et al.(2005b)]{rup05b} Rupke, D. S., Veilleux, S., \&
    Sanders, D. B. 2005b, \apjs, 160, 87
  \bibitem[Rupke et al.(2005c)]{rup05c} Rupke, D. S., Veilleux, S., \&
    Sanders, D. B. 2005c, \apjs, 160, 115
  \bibitem[Rupke \& Veilleux(2013a)]{rup13a} Rupke, D. S., \& Veilleux,
    S. 2013a, \apj, 768, 75
  \bibitem[Rupke \& Veilleux(2013b)]{rup13b} Rupke, D. S., \& Veilleux,
    S. 2013b, \apj, 775, L15
  \bibitem[Sakamoto et al.(2009)]{sak09} Sakamoto, K., Aalto, S., Wilner,
    D. J., et al. 2009, ApJL, 700, L104 
  \bibitem[Sakamoto et al.(2013)]{sak13} Sakamoto, K., Aalto, S., Costagliola,
    F., Mart\'{\i}n, S., Ohyama, Y., Wiedner, M. C., \& Wilner, D. J. 2013,
    \apj, 764, 42
  \bibitem[Sanders(1989)]{san89} Sanders, R. H. 1989, IAUS, 136, 77
  \bibitem[Sanders et al.(1988)]{san88} Sanders, D. B., Soifer, B. T., Elias,
    J. H., Madore, B. F., Matthews, K., Neugebauer, G., \& Scoville,
    N. Z. 1988, \apj, 325, 74
  \bibitem[Sanders et al.(2003)]{san03} Sanders, D. B., Mazzarella, J. M.,
    Kim, D.-C., Surace, J. A., \& Soifer, B. T. 2003, AJ, 126, 1607
  \bibitem[Sani \& Nardini(2012)]{san12} Sani, E., \& Nardini, E. 2012,
    AdAst, 783451
  \bibitem[Schawinski et al.(2014)]{sch14} Schawinski, K., Urry, C. M.,
    Simmons, B. D., et al. 2014, MNRAS, 440, 889
  \bibitem[Scoville et al.(2000)]{sco00} Scoville, N. Z., Evans, A. S.,
    Thompson, R., Rieke, M., Hines, D. C., Low, F. J., Dinshaw, N., Surace,
    J. A., \& Armus, L. 2000, \aj, 119, 991 
  \bibitem[Severgnini \& Risaliti(2000)]{sev00} Severgnini, P., \& Risaliti,
    G. 2000, Memorie della Societ\`a Astronomica Italiana, 72, 63
  \bibitem[Shankar et al.(2016)]{sha16} Shankar, F., Bernardi, M., Sheth,
    R. K., et al. 2016, MNRAS, 460, 3119
  \bibitem[Shih \& Rupke(2010)]{shi10} Shih, H.-Y., \& Rupke, D. S. N. 2010,
    \apj, 724, 1430	
  \bibitem[Shirahata et al.(2013)]{shi13} Shirahata, M., Nakagawa, T., Usuda,
    T., Goto, M., Suto, H., Geballe, T. R. 2013, PASJ, 65, 5
  \bibitem[Silk \& Rees(1998)]{sil98} Silk, J., \& Rees, M. J. 1998, A\&A,
    331, L1
  \bibitem[Solomon et al.(1997)]{sol97} Solomon, P. M., Downes, D., Radford,
    S. J. E., \& Barrett, J. W. 1997, \apj, 478, 144
  \bibitem[Spoon \& Holt(2009)]{spo09} Spoon, H. W. W.,  \& Holt, J. 2009,
    \apj, 702, L42
  \bibitem[Spoon et al.(2013)]{spo13} Spoon, H. W. W.,  Farrah, D.,
  Lebouteiller, V., et al. 2013, ApJ, 775, 127 
  \bibitem[Springel et al.(2005)]{spr05} Springel, V., Di Matteo, T., \&
    Hernquist, L. 2005, ApJ, 620, L79
  \bibitem[Stern et al.(2016)]{ste16} Stern, J., Faucher-Gigu\`ere, C.-A.,
    Zakamska, N. L., \& Hennawi, J. F. 2016, ApJ, 819, 130
  \bibitem[Sternberg \& Dalgarno(1995)]{ste95} Sternberg, A. \& Dalgarno,
    A. 1995, \apjs, 99, 565
  \bibitem[Stierwalt et al.(2013)]{sti13} Stierwalt, S., Armus, L., Surace,
    J. A., et al. 2013, \apjs, 206, 1
  \bibitem[Stone et al.(2016)]{sto16} Stone, M., Veilleux, S., Mel\'endez, M.,
    et al. 2016, \apj, 826, 111
  \bibitem[Strateva et al.(2001)]{str01} Strateva, I., Ivezi\'c, Z., Knapp,
    G., et al. 2001, AJ, 122, 1861
  \bibitem[Sturm et al.(2011)]{stu11} Sturm, E., Gonz\'alez-Alfonso, E.,
    Veilleux, S., et al. 2011, ApJ, 733, L16 
  \bibitem[Surace et al.(2004)]{sur04} Surace, J. A., Sanders, D. B., \&
    Mazzarella, J. M. 2004, AJ, 127, 3235
  \bibitem[Tacchella et al.(2015)]{tac15} Tacchella, S., Carollo, C. M.,
    Renzini, A., et al. 2015, Science, 348, 314
  \bibitem[Tacconi et al.(2002)]{tac02} Tacconi, L. J., Genzel, R., Lutz, D.,
    Rigopoulou, D., Baker, A. J., Iserlohe, C., \& Tecza, M. 2002, \apj, 580,
    73 
  \bibitem[Teng et al.(2015)]{ten15} Teng, S. H.; Rigby, J. R.; Stern, D., et
    al. 2015, \apj, 814, 56
  \bibitem[Thompson et al.(2005)]{tho05} Thompson, T. A., 
    Quataert, E., \& Murray, N. 2005, \apj, 630, 167 
  \bibitem[Thompson et al.(2015)]{tho15} Thompson, T. A., Fabian, A. C.,
    Quataert, E., \& Murray, N. 2015, MNRAS, 449, 147
  \bibitem[Tombesi et al.(2015)]{tom15} Tombesi, F., Mel\'endez, M., Veilleux,
    S., Reeves, J. N., Gonz\'alez-Alfonso, E., \& Reynolds, C. S. 2015,
    Nature, 519, 436
  \bibitem[Tremaine et al.(2002)]{tre02} Tremaine, S., Gebhardt, K., Bender, 
    R., et al. 2002, ApJ, 574,740
  \bibitem[Tunnard et al.(2015)]{tun15} Tunnard, R., Greve, T. R.,
    Garc\'{\i}a-Burillo, S., et al. 2015, \apj, 800, 25 
  \bibitem[Veilleux et al.(1995)]{vei95} Veilleux, S., Kim, D.-C., Sanders,
    D. B., Mazzarella, J. M., \& Soifer, B. T. 1995, \apjs, 98, 171
  \bibitem[Veilleux et al.(1997)]{vei97} Veilleux, S., Sanders, D. B., 
    \& Kim, D.-C. 1997, \apj, 484, 92
  \bibitem[Veilleux et al.(1999)]{vei99} Veilleux, S., Kim, D.-C., \& Sanders,
    D. B. 1999, \apj, 522, 113
  \bibitem[Veilleux et al.(2005)]{vei05} Veilleux, S., Cecil, G., \&
    Bland-Hawthorn, J. 2005, ARA\&A, 43, 769
  \bibitem[Veilleux et al.(2009)]{vei09} Veilleux, S., Rupke, D. S. N., Kim,
    D.-C., et al. 2009, \apjs, 182, 628  
  \bibitem[Veilleux et al.(2013)]{vei13} Veilleux, S., Mel\'endez, M.; Sturm,
    E., et al. 2013, ApJ, 776, 27 
  \bibitem[V\'eron-Cetty \& V\'eron(2006)]{ver06} V\'eron-Cetty,
    M.-P. \& V\'eron, P. 2006, A\&A, 455, 773 
  \bibitem[Weaver et al.(1977)]{wea77} Weaver, R., McCray, R., Castor, J.,
    Shapiro, P., \& Moore, R. 1977, \apj, 218, 377 
  \bibitem[Zubovas \& King(2012)]{zub12} Zubovas, K., \& King, A. 2012, \apj,
    745, L34 
  \bibitem[Zubovas \& Nayakshin(2014)]{zub14} Zubovas, K., \& Nayakshin,
    S. 2014, MNRAS, 440, 2625
\end{thebibliography}
\end{document}